\documentclass[%
 aip,
 pof,
 amsmath,amssymb,
 reprint,%
]{revtex4-2} 

\usepackage{blindtext}
\usepackage{graphicx}
\usepackage{caption}
\usepackage{subcaption}
\usepackage{dcolumn}
\usepackage{bm}
\usepackage[utf8]{inputenc}
\usepackage[T1]{fontenc}
\usepackage{mathptmx}
\usepackage{etoolbox}
\usepackage{upgreek} 
\usepackage[normalem]{ulem} 

\usepackage{amsmath}
\usepackage{amsthm}
\usepackage{amssymb}
\usepackage{esint}
\usepackage[linkcolor=blue,colorlinks=true]{hyperref}
\usepackage{multirow}

\makeatletter 

\newenvironment{newnumbering}[1][Alph]
 {
  \def\jr@counter{#1}%
  \jr@setup@numbering{#1}{1}%
 }
 {%
  \setcounter{jr@\jr@counter @equation}{\value{equation}}%
  \setcounter{equation}{\value{jr@equation@save}}%
  \ignorespacesafterend
 }
\newenvironment{newnumbering*}[1][Alph]
 {%
  \def\jr@counter{#1}%
  \jr@setup@numbering{#1}{0}%
 }
 {%
  \setcounter{jr@\jr@counter @equation}{\value{equation}}%
  \setcounter{equation}{\value{jr@equation@save}}%
  \ignorespacesafterend
 }

\newcounter{jr@equation@save}
\newcommand{\jr@setup@numbering}[2]{%
  \@ifundefined{c@jr@#1@equation}{\newcounter{jr@#1@equation}}{}%
  \setcounter{jr@equation@save}{\value{equation}}%
  \setcounter{equation}{%
    \ifnum#2>0
      \value{jr@#1@equation}%
    \else
      0%
    \fi
  }%
  \renewcommand{\theequation}{\csname#1\endcsname{equation}}%
}
\makeatother

\makeatletter
\def\@email#1#2{%
 \endgroup
 \patchcmd{\titleblock@produce}
  {\frontmatter@RRAPformat}
  {\frontmatter@RRAPformat{\produce@RRAP{*#1\href{mailto:#2}{#2}}}\frontmatter@RRAPformat}
  {}{}
}%
\makeatother

\begin{document}

\title[Variational principles for the hydrodynamics of the classical one-component plasma]{Variational principles for the hydrodynamics of the classical one-component plasma}
\author{Daniels Krimans}
\email{danielskrimans@physics.ucla.edu}
\affiliation{Physics and Astronomy Department, University of California Los Angeles, Los Angeles, California 90095, USA}
\author{Seth Putterman}
\affiliation{Physics and Astronomy Department, University of California Los Angeles, Los Angeles, California 90095, USA}

\date{\today}

\begin{abstract} 
Hydrodynamic equations for a one-component plasma are derived as a unification of the Euler equations with long-range Coulomb interaction.
By using a variational principle, these equations self-consistently unify thermodynamics, dispersion laws, nonlinear motion, and conservation laws. 
In the moderate and strong coupling limits, it is argued that these equations work down to the length scale of the interparticle spacing. 
Use of a variational principle also ensures that closure is achieved self-consistently.
Hydrodynamic equations are evaluated in both the Eulerian frame, where the fluid variables depend on the position in the laboratory, and the Lagrangian frame, where they depend on the position in some reference state, such as the initial position. 
Each frame has its advantages and our final theory combines elements of both. 
The properties of longitudinal and transverse dispersion laws are calculated for the hydrodynamic equations. 
A simple step function approximation for the pair distribution function enables simple calculations that reveal the structure of the equations of motion. 
The obtained dispersion laws are compared to molecular dynamics simulations and the theory of quasilocalized charge approximation. 
The action, which gives excellent agreement for both longitudinal and transverse dispersion laws for a wide range of coupling strengths, is elucidated. 
Agreement with numerical experiments shows that such a hydrodynamic approach can be used to accurately describe a one-component plasma at very small length scales comparable to the average interparticle spacing. 
The validity of this approach suggests considering nonlinear flows and other systems with long-range interactions in the future.
\newline
\newline
\textit{This article may be downloaded for personal use only. Any other use requires prior permission of the author and AIP Publishing. This article appeared in D. Krimans and S. Putterman, "Variational principles for the hydrodynamics of the classical one-component plasma," Phys. Fluids }\textbf{36}\textit{, 037131 (2024) and may be found at https://doi.org/10.1063/5.0194352.}
\end{abstract}

\maketitle

\section{Introduction}
The purpose of this paper is to apply a variational principle to achieve hydrodynamic equations for a classical system with a long-range interaction. A well-known example of such a system is the one-component plasma (OCP), which consists of particles of charge $q$ that interact pairwise using the Coulomb potential $\phi(r) = q^2/4 \pi \varepsilon_0 r$, where $r$ is the distance between them, in the presence of a stationary neutralizing background. For the OCP, the strength of the coupling is determined by the dimensionless ratio of the average potential energy to the thermal kinetic energy \cite{doi:10.1063/1.1727895, PhysRevA.8.3096}, $\Gamma = q^2 / 4 \pi \varepsilon_0 a k_B T$, where $T$ is the temperature and $a$ is the average interparticle spacing so that the number density is given by $n = 3/4 \pi a^3$. We will be especially interested in cases where the coupling between particles is significant, and thus, $\Gamma \geq 1$. \par

Such a system is important for understanding white dwarfs in astrophysics \cite{1968ApJ151227V, VanHorn_2019}. The linear regime can be probed using scattering experiments on molten salts \cite{DEMMEL200598, Demmel_2021}, as molten salts are believed to be strongly coupled plasmas \cite{PhysRevA.11.2111}. Nonlinear effects can be experimentally tested through the expansion of plasma during laser breakdown \cite{Bataller2019DynamicsOS}, the expansion of ultracold neutral plasma \cite{KILLIAN200777}, and by considering photoelectron sources, where the emitted electrons are strongly coupled right after emission \cite{Maxson_2013}. Additionally, the OCP can be thought of as a limit of more complicated Yukawa fluids, where particles interact using the Yukawa potential. Such Yukawa fluids can be experimentally realized as dusty plasmas \cite{RevModPhys.81.1353}. \par

If you are describing turbulence in the ocean, would you consider solving coupled equations of motion for the distribution functions of the individual water molecules or would you use hydrodynamics which is a contracted and closed description in terms of a few continuous variables, for example\cite{Landau1987Fluid}, the number density $n$, specific entropy $s$ and three components of velocity $\vec{v}$? With this perspective in mind, we have developed a contracted hydrodynamic theory that is suitable for describing the complex nonlinear motion of the OCP. \par

In analogy with the Euler and Navier-Stokes equations, we will present a hydrodynamic theory of the OCP that has the same set of hydrodynamic variables. The theory is structurally different because it is in essence nonlocal due to the appearance of the pair distribution function $g(r)$ that describes the probability of two particles being distance $r$ apart. As with the local hydrodynamics, a key issue is the closure of the theory and whether the assumed hydrodynamic variables form a complete description. The range of validity of the Euler and Navier-Stokes equations is generally determined by the transport processes, which are characterized by the mean free path of the excitations \cite{Landau1987Fluid}. Motions on a shorter length scale fall outside the regime of classical hydrodynamics' validity. In the case of the strongly coupled OCP, we propose that the range of validity for the continuum equations is even greater than that for classical fluids. In particular, we conjecture that the stresses introduced by strong coupling lead to hydrodynamics equations at the level of description of the Euler equations that are valid even when the wavelength of the excitation is smaller than $a$. This is confirmed in this paper by comparing the results to molecular dynamics simulations and will be determined in the future through experiments testing our theory. \par

OCP has been numerically investigated in thermal equilibrium \cite{doi:10.1063/1.1727895, PhysRevA.8.3096, doi:10.1063/1.4963388, refId0, doi:10.1063/1.4897386}. Due to the long-range interaction, in addition to the ideal gas terms, there are nonlocal contributions to the energy and pressure that depend not only on the interaction potential but also on the pair distribution function. For example, this leads to a phase transition at around $\Gamma = 172$ \cite{10.1063/1.472802} between the fluid and body-centered cubic crystalline phases. As we are specifically interested in the fluid phase, our paper will only examine the region with $\Gamma < 172$. \par

Using numerical simulations, one can also investigate out-of-equilibrium properties of the OCP, such as the nonlinear effects on short time scales \cite{PhysRevLett.96.165001, Acciarri_2022} and a simpler problem of linear behavior close to equilibrium by computing dispersion curves \cite{PhysRevA.11.1025, https://doi.org/10.1002/ctpp.201400098}. Accurate numerical simulations using molecular dynamics require vast computational resources, especially for large systems and long observation times. In comparison, hydrodynamic equations are independent of the number of particles in the plasma, and the equations are usually quickly solved numerically. This is evident for some nonlinear problems, such as the expansion of the plasma \cite{KILLIAN200777}. Other examples of nonlinear phenomena that interest us are solid body rotation, vortex motion, and shock waves. Finally, working in terms of a few macroscopic hydrodynamic variables provides a clearer physical picture. The goal of the proposed variational approach is to generate equations for the reversible hydrodynamics analogous to Euler equations, which would generalize the equilibrium state of the OCP to motion that depends on both space and time. \par

In our paper, the variational principles should be understood in the sense of the mechanical least action principle, as discussed in classical mechanics \cite{Landau1976Mechanics} and classical field theory \cite{Landau1980Classical}. This can be contrasted with variational approaches that focus on obtaining thermodynamic properties of a system, for example, by minimizing the Helmholtz free energy \cite{DEWITT197979}. The mechanical variational principle has been shown to be useful in many areas of physics \cite{herivel_1955, doi:10.1098/rspa.1968.0103, lin_1963, PUTTERMAN1982146}. However, we are not aware of it being used to describe the OCP. \par

The first variational approach explored in this paper is given in the Eulerian frame, where hydrodynamic variables depend on positions $\vec{x}, \vec{x}'$ in the laboratory. The structure of the action $S$ in this approach is encapsulated by the following equation:
\begin{newnumbering}
\begin{equation}
\label{eq_A}
\begin{gathered}
S =  \iint \left( \frac{1}{2} m n \vec{v}^2 - n f(n, s) \right) \mathrm{d}\vec{x} \mathrm{d}t 
- \frac{1}{2} \iiint H \left(n, n', s, s', |\vec{x}-\vec{x}'| \right) \mathrm{d}\vec{x} \mathrm{d}\vec{x}' \mathrm{d}t \\
+ \textrm{ (constraint terms) } 
+ \textrm{ (terms for the stationary neutralizing background)},
\end{gathered}
\end{equation}
\end{newnumbering}
where $m$ is the mass of the particle with charge $q$ and $n' = n(\vec{x}',t)$, $s' = s(\vec{x}', t)$. Here, function $H$ determines the nonlocal contribution to the action which is the topic of this paper. The action includes constraint terms that ensure continuity equations for both number density and specific entropy. It also includes terms that correspond to the interaction with and within the stationary neutralizing background which are important for the convergence of equations of motion and energy. \par

Application of the extremum principle to $S$ yields a generalization of Euler equations that include nonlocal interaction within the OCP:
\begin{newnumbering}
\begin{equation}
\begin{gathered}
mn \left( \frac{\partial \vec{v}}{\partial t} + (\vec{v} \cdot \vec{\nabla})\vec{v} \right)= - \vec{\nabla} \left( n^2 \frac{\partial f}{\partial n}\Big|_s \right) \\
- n \vec{\nabla} \left( \int \frac{\partial H}{\partial n} \Big|_{n',s,s',|\vec{x}-\vec{x}'|} \mathrm{d}\vec{x}' \right) - n \vec{\nabla} \int  n'_- \phi_{+-}(|\vec{x}-\vec{x}'|) \mathrm{d}\vec{x}',
\end{gathered}
\end{equation}
\end{newnumbering}
where $H$ is assumed to be symmetric concerning $\vec{x}$ and $\vec{x}'$, and the last term describes the interaction with the stationary neutralizing background. Here, $n'_- = n_-(\vec{x}')$ represents the time-independent number density of particles constituting the background, and $\phi_{+-}$ is the Coulomb potential between oppositely charged particles. In Section \ref{section_Eulerian}, we explore these equations of motion and derive some of their properties. For instance, this dynamic equation does not permit the propagation of small amplitude transverse waves. This can be seen by dividing the equation by $n$ and taking the curl, which demonstrates that for Eulerian flow the new nonlocal terms do not contribute to the dynamics of vorticity. \par

Extensive data on the dispersion curves of the OCP obtained from molecular dynamics simulations are available. This provides a perfect opportunity to test our general theoretical framework in the linear regime. For instance, propagating transverse waves are observed at sufficiently short wavelength \cite{10.1063/1.5088141}. Furthermore, at longer wavelengths, there is an onset of negative dispersion for longitudinal waves as $\Gamma$ is increased above a critical value \cite{doi:10.1063/1.3679586}. \par

To formulate a theory that includes transverse waves, we note that in hydrodynamics, there are two coordinate systems for describing the fluid: Eulerian, where hydrodynamic quantities depend on the position in the laboratory frame, and Lagrangian, where hydrodynamic quantities depend on positions in some reference state, such as the initial or equilibrium positions of the particles forming the fluid. It is known that the Euler equations of hydrodynamics, which successfully explain many hydrodynamic phenomena, can be obtained from variational principles in both Eulerian and Lagrangian approaches \cite{herivel_1955, doi:10.1098/rspa.1968.0103, lin_1963}. However, we are not aware of generalizations for systems with a nonlocal interaction. In this paper, we present variational principles for the classical OCP in both Eulerian and Lagrangian approaches. Although they only differ by a choice of coordinates, different assumptions about the out-of-equilibrium behavior of the pair distribution function are more natural in each of them. This facilitates different equations of motion with distinct properties and enables a theory that unifies longitudinal waves, transverse waves, and thermodynamics. \par

We will denote the reference or Lagrangian coordinates as $\vec{a}$ and $\vec{a}'$ and think of them as the reference locations of fluid particles, which, at time $t$, are located at $\vec{x}$ and $\vec{x}'$ in the laboratory frame. The Eulerian or laboratory frame coordinates are simply $\vec{x}$ and $\vec{x}'$. In the Eulerian approach, the general form for the nonlocal function $H$, as defined in Eq.~\eqref{eq_A}, that is consistent with thermodynamic equilibrium \cite{doi:10.1063/1.1727895, PhysRevA.8.3096}, is the following. Here, $g$ is the out-of-equilibrium generalization of the pair distribution function. 
\begin{newnumbering}
\begin{equation}
\label{eulerian_H}
\begin{gathered}
H = n n' \phi(|\vec{x}-\vec{x}'|) g(n, n', s, s', |\vec{x}-\vec{x}'|)
\end{gathered}
\end{equation}
\end{newnumbering}\par

To obtain a system that can dynamically display propagating transverse waves, it is essential to consider the system from the perspective of Lagrangian coordinates. We consider the action given in Eq.~\eqref{eq_A}, rewritten in Lagrangian coordinates, where the dependence of the out-of-equilibrium pair distribution function on coordinates is instead fixed to the reference values, with the dependence on number density and specific entropy remaining Eulerian. In this case, the pair distribution function appearing in Eq.~\eqref{eulerian_H} is modified as follows.
\begin{newnumbering}
\begin{equation}
\label{lagrangian_H}
\begin{gathered}
g = g(n, n', s, s', |\vec{a}-\vec{a}'|)
\end{gathered}
\end{equation}
\end{newnumbering}\par
 
In Section \ref{section_Eulerian}, we develop the theory of the purely Eulerian system, where Eq.~\eqref{eulerian_H} applies in equilibrium as well as in the dynamical state. In Section \ref{section:lagrangian_approach}, we use the variational method to develop the theory corresponding to Eq.~\eqref{lagrangian_H}, where the reference variables play a key role. We discuss equations of motion, momentum, and energy conservation laws, and, finally, both longitudinal and transverse dispersion laws. In the case of Eq.~\eqref{lagrangian_H}, the dispersion law will be shown to have transverse waves, along with a transition to negative dispersion for the longitudinal waves. Additionally, the dispersion laws that follow from the latter equation meet the Einstein frequency \cite{doi:10.1063/1.4942169} as wave number and $\Gamma$ become large. \par

\begin{figure*}
\includegraphics[width=4.8in, height=4.0in]{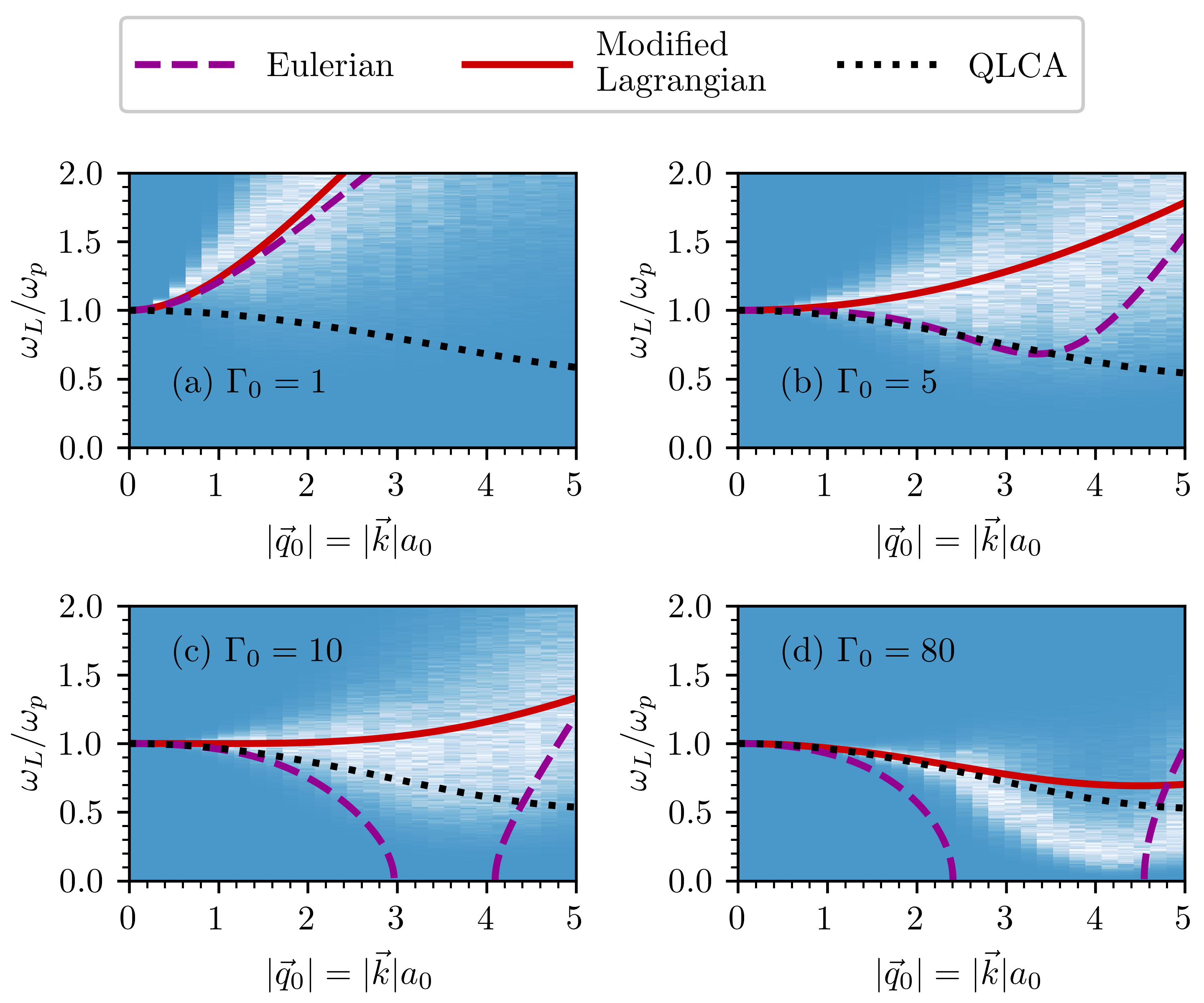} 
\caption{Longitudinal dispersion law in the normalized variables for different values of $\Gamma_0$. The results from the variational approaches are: Eulerian (dashed purple line, as described in Section~\ref{section:eulerian_dispersion_laws}) and modified Lagrangian (solid red line, as described in Section~\ref{section:dispersion_laws_modified_lagrangian}). The step function approximation for the pair distribution function is assumed, and $u(\Gamma) = -0.9\Gamma + 0.5944\Gamma^{1/3} - 0.2786$ is defined in Eq.~\eqref{equilibrium_energy_and_pressure} and taken from Ref.~\onlinecite{doi:10.1063/1.4897386}. For comparison, we consider the result from QLCA \cite{doi:10.1063/1.4942169} (dotted black line) and values of the longitudinal current fluctuation spectrum obtained from molecular dynamics simulations provided by the authors of Ref.~\onlinecite{https://doi.org/10.1002/ctpp.201400098} (colored background with large values being white and small values being blue), where the dispersion law is identified by the peaks of the spectrum. Note that the QLCA dispersion law shows negative dispersion for all $\Gamma_0$, whereas the proposed modified Lagrangian and simulations show a transition to negative dispersion around $\Gamma_0=10$. Additionally, the purely Eulerian theory shows instability, where $\omega_L^2 < 0$, for $\Gamma_0 \geq 7.9$.}
\label{fig:comparison_of_longitudinal}
\end{figure*}

\begin{figure*}
\includegraphics[width=4.8in, height=4.0in]{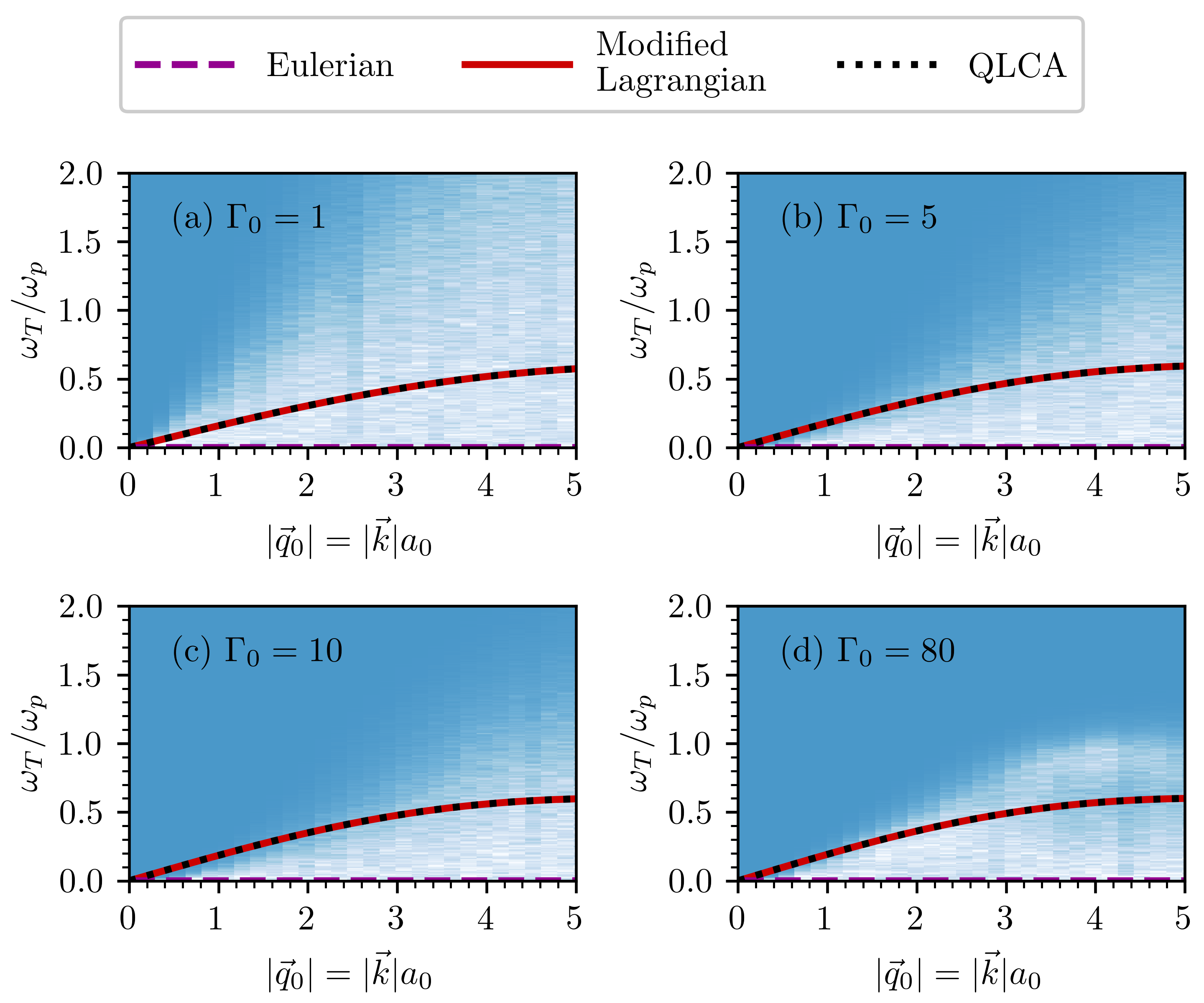} 
\caption{Transverse dispersion law in the normalized variables for different values of $\Gamma_0$. The results from the variational approaches are: Eulerian (dashed purple line, as described in Section~\ref{section:eulerian_dispersion_laws}) and modified Lagrangian (solid red line, as described in Section~\ref{section:dispersion_laws_modified_lagrangian}). The step function approximation for the pair distribution function is assumed, and $u(\Gamma) = -0.9\Gamma + 0.5944\Gamma^{1/3} - 0.2786$ is defined in Eq.~\eqref{equilibrium_energy_and_pressure} and taken from Ref.~\onlinecite{doi:10.1063/1.4897386}. For comparison, we consider the result from QLCA \cite{doi:10.1063/1.4942169} (dotted black line) and values of the transverse current fluctuation spectrum obtained from molecular dynamics simulations provided by the authors of Ref.~\onlinecite{https://doi.org/10.1002/ctpp.201400098} (colored background with large values being white and small values being blue), where the dispersion law is identified by the peaks of the spectrum.}
\label{fig:comparison_of_transverse}
\end{figure*}

To obtain the longitudinal and transverse dispersion laws, the equations of motion are linearized, with various coefficients being related to the local and nonlocal thermodynamic parameters. To have simple and practical expressions, we use a simple step function approximation for the pair distribution function, as has been done for the quasilocalized charge approximation (QLCA) \cite{doi:10.1063/1.4942169}. Such calculations are performed for the different actions to which the variational principle is applied. 
The results are also compared to the theoretical calculations using QLCA \cite{doi:10.1063/1.4942169} and to the numerical results of molecular dynamics \cite{https://doi.org/10.1002/ctpp.201400098, 10.1063/1.5088141, doi:10.1063/1.3679586}. The key results can be seen in Figs.~\ref{fig:comparison_of_longitudinal}, \ref{fig:comparison_of_transverse}, which show that different variational principles with different assumptions about the out-of-equilibrium behavior of the pair distribution function give significantly different results. For example, one can observe that in the Eulerian approach, there are no transverse modes, unlike in the modified Lagrangian approach. Additionally, we see that the modified Lagrangian approach, given by Eq.~\eqref{lagrangian_H} and discussed in Section~\ref{section:dispersion_laws_modified_lagrangian}, provides the best results among all considered variational approaches for both longitudinal and transverse dispersion curves for all equilibrium values of the coupling parameter, $\Gamma_0$. The results are accurate even at such small wavelengths that are comparable to the average distance between particles. Moreover, it was numerically confirmed that in the strong coupling limit, where $\Gamma_0 \to \infty$, the results of the modified Lagrangian variational principle tend to the QLCA result. Thus, it correctly predicts the finite values of both longitudinal and transverse frequencies as the wavelength tends to zero \cite{doi:10.1063/1.4942169}. \par

Important work has already been carried out to derive and interpret the simulated dispersion curves \cite{PhysRevA.11.1025, https://doi.org/10.1002/ctpp.201400098, 10.1063/1.5088141, doi:10.1063/1.3679586, doi:10.1063/1.4942169}. For example, one can use the generalized Langevin equation with memory functions \cite{Hansen1981} or QLCA \cite{PhysRevA.41.5516, doi:10.1063/1.873814}. However, these advances rely on the approximation of very strong coupling, $\Gamma \gg 1$, and are not easily extended for the analysis of nonlinear effects. Moreover, QLCA does not include thermodynamic effects that would result in the usual speed of sound term at weak coupling \cite{doi:10.1063/1.873814}. While it is possible to add such a term phenomenologically \cite{10.1063/1.4965903, PhysRevE.102.033207}, we believe it lacks rigorous justification. In contrast, our variational approach is consistent with thermodynamics, capable of predicting nonlinear effects, and does not rely on the assumption of very strong coupling. 

Both the linear and nonlinear dynamics of the OCP can be analyzed using the framework of kinetic theory and generalized hydrodynamics \cite{doi:10.1063/1.873037, doi:10.1063/1.873073, PhysRevE.92.013107}. Such an approach has the advantage of including the effects of heat transfer, viscosity, and relaxation. However, the challenge is to find an appropriate closure to the hierarchy of equations that would make them simple enough for computations and, at the same time, consistent with thermodynamics. In the development of our variational approach, our closure assumption is that a general description of the OCP can be done in terms of a few macroscopic hydrodynamic variables. Then, the variational principle guarantees that the equations are closed and have the conservation laws and symmetries built in. We hope that such an alternative would, in some cases, provide simpler and more practical results. Even though our approach does not include heat transfer, viscosity, or relaxation, one can later introduce these effects, as in the case of Navier-Stokes equations \cite{Landau1987Fluid}, generalized hydrodynamics \cite{PhysRevA.31.2502, PhysRevE.92.013107}, or by considering linear in velocity drag force between moving particles and the stationary background, as in the case of two-fluid plasma equations \cite{chen2015introduction}.

\section{Eulerian approach}
\label{section_Eulerian}
\subsection{Equations of motion}

In this section, we present and discuss the key results obtained from the Eulerian variational approach. The detailed derivations and the local differential forms of the conservation laws can be found in Appendix \ref{appendix_Eulerian}. \par

Let us assume that we are working with the OCP consisting of two species of charged particles. The particles that move have a charge $q_+$, while the particles forming a stationary neutralizing background have a charge $q_-$. In the Eulerian approach, each of the hydrodynamic functions is assumed to depend on the position in the laboratory frame, $\vec{x}$, and time, $t$. Similar to the case of Euler equations \cite{herivel_1955, doi:10.1098/rspa.1968.0103, lin_1963}, we assume that the complete set of variables for the hydrodynamic motion of moving particles consists of velocity, $\vec{v}$, number density, $n$, and specific entropy, $s$. The latter two satisfy continuity equations. 
\begin{equation}
\label{eulerian_continuity_equations}
\begin{gathered}
\frac{\partial n}{\partial t} + \vec{\nabla} \cdot (n \vec{v}) = 0, \quad \frac{\partial s}{\partial t} + \vec{v} \cdot \vec{\nabla}s = 0.
\end{gathered}
\end{equation} \par

Therefore, our goal is to write a variational principle with a nonlocal Lagrangian density for the OCP that would include the effects of strong coupling, from which the equation of motion for $\vec{v}$ could be found. \par

To satisfy the continuity equations given by Eq.~\eqref{eulerian_continuity_equations}, we introduce additional Lagrange multiplier fields $\alpha$ and $\beta$, as in the case of usual Euler equations. In general, one should also introduce an additional Lagrange multiplier field to ensure that the initial coordinates of fluid particles do not change along a particle's path \cite{doi:10.1098/rspa.1968.0103, lin_1963, PUTTERMAN1982146}. However, we do not explicitly consider it in our calculations, as it does not change the obtained results. \par

Next, we motivate a Lagrangian, see Eq.~\eqref{eulerian_lagrangian} below, for the generalized model of the OCP in the Eulerian approach, where the variation is performed for $n, \vec{v}, s$, and also for the functions $\alpha, \beta$. The local terms, which are integrated only over $\vec{x}$, correspond exactly to the Lagrangian for Euler equations. The first local term, including the mass $m$ of a single moving particle, represents kinetic energy, and the second local term represents the local internal energy that generates the local pressure. \par

On the other hand, the nonlocal terms, which are integrated over both $\vec{x}$ and $\vec{x}'$, represent nonlocal interaction between the particles. The first nonlocal term describes the interaction between the moving particles and, without any loss of generality, is written as an arbitrary function $F(n, n', s, s', |\vec{x}-\vec{x}'|)$, where we use the notation $n', s'$ to indicate that the functions are evaluated at $\vec{x}'$ instead of $\vec{x}$. For this term, we assume that the dependence on coordinates is solely through the combination $|\vec{x}-\vec{x}'|$, ensuring conservation laws for both linear and angular momentum. \par

The second nonlocal term represents the interaction between oppositely charged particles, which is necessary to ensure that the equations have an equilibrium solution. The third nonlocal term represents the interaction between the particles that form the stationary neutralizing background, necessary to ensure that the expression for the energy of the OCP is well-defined in the thermodynamic limit. Here, we assume that the interaction between charged species is due to the Coulomb potential, $\phi_{ij}(r) = q_i q_j/4 \pi \varepsilon_0 r$, where $i, j$ correspond to the species of the particles, and $r$ is the distance between particles. Additionally, $n_-$ represents the time-independent number density of the particles forming the neutralizing background.  
\begin{equation}
\label{eulerian_lagrangian}
\begin{gathered}
L = \int \mathcal{L}_1 \mathrm{d}\vec{x} + \iint \mathcal{L}_2 \mathrm{d}\vec{x}\mathrm{d}\vec{x}'
= \int \bigg( \frac{1}{2} m n \vec{v}^2 - n f(n, s) + \alpha \left( \frac{\partial n}{\partial t} + \vec{\nabla} \cdot (n \vec{v})  \right)  + \beta \left( \frac{\partial s}{\partial t} + \vec{v} \cdot \vec{\nabla} s  \right) \bigg) \mathrm{d}\vec{x} \\
+ \iint \bigg( - n n' F(n, n', s, s', |\vec{x}-\vec{x}'|)  - n n'_- \phi_{+-}(|\vec{x} - \vec{x}'|) 
- \frac{1}{2} n_- n'_- \phi_{--}(|\vec{x}-\vec{x}'|)\bigg) \mathrm{d}\vec{x} \mathrm{d}\vec{x}' 
\end{gathered}
\end{equation}

Performing variations with respect to fields $\alpha, \beta$ gives the continuity equations in Eq.~\eqref{eulerian_continuity_equations}, as intended. However, for variations with respect to the remaining functions, one must be careful when analyzing the nonlocal terms. It is useful to introduce, for each function $f$, the notation $f^T$ to mean taking $f$ but replacing all occurrences of $\vec{x}$ with $\vec{x}'$ and all occurrences of $\vec{x}'$ with $\vec{x}$. For example, $f^T(n, n') = f(n', n)$. The final equation of motion for $\vec{v}$ is as follows, but at this point, functions $f, F$ are still arbitrary.
\begin{equation}
\label{eulerian_final_EOM}
\begin{gathered}
mn \left( \frac{\partial \vec{v}}{\partial t} + (\vec{v} \cdot \vec{\nabla})\vec{v} \right)= - \vec{\nabla} \bigg( n^2 \frac{\partial f}{\partial n}\Big|_s 
+ \int n^2 n' \frac{\partial (F+F^T)}{\partial n}\Big|_{n', s, s', |\vec{x}-\vec{x}'|} \mathrm{d}\vec{x}' \bigg)\\
- \int \Bigg( nn'\frac{\partial (F+F^T)}{\partial \vec{x}}\Big|_{n,n',s,s',\vec{x}'} + nn'_- \frac{\partial \phi_{+-}}{\partial \vec{x}}\Big|_{\vec{x}'} \Bigg) \mathrm{d}\vec{x}'
\end{gathered}
\end{equation}

\subsection{Conservation laws}

Let us consider the momentum and energy conservation laws of our theory in the Eulerian approach. In particular, energy conservation will be important because the expression for the conserved energy coming from the variational principle that depends on functions $f, F$ can be computed in equilibrium and compared to the expression found in thermodynamics. This ensures both the consistency of our theory with thermodynamics and allows us to identify what functions $f, F$ should be in terms of the thermodynamic quantities. \par

We now examine the momentum conservation law in each of the directions $i = 1,2,3$. The momentum conservation law relies on the observation that in Eq.~\eqref{eulerian_lagrangian}, the local part of the Lagrangian density, $\mathcal{L}_1$, does not depend explicitly on $\vec{x}$, and the term with $F$ in the nonlocal part of this Lagrangian density, $\mathcal{L}_2$, as well as $\phi_{+-}, \phi_{--}$, all depend explicitly on $\vec{x}, \vec{x}'$ only in a translationally invariant combination $\vec{x} - \vec{x}'$. It is important to note that it is not true that $\mathcal{L}_2$ depends explicitly on $\vec{x}, \vec{x}'$ in a translationally invariant way, as the number density of the stationary neutralizing background, $n_-$, might depend on $\vec{x}$. 
\par

The momentum conservation law allows us to define the full momentum $\vec{P}$ of the particles that move in the OCP and consider how it changes over time. The time evolution for the momentum $\vec{P}$ in each direction is given as follows. It can be observed that the momentum of moving particles is not necessarily conserved and can change due to the force from the stationary neutralizing background. 
\begin{equation}
\label{eulerian_momentum_equation}
\begin{gathered}
\frac{\mathrm{d}P_i}{\mathrm{d}t} = \frac{\mathrm{d}}{\mathrm{d}t} \left( \int m n v_i \mathrm{d}\vec{x} \right) = - \iint nn'_- \frac{\partial \phi_{+-}}{\partial x_i}\Big|_{\vec{x}'}  \mathrm{d}\vec{x} \mathrm{d}\vec{x}'
\end{gathered}
\end{equation} \par

Notice that the momentum of moving particles is conserved in the case when the neutralizing background is both uniform and infinite, as can be shown by Eq.~\eqref{eulerian_momentum_equation} by using that $\phi_{+-}$ depends on $\vec{x}, \vec{x}'$ only through combination $\vec{x}-\vec{x}'$, and by integrating by parts. \par

Let us now explore the energy conservation law that will be used to match functions in the variational principle to the thermodynamic quantities. Energy conservation law relies on the observation that in the proposed Lagrangian in Eq.~\eqref{eulerian_lagrangian} both the local part of the Lagrangian density, $\mathcal{L}_1$, and the nonlocal part, $\mathcal{L}_2$, do not depend explicitly on time $t$. \par

The energy conservation law allows us to define the energy $E$ of the OCP and consider how it changes over time. 
\begin{equation}
\label{eulerian_energy_equation}
\begin{gathered}
\frac{\mathrm{d}E}{\mathrm{d}t} = \frac{\mathrm{d}}{\mathrm{d}t} \Bigg( \int \left( \frac{1}{2} m n \vec{v}^2 + nf \right) \mathrm{d}\vec{x} + \iint \Bigg( \frac{n n'}{2}(F+F^T) \\
+ \frac{nn'_-}{2} \phi_{+-} + \frac{n'n_-}{2} \phi_{+-} + \frac{n_- n'_-}{2} \phi_{--} \Bigg) \mathrm{d}\vec{x} \mathrm{d}\vec{x}' \Bigg) = 0
\end{gathered}
\end{equation} \par

We would like to analyze the expression for the energy, $E$, in the thermodynamic equilibrium. In terms of our variational principle, this means that we consider the hydrodynamic functions $n, s, \vec{v}$ to be uniform and time-independent with values $n = n_0, s = s_0, \vec{v} = \vec{0}$, which correspond to no macroscopic motion. To fully specify the value of the energy, we also assume that the number density of the stationary neutralizing background is uniform, with a value $n_- = n_{-,0}$. To relate $n_0, n_{-,0}$, we use an assumption that the total charge of the OCP is zero. This assumption is relevant to strongly coupled plasma experiments \cite{1968ApJ151227V, VanHorn_2019, DEMMEL200598, Demmel_2021, Bataller2019DynamicsOS, KILLIAN200777, RevModPhys.81.1353}. In terms of equilibrium number densities, this charge neutrality condition becomes $q_+ n_0 + q_- n_{-,0} = 0$. \par

In the thermodynamic limit, where the number of moving particles $N \to \infty$ and the volume of the system $V \to \infty$ while the number density is fixed, the equilibrium energy of the OCP diverges. Therefore, instead, we consider the equilibrium energy per particle, $E_0/N$. From Eq.~\eqref{eulerian_energy_equation}, we obtain the following expression.
\begin{equation}
\label{eulerian_equilibrium_energy}
\begin{gathered}
\frac{E_0}{N} = f(n_0, s_0) + \frac{n_0}{2} \int \bigg( (F+F^T)(n_0, n_0, s_0, s_0, |\vec{r}|) 
- \frac{q^2_+}{4 \pi \varepsilon_0 |\vec{r}|} \bigg) \mathrm{d}\vec{r}
\end{gathered}
\end{equation}\par

This expression can be compared to the expression from the thermodynamics of the thermodynamic equilibrium energy of the OCP per moving particle, $E_{\mathrm{th}}/N$ \cite{doi:10.1063/1.1727895, PhysRevA.8.3096}, where $T$ is the temperature and $g$ is the equilibrium pair distribution function.
\begin{equation}
\label{thermodynamic_equilibrium_energy}
\begin{gathered}
\frac{E_{\mathrm{th}}}{N} = \frac{3}{2} k_B T(n_0, s_0)+ \frac{n_0}{2} \int \frac{q^2_+}{4 \pi \varepsilon_0 |\vec{r}|}\left( g(n_0, s_0, |\vec{r}|) - 1 \right) \mathrm{d}\vec{r}
\end{gathered}
\end{equation} \par

By comparing Eqs.~\eqref{eulerian_equilibrium_energy} and \eqref{thermodynamic_equilibrium_energy}, we have the following constraints on functions $f, F$. Here, we use that $F^T(n_0, n_0, s_0, s_0, |\vec{r}|) = F(n_0, n_0, s_0, s_0, |\vec{r}|)$ directly from the definition.
\begin{equation}
\label{eulerian_energy_constraint_1}
\begin{gathered}
f(n_0, s_0) = \frac{3}{2} k_B T(n_0, s_0), \quad
F(n_0, n_0, s_0, s_0, |\vec{r}|) = \frac{q^2_+}{8 \pi \varepsilon_0 |\vec{r}|}  g(n_0, s_0, |\vec{r}|).
\end{gathered}
\end{equation}\par

This constraint uniquely determines function $f$, but not $F$. This is because $F$ is specified by this constraint only for values when $n$ is equal to $n'$, and when $s$ is equal to $s'$. To resolve this issue, from now on we replace Lagrangian in Eq.~\eqref{eulerian_lagrangian} with a less general Lagrangian, where $F$ is assumed to be a function of only $n, s, |\vec{x} - \vec{x}'|$. Notice that this is not a unique choice as one could also take, for example, $F = F( (n+n')/2, (s+s')/2, |\vec{x} - \vec{x}'|)$ or $F = F( \sqrt{nn'}, \sqrt{ss'}, |\vec{x}-\vec{x}'|)$. However, in some sense, our choice is the simplest possible, and then $F$ is uniquely determined from the constraint in Eq.~\eqref{eulerian_energy_constraint_1}. Therefore, we have the following result. 
\begin{equation}
\label{eulerian_energy_constraint_final}
\begin{gathered}
f(n, s) = \frac{3}{2} k_B T(n, s), \quad F(n, s, |\vec{r}|) = \frac{q^2_+}{8 \pi \varepsilon_0 |\vec{r}|}  g(n, s, |\vec{r}|).
\end{gathered}
\end{equation}\par

Such identification allows us to rewrite the equation of motion for velocity given by Eq.~\eqref{eulerian_final_EOM} in the following way in terms of thermodynamic functions. Here, we use $\phi_{ij}(r) = q_i q_j/4 \pi \varepsilon_0 r$, where $i, j$ are the corresponding species of the particles, and $r$ is the distance between the particles.
\begin{equation}
\label{eulerian_final_with_thermo}
\begin{gathered}
mn \left( \frac{\partial \vec{v}}{\partial t} + (\vec{v} \cdot \vec{\nabla})\vec{v} \right)= - \vec{\nabla} \Bigg( \frac{3}{2} k_B n^2 \frac{\partial T}{\partial n}\Big|_s 
+ \frac{q^2_+}{8 \pi \varepsilon_0} \int \frac{n^2  n'}{|\vec{x}-\vec{x}'|} \frac{\partial g}{\partial n}\Big|_{s, |\vec{x}-\vec{x}'|}\mathrm{d}\vec{x}' \Bigg)\\
- \frac{q_+ n}{4 \pi \varepsilon_0} \int \Bigg( q_+ n'\frac{\partial}{\partial \vec{x}} \left( \frac{1}{|\vec{x}-\vec{x}'|} \frac{(g + g^T)}{2} \right)\Big|_{n,n',s,s',\vec{x}'} 
+ q_- n'_- \frac{\partial }{\partial \vec{x}}\left( \frac{1}{|\vec{x}-\vec{x}'|}\right)\Big|_{\vec{x}'} \Bigg) \mathrm{d}\vec{x}'
\end{gathered}
\end{equation}

\subsection{Dispersion laws}
\label{section:eulerian_dispersion_laws}

While discussing the energy conservation law, we considered equilibrium solutions in which hydrodynamic functions, $n, s, \vec{v}$, are uniform and time-independent, with values $n = n_0, s = s_0, \vec{v} = \vec{0}$. In this context, the stationary neutralizing background has a uniform number density with the value $n_- = n_{-,0}$, which is related to $n_0$ by using the charge neutrality assumption, $q_+ n_0 + q_- n_{-,0} = 0$. \par

Now, we would like to examine the linearized equations of motion that correspond to the OCP being close to thermodynamic equilibrium. In this case, we assume the following form for the time-dependent functions, where $n_1, s_1, \vec{v}_1$ denote first-order corrections.
\begin{equation}
\begin{gathered}
n = n_0 + n_1(\vec{x}, t), \quad s = s_0 + s_1(\vec{x}, t), \quad \vec{v} = \vec{v}_1(\vec{x}, t).
\end{gathered}
\end{equation}\par

With such assumptions regarding $n, s, \vec{v}$, we expand equations of motion given by Eqs.~\eqref{eulerian_continuity_equations}, \eqref{eulerian_final_with_thermo} up to the first order. We first checked that our proposed equilibrium solutions indeed satisfy these equations of motion. To analyze the first-order equations, we use the Fourier transform with respect to the spatial coordinates, $\vec{x}$. The intermediate steps of the derivation are found in Appendix~\ref{appendix_eulerian_dispersion}. We find that the transverse dispersion law is $\omega_T(|\vec{k}|) = 0$. For the longitudinal direction, we have the following dispersion law, with the subscript "$0$" indicating that a function is evaluated at equilibrium values.
\begin{equation}
\label{longitudinal_dispersion_Eulerian}
\begin{gathered} 
\omega^2_L(|\vec{k}|) = \frac{q_+^2 n_0}{m \varepsilon_0} + |\vec{k}|^2 \Bigg( \Bigg[ \frac{\partial}{\partial n} \Big( \frac{3 k_B}{2 m}  n^2 \frac{\partial T}{\partial n}\Big|_s \Big)\Big|_s \Bigg]_0
+  \frac{q^2_+ n_0}{2 m \varepsilon_0} \int_0^{\infty} r \Bigg[ \frac{\partial}{\partial n} \Big( n^2 \frac{\partial (g-1)}{\partial n}\Big|_{s, r} \Big)\Big|_{s, r} \Bigg]_0 \mathrm{d}r \Bigg) \\
+ \frac{q^2_+ n_0 |\vec{k}|}{m \varepsilon_0} \int_0^{\infty} \sin(|\vec{k}| r) \Bigg( (g - 1)_0 + n_0 \Bigg[ \frac{\partial (g - 1)}{\partial n}\Big|_{s, r}\Bigg]_0  \Bigg) \mathrm{d}r
\end{gathered}
\end{equation}\par

One can see that the longitudinal dispersion relation does not only depend on the equilibrium pair distribution function but also on its adiabatic derivatives, in other words, derivatives with respect to the number density at constant entropy. In the usual case of Euler equations, the dispersion relation depends on the adiabatic derivatives of the local energy \cite{Landau1987Fluid}. In our case, the nonlocal contribution to the energy of the OCP depends on the pair distribution function, as can be seen in Eq.~\eqref{thermodynamic_equilibrium_energy}. Therefore, the adiabatic derivatives of the pair distribution function in the dispersion relation are related to the adiabatic derivatives of the nonlocal energy. This means that our dispersion relation accounts for both local and nonlocal contributions to the total energy of the OCP in a way consistent with thermodynamics. \par

In thermal equilibrium, it is convenient to work with a parameter that describes the ratio between the average potential Coulomb energy of the moving particles and their average thermal kinetic energy,  $\Gamma = q_+^2/4 \pi \varepsilon_0 a k_B T$, where $T$ is the temperature, and $a$ is the average interparticle distance defined by $4 \pi a^3/3 = 1/n$ \cite{doi:10.1063/1.1727895, PhysRevA.8.3096}. It is known that $g(n, s, |\vec{r}|) = g(\Gamma, |\vec{r}|/a)$, and the expressions for the equilibrium energy $E_0$ and equilibrium pressure $p_0$ are as follows, where $u$ is the excess internal energy describing nonlocal contributions.
\begin{equation}
\label{equilibrium_energy_and_pressure}
\begin{gathered}
\frac{E_0}{N k_B T} = \frac{3}{2} + u(\Gamma), \quad \frac{p_0}{n k_B T} = 1 + \frac{u(\Gamma)}{3}.
\end{gathered}
\end{equation} \par

When analyzing the dispersion relation of the OCP, the newly introduced variables $\Gamma$ and $a$ will also oscillate around their respective equilibrium values, $\Gamma_0, a_0$, due to their dependence on $n, s$. Nevertheless, it is convenient to rewrite the longitudinal dispersion law in Eq.~\eqref{longitudinal_dispersion_Eulerian} in terms of normalized variables, so that the dispersion law is parametrized only in terms of $\Gamma_0$, similar to $E_0, p_0$. To do this, one introduces the plasma frequency of the moving particles, $\omega_p$, as $\omega_p^2 = q_+^2 n_0 / m \varepsilon_0$, and the normalized wavevector $\vec{q}_0 = \vec{k}a_0$ \cite{PhysRevA.11.1025, https://doi.org/10.1002/ctpp.201400098, doi:10.1063/1.3679586}. To compute the term in Eq.~\eqref{longitudinal_dispersion_Eulerian} that is related to the temperature derivatives, we use the following result from thermodynamics \cite{landau2013statistical}. Note that to compute it, we need knowledge of both $E_0$ and $p_0$.
\begin{equation}
\begin{gathered}
\frac{\partial T}{\partial V}\Big|_s = -\frac{(\partial E_0/ \partial V|_T + p_0)}{\partial E_0/ \partial T|_V}
\end{gathered}
\end{equation}\par

As mentioned in the discussion of the energy law, we are interested in the thermodynamic limit, where the number of moving particles $N \to \infty$ and the volume of the system $V \to \infty$. Therefore, it is convenient to work with densities and $E_0/N$. One can use the chain rule to rewrite derivatives of $V$ in terms of $n$ in the previous equation. Combining this with Eq.~\eqref{equilibrium_energy_and_pressure} leads to the following.
\begin{equation}
\label{eulerian_f1}
\begin{gathered}
\frac{n}{T} \frac{\partial T}{\partial n}\Big|_s = \frac{2}{3} \frac{\Big( 1 - (\Gamma^2/3)\mathrm{d}(u/\Gamma)/\mathrm{d}\Gamma \Big)}{\Big(1 - (2\Gamma^2/3)\mathrm{d}(u/\Gamma)/\mathrm{d}\Gamma \Big)} = \frac{2}{3} f_1(\Gamma)
\end{gathered}
\end{equation}\par

This result also allows us to use the following identity, which holds for an arbitrary function $f(\Gamma, |\vec{q}|)$, to perform number density derivatives at constant $s$. 
\begin{equation}
\label{constant_s_derivative}
\begin{gathered}
n \frac{\partial f}{\partial n}\Big|_s (\Gamma, |\vec{q}|) = \frac{1}{3} \left( \frac{\partial f}{\partial \Gamma} \Big|_{|\vec{q}|}  \Gamma \Big( 1 - 2f_1(\Gamma) \Big) - \frac{\partial f}{\partial |\vec{q}|} \Big|_{\Gamma} |\vec{q}| \right)
\end{gathered}
\end{equation}\par

Combining all of the computations, we obtain the following longitudinal dispersion relation in Eq.~\eqref{longitudinal_dispersion_Eulerian_normalized} in the normalized variables, which indeed depends solely on the equilibrium value $\Gamma_0$. The terms appearing in the dispersion relation are given by Eqs.~\eqref{eulerian_f1}, \eqref{eulerian_f2}, \eqref{eulerian_js}, \eqref{eulerian_hs}, and the derivative at a constant value of $s$ is given by Eq.~\eqref{constant_s_derivative}.
\begin{equation}
\label{longitudinal_dispersion_Eulerian_normalized}
\begin{gathered} 
\left( \frac{\omega_L}{\omega_p} \right)^2 (\Gamma_0, |\vec{q}_0|) = 1 + \frac{|\vec{q}_0|^2}{\Gamma_0}f_2(\Gamma_0) + \left( n \frac{\partial j}{\partial n}\Big|_s \right) (\Gamma_0, |\vec{q}_0|) \\
+ \left( n \frac{\partial }{\partial n} \left( n \frac{\partial j}{\partial n}\Big|_s \right)\Big|_s \right) (\Gamma_0, |\vec{q}_0|) + h(\Gamma_0, |\vec{q}_0|)
 + \left( n \frac{\partial h}{\partial n}\Big|_s \right) (\Gamma_0, |\vec{q}_0|),
\end{gathered}
\end{equation}
\begin{equation}
\label{eulerian_f2}
\begin{gathered}
f_2(\Gamma) = \frac{1}{3} \left( f_1(\Gamma) + \frac{2}{3}f_1^2(\Gamma) + \frac{\Gamma}{3} \frac{\mathrm{d}f_1}{\mathrm{d}\Gamma}(\Gamma) \Big(1 - 2f_1(\Gamma) \Big) \right),
\end{gathered}
\end{equation}
\begin{equation}
\label{eulerian_js}
\begin{gathered}
j(\Gamma, |\vec{q}|) = \frac{|\vec{q}|^2}{2} \int_0^{\infty} x \Big( g - 1 \Big)(\Gamma, x) \mathrm{d}x,
\end{gathered}
\end{equation}
\begin{equation}
\label{eulerian_hs}
\begin{gathered}
h(\Gamma, |\vec{q}|) = |\vec{q}| \int_0^{\infty} \Big( g - 1 \Big)(\Gamma, x) \sin(|\vec{q}|x) \mathrm{d}x.
\end{gathered}
\end{equation}\par

In principle, the longitudinal dispersion relation can now be computed for an arbitrary pair distribution function, $g(\Gamma, |\vec{r}|/a)$. However, it is important to notice that it depends not only on the integrals with respect to the variable $x = |\vec{r}|/a$ but also on the $\Gamma$ derivatives of such integrals. This implies that, for a precise evaluation of the dispersion law, a pair distribution function should be known precisely as a function of $\Gamma$. Precise fits for the excess internal energy in Eq.~\eqref{equilibrium_energy_and_pressure}, $u(\Gamma)$, are known \cite{refId0, doi:10.1063/1.4897386}. However, we are not aware of a precise numerical fit for $g(\Gamma, |\vec{r}|/a)$. For example, a multiparameter fit for the pair distribution function that is accurate for a wide range of $\Gamma$ values as a function of $x = |\vec{r}|/a$ is known \cite{doi:10.1063/1.4963388}. Nevertheless, the $\Gamma$ derivative of such a fit is imprecise, as can be confirmed when using it to numerically compute the thermodynamic heat capacity because the obtained results contradict the direct results from Monte Carlo simulations \cite{doi:10.1063/1.1727895, PhysRevA.8.3096, refId0}. For other theoretical approaches, such as QLCA \cite{PhysRevA.41.5516, doi:10.1063/1.873814}, this is not an issue as their predictions do not include $\Gamma$ derivatives. \par

To address this issue, we will use an approximation for the pair distribution function that has previously been employed in QLCA and has yielded good results when compared to using the pair distribution function without any approximations \cite{doi:10.1063/1.4942169}. We assume that the pair distribution function, $g(\Gamma, |\vec{r}|/a)$, takes the form of a step function with a value of $0$ when $|\vec{r}|/a < (R/a)(\Gamma)$ and $1$ when $|\vec{r}|/a > (R/a)(\Gamma)$. Notice that these values have been chosen to be consistent with the behavior of the pair distribution function as $|\vec{r}| \to 0$ and $|\vec{r}| \to \infty$ \cite{doi:10.1063/1.1727895, PhysRevA.8.3096, doi:10.1063/1.4963388}. More complicated approximations for the pair distribution function are also possible, such as the two-step approximation that has been used to analyze strongly coupled Yukawa fluids \cite{FAIRUSHIN2020103359}. However, in that case, it was noticed that it does not provide a significant improvement over the single-step function approximation \cite{doi:10.1063/1.4942169}. Because of that, in our calculations, we decided to use the single-step function approximation.\par

To establish the correct dependence $(R/a)(\Gamma)$, we compute the equilibrium energy in Eq.~\eqref{thermodynamic_equilibrium_energy} using this approximation and use Eq.~\eqref{equilibrium_energy_and_pressure} to establish the relationship between $(R/a)(\Gamma)$ and $u(\Gamma)$. The latter is known with precision.
\begin{equation}
\label{R_over_a}
\left( \frac{R}{a} \right)(\Gamma) = \sqrt{ - \frac{4}{3} \frac{u(\Gamma)}{\Gamma}}
\end{equation}\par

This step function approximation also simplifies the calculation of the dispersion law. The integrals in Eqs.~\eqref{eulerian_js}, \eqref{eulerian_hs} can now be evaluated exactly, leading to simpler and more practical expressions while avoiding additional numerical errors resulting from the computation of the integral. 
\begin{equation}
\label{simplified_j_h_Eulerian}
\begin{gathered}
j(\Gamma, |\vec{q}|) = -\frac{|\vec{q}|^2}{4} \left( \frac{R}{a} \right)^2(\Gamma), \quad
h(\Gamma, |\vec{q}|) = \cos\left( |\vec{q}| \left( \frac{R}{a} \right) (\Gamma) \right) - 1.
\end{gathered}
\end{equation}\par

First, let us consider a simple fit $u(\Gamma) = -0.9 \Gamma$, which is accurate for very strong coupling where $\Gamma \gg 1$ \cite{doi:10.1063/1.4897386}. For this particular fit, the longitudinal dispersion law in Eq.~\eqref{longitudinal_dispersion_Eulerian_normalized}, complemented with Eq.~\eqref{eulerian_f2} and the simplified expressions in Eq.~\eqref{simplified_j_h_Eulerian}, takes the following form. The first term, proportional to the $|\vec{q}_0|^2$, represents the local and nonlocal parts of the pressure, which are the terms under the gradient in Eq.~\eqref{eulerian_final_with_thermo}. The last two terms arise from the remaining force terms in Eq.~\eqref{eulerian_final_with_thermo}. 
\begin{equation}
\label{simplified_Eulerian_dispersion_simple_fit}
\begin{gathered}
\left( \frac{\omega_L}{\omega_p} \right)^2 (\Gamma_0, |\vec{q}_0|) = \left( \frac{5}{9\Gamma_0} + \frac{1}{15} \right) |\vec{q}_0|^2
+ \cos \left( \sqrt{\frac{6}{5}}|\vec{q}_0| \right)  + \sqrt{\frac{2}{15}} |\vec{q}_0| \sin \left( \sqrt{\frac{6}{5}}|\vec{q}_0| \right)
\end{gathered}
\end{equation}\par

\begin{figure*}
\includegraphics[width=5.4in, height=1.8in]{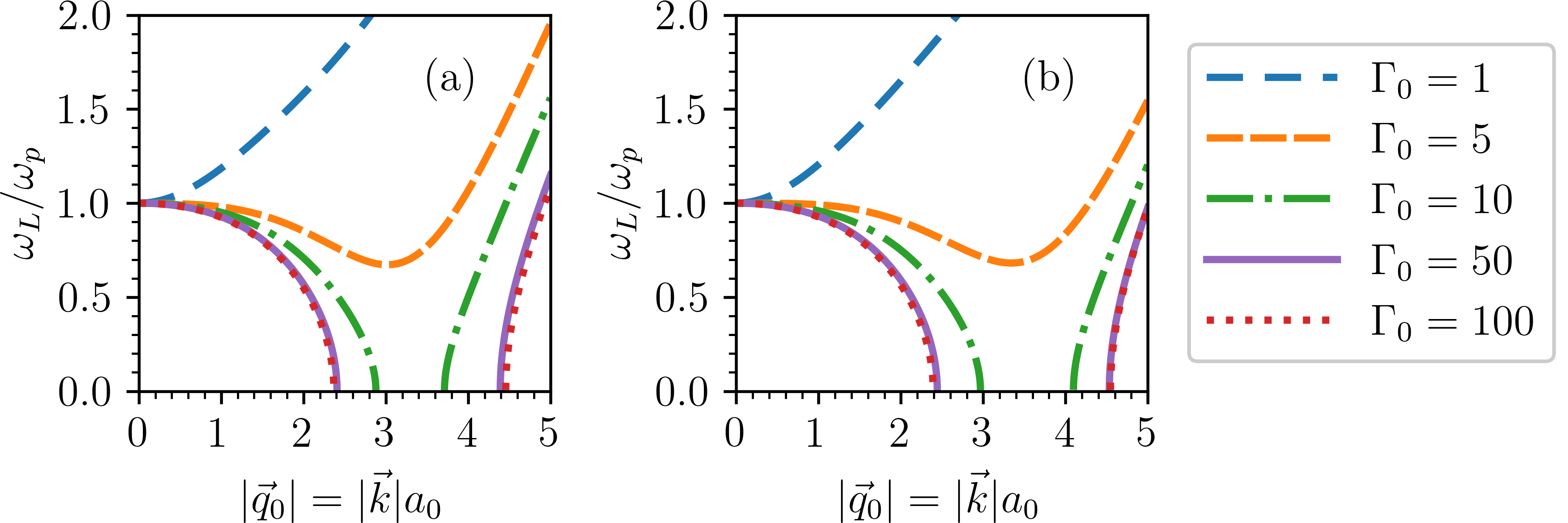} 
\caption{The longitudinal dispersion law in the Eulerian approach using normalized variables for different values of $\Gamma_0$, as given by Eq.~\eqref{longitudinal_dispersion_Eulerian_normalized}. The step function approximation for the pair distribution function is assumed, and we use Eqs.~\eqref{eulerian_f1}, \eqref{eulerian_f2}, \eqref{R_over_a}, \eqref{simplified_j_h_Eulerian} for a given $u(\Gamma)$ defined in Eq.~\eqref{equilibrium_energy_and_pressure}. In (a), we set $u(\Gamma) = -0.9\Gamma$, and the corresponding result is given by Eq.~\eqref{simplified_Eulerian_dispersion_simple_fit}. In (b), we use $u(\Gamma) = -0.9\Gamma + 0.5944\Gamma^{1/3} - 0.2786$.}
\label{fig:eulerian_longitudinal}
\end{figure*}

The results for different values of $\Gamma_0$ are presented in Fig.~\ref{fig:eulerian_longitudinal}(a). Several key properties can be observed from the figure. For weak coupling, where $\Gamma_0 = 1$, the dispersion relation resembles the behavior of an ideal gas. However, as $\Gamma_0$ increases, the behavior undergoes a dramatic change. At a critical value of $\Gamma_0 = 4.2$, we observe the onset of negative dispersion. In other words, for small values of $|\vec{q}_0|$, it holds true that $\omega_L \leq \omega_p$ instead of $\omega_L \geq \omega_p$. Additionally, as $\Gamma_0$ increases, an unstable region emerges, where $\omega_L^2 < 0$. 

The simpler fit for $u(\Gamma)$ does not incorporate thermal effects for the nonlocal interaction \cite{doi:10.1063/1.4897386}. Due to this limitation, it is interesting to consider a more precise fit, $u(\Gamma) = -0.9\Gamma + 0.5944\Gamma^{1/3} - 0.2786$, which is accurate for $\Gamma \geq 1$ and includes thermal effects \cite{doi:10.1063/1.4897386}. The results for different values of $\Gamma_0$ are presented in Fig.~\ref{fig:eulerian_longitudinal}(b), and one can observe that the results are qualitatively similar to those obtained with the simpler fit, with small quantitative differences. For instance, the onset of negative dispersion is now observed at $\Gamma_0 = 4.9$.  \par

The results for the more precise fit are also compared to the results obtained by QLCA \cite{doi:10.1063/1.4942169} using the same pair distribution function approximation. Additionally, we compare them to the current fluctuation spectra obtained from molecular dynamics simulations, provided by the authors of Ref.~\onlinecite{https://doi.org/10.1002/ctpp.201400098}, where dispersion laws are identified by the peaks of the spectra. The comparison for the longitudinal dispersion law is presented in Fig.~\ref{fig:comparison_of_longitudinal}. It can be observed that for small values of $\Gamma_0$, where $\Gamma_0 = 1$ or $\Gamma_0 = 5$, the predicted longitudinal dispersion law reasonably agrees with the results of molecular dynamics simulations, unlike the QLCA result. However, as $\Gamma_0$ increases, the predicted dispersion law starts to diverge from the results of the molecular dynamics simulations. \par

In particular, the latter do not exhibit regions of instability where $\omega_L^2 < 0$. Additionally, the onset of negative dispersion for the predicted dispersion law occurs at $\Gamma_0 = 4.9$, in contrast to the range of values from $\Gamma_0 = 9.5$ to $\Gamma_0 = 10.0$ estimated by the molecular dynamics simulations \cite{https://doi.org/10.1002/ctpp.201400098, doi:10.1063/1.3679586}. Furthermore, the predicted longitudinal dispersion law in the limit as $\Gamma_0 \to \infty$ does not converge to the QLCA result, which is considered accurate in such a limit \cite{doi:10.1063/1.4942169}. \par

The comparison for the transverse dispersion law is presented in Fig.~\ref{fig:comparison_of_transverse}. It correctly predicts that for small values of the wavevector, there are no transverse waves. However, it fails to predict a result that is known numerically, where at large values of the wavevector, transverse waves appear and are well-described by QLCA \cite{10.1063/1.5088141}.\par

The issues found for the predicted dispersion laws and their comparison with the results from QLCA and molecular dynamics simulations motivate us to consider an alternative variational approach, where we consider formulating it in the Lagrangian coordinates. It is discussed in the next Section~\ref{section:lagrangian_approach}.

\section{Lagrangian approach}
\label{section:lagrangian_approach}
\subsection{Equations of motion}

In this section, we present and discuss the key results obtained from the Lagrangian variational approach. The detailed derivations and the local differential forms of the conservation laws can be found in Appendix \ref{appendix_Lagrangian}. \par

Let us assume once again that we are dealing with the OCP, which consists of two species of charged particles. The particles in motion have a charge $q_+$, while the particles forming the stationary neutralizing background have a charge $q_-$. However, in this case, we consider a Lagrangian approach that employs Lagrangian coordinates. Similar to the case of Euler equations \cite{herivel_1955, doi:10.1098/rspa.1968.0103}, we assume that the complete set of variables for the hydrodynamic motion comprises solely the displacement field $\vec{x}$. This field represents the current location of a particle in the fluid at time $t$, given that its position in some conveniently chosen reference state was $\vec{a}$. The choice of a reference state usually depends on the specific problem. For example, one can select the reference positions as the initial positions at the start of the experiment. Alternatively, as is useful in the case of the dispersion law, one can choose them to be the corresponding positions of the particles in thermal equilibrium.
\par

In the Lagrangian coordinates, the number density $n$ and specific entropy $s$ are not independent hydrodynamic functions that need to be solved. Instead, they can be expressed in terms of the displacement field $\vec{x}$ and time-independent number density $n^{\mathrm{ref}}$ and specific entropy $s^{\mathrm{ref}}$ that correspond to the reference state. The continuity equations, as given in Eq.~\eqref{eulerian_continuity_equations}, are rewritten in Lagrangian variables as follows \cite{herivel_1955, doi:10.1098/rspa.1968.0103}.
\begin{equation}
\label{lagrangian_continuity_equations}
\begin{gathered}
n(\vec{x}, t) = \frac{n^{\mathrm{ref}}}{\mathrm{det} \left( \partial \vec{x} / \partial \vec{a} \right)}, \quad s(\vec{x}, t) = s^{\mathrm{ref}}.
\end{gathered}
\end{equation}\par

As discussed in the Eulerian approach, our objective is to formulate a variational principle with a nonlocal Lagrangian density for the OCP that includes the effects of strong coupling. In the case of Lagrangian coordinates, this formulation should yield an equation of motion for $\vec{x}$. \par

As in the case of the Eulerian approach, we may consider the Lagrangian given in Eq.~\eqref{eulerian_lagrangian}, which can be rewritten in Lagrangian coordinates by changing the variables under the integral from $\vec{x}, \vec{x}'$ to the reference state coordinates $\vec{a}, \vec{a}'$. For the stationary particles, we can choose an independent reference state. For convenience, we will choose it to be the one where $\vec{a}$ represents the initial position of the stationary particle, making $\vec{x} = \vec{a}$. Additionally, it is useful to introduce the number density of the stationary neutralizing background in the reference state, $n_-^{\mathrm{ref}}$. \par

In the case of Lagrangian variables, the number density $n$ and the specific entropy $s$ are not independent variables, eliminating the need to introduce constraint fields $\alpha, \beta$. So, the Lagrangian used in the Eulerian approach is as follows when rewritten in Lagrangian coordinates, where the notation $(n^{\mathrm{ref}})'$ now indicates that the function $n^{\mathrm{ref}}$ is evaluated at $\vec{a}'$ instead of $\vec{a}$. 
\begin{equation}
\label{eulerian_lagrangian_lagrangian_coordinates}
\begin{gathered}
L = \int \mathcal{L}_1 \mathrm{d}\vec{a} + \iint \mathcal{L}_2 \mathrm{d}\vec{a}\mathrm{d}\vec{a}'
= \int \left( \frac{1}{2} m n^{\mathrm{ref}} \left( \frac{\partial \vec{x}}{\partial t} \right)^2 - n^{\mathrm{ref}} f\left( \frac{n^{\mathrm{ref}}}{\mathrm{det} \left( \partial \vec{x} / \partial \vec{a} \right)}, s^{\mathrm{ref}}  \right) \right) \mathrm{d}\vec{a} \\
+ \iint \bigg( - n^{\mathrm{ref}} (n^{\mathrm{ref}})' F\bigg( \frac{n^{\mathrm{ref}}}{\mathrm{det} \left( \partial \vec{x} / \partial \vec{a} \right)}, \frac{(n^{\mathrm{ref}})'}{\mathrm{det} \left( \partial \vec{x}' / \partial \vec{a}' \right)},  s^{\mathrm{ref}}, (s^{\mathrm{ref}})', |\vec{x}-\vec{x}'| \bigg) \\
- n^{\mathrm{ref}} (n_-^{\mathrm{ref}})' \phi_{+-}(|\vec{x} - \vec{a}'|) - \frac{1}{2} n_-^{\mathrm{ref}} (n_-^{\mathrm{ref}})' \phi_{--}(|\vec{a}-\vec{a}'|)\bigg) \mathrm{d}\vec{a} \mathrm{d}\vec{a}' 
\end{gathered}
\end{equation}\par

It is not surprising that when one performs the variations, the equation of motion for $\vec{x}$, which arises from the Lagrangian in Eq.~\eqref{eulerian_lagrangian_lagrangian_coordinates} and is subsequently rewritten in Eulerian variables, is identical to Eq.~\eqref{eulerian_final_EOM}. This is because the only difference in the calculations lies in using a different coordinate system, while the form of the Lagrangian remains the same. However, in Lagrangian variables, it becomes evident how one can generalize the Lagrangian shown in Eq.~\eqref{eulerian_lagrangian_lagrangian_coordinates}. In particular, each particle has two positions available: the position of the particle in the laboratory frame $\vec{x}$ and its reference position $\vec{a}$. Therefore, one can assume that the function $F$ also depends on $\vec{a}, \vec{a}'$. This extension does not contradict any of the conservation laws but allows for a different model for the OCP, as we will discuss next. \par

As an alternative to the Lagrangian given in Eq.~\eqref{eulerian_lagrangian_lagrangian_coordinates}, we propose the following Lagrangian.
\begin{equation}
\label{lagrangian_lagrangian}
\begin{gathered}
L^{\mathrm{L}} = \int \mathcal{L}^{\mathrm{L}}_1 \mathrm{d}\vec{a} + \iint \mathcal{L}^{\mathrm{L}}_2 \mathrm{d}\vec{a}\mathrm{d}\vec{a}'
= \int \left( \frac{1}{2} m n^{\mathrm{ref}} \left( \frac{\partial \vec{x}}{\partial t} \right)^2 - n^{\mathrm{ref}} f\left( \frac{n^{\mathrm{ref}}}{\mathrm{det} \left( \partial \vec{x} / \partial \vec{a} \right)}, s^{\mathrm{ref}}  \right) \right) \mathrm{d}\vec{a} \\
+ \iint \bigg( - n^{\mathrm{ref}} (n^{\mathrm{ref}})' F\bigg( \frac{n^{\mathrm{ref}}}{\mathrm{det} \left( \partial \vec{x} / \partial \vec{a} \right)}, \frac{(n^{\mathrm{ref}})'}{\mathrm{det} \left( \partial \vec{x}' / \partial \vec{a}' \right)},  s^{\mathrm{ref}}, (s^{\mathrm{ref}})', |\vec{x}-\vec{x}'|, \vec{a}, \vec{a}' \bigg) \\
- n^{\mathrm{ref}} (n_-^{\mathrm{ref}})' \phi_{+-}(|\vec{x} - \vec{a}'|) - \frac{1}{2} n_-^{\mathrm{ref}} (n_-^{\mathrm{ref}})' \phi_{--}(|\vec{a}-\vec{a}'|)\bigg) \mathrm{d}\vec{a} \mathrm{d}\vec{a}' 
\end{gathered}
\end{equation}\par

It is once again useful to introduce the notation $f^T$ for each function $f$ that indicates taking $f$ but now, as we are working in Lagrangian coordinates, replacing all occurrences of $\vec{a}$ with $\vec{a}'$ and all occurrences of $\vec{a}'$ with $\vec{a}$. For example, $f^T(\vec{x}, \vec{x}') = f(\vec{x}', \vec{x})$. The final equation of motion for each of the directions $i = 1,2,3,$ is as follows, but at this point, functions $f, F$ are still arbitrary.
\begin{equation}
\label{lagrangian_final_EOM}
\begin{gathered}
m n^{\mathrm{ref}} \frac{\partial^2 x_i}{\partial t^2} = - \sum \limits_{j=1}^3 \frac{\partial (\mathrm{det} \left( \partial \vec{x} / \partial \vec{a} \right))}{\partial (\partial_j x_i)} \frac{\partial}{\partial a_j} \Bigg( n^2 \frac{\partial f}{\partial n}\Big|_s 
+ \int n^2 (n^{\mathrm{ref}})' \frac{\partial (F+F^T)}{\partial n}\Big|_{n', s, s', |\vec{x}-\vec{x}'|, \vec{a}, \vec{a}'} \mathrm{d}\vec{a}' \Bigg)\\
- \int \bigg( n^{\mathrm{ref}}(n^{\mathrm{ref}})'\frac{\partial (F+F^T)}{\partial x_i}\Big|_{n,n',s,s',\vec{x}', \vec{a}, \vec{a}'}
+ n^{\mathrm{ref}}(n_-^{\mathrm{ref}})' \frac{\partial \phi_{+-}}{\partial x_i}\Big|_{\vec{a}'} \bigg) \mathrm{d}\vec{a}'
\end{gathered}
\end{equation} \par

To check the consistency with the Eulerian approach, it can be demonstrated that if $F$ does not depend on $\vec{a}, \vec{a}'$, then Eq.~\eqref{lagrangian_final_EOM} can be rewritten in Eulerian variables by changing coordinates from $(\vec{a}, \vec{a}')$ to $(\vec{x}, \vec{x}')$. This transformation results in exactly Eq.~\eqref{eulerian_final_EOM}. 

\subsection{Conservation laws}

Let us consider the momentum and energy conservation laws of our theory in the Lagrangian approach. As in the Eulerian case, the energy conservation law will be important because the expression for the conserved energy coming from the variational principle can be computed in equilibrium and compared to the expression found in thermodynamics. This ensures that the Lagrangian approach is also consistent with thermal equilibrium and allows to identify functions in the variational principle in terms of the thermodynamic quantities. Such comparison also shows how the Lagrangian approach is different from the previously described Eulerian approach. \par

We now examine the momentum conservation law in each of the directions, $i = 1,2,3$. As in the Eulerian case, the momentum conservation law now also relies on the observation that in Eq.~\eqref{lagrangian_lagrangian}, the local part of the Lagrangian density, $\mathcal{L}_1^{\mathrm{L}}$, does not depend on $\vec{x}$ explicitly but only on its derivatives. Furthermore, the term with $F$ in the nonlocal part of this Lagrangian density, $\mathcal{L}_2^{\mathrm{L}}$, depends on $\vec{x}, \vec{x}'$ only in a translationally invariant combination, $\vec{x}-\vec{x}'$, as well as on their derivatives. It is important to note that it is not true that $\mathcal{L}_2^{\mathrm{L}}$ depends explicitly on $\vec{x}, \vec{x}'$ in a translationally invariant way, as $\phi_{+-}$ depends on $\vec{x}-\vec{a}'$. \par

The momentum conservation law allows us to define the total momentum $\vec{P}$ of the particles moving in the OCP and examine how it changes over time. The time evolution of the momentum $\vec{P}$ in each direction is provided below. Here, it is evident that the momentum of moving particles may not be conserved and can change due to the force exerted by the stationary neutralizing background. 
\begin{equation}
\label{lagrangian_momentum_equation}
\begin{gathered}
\frac{\mathrm{d}P_i}{\mathrm{d}t} = \frac{\mathrm{d}}{\mathrm{d}t} \left( \int m n^{\mathrm{ref}} \frac{\partial x_i}{\partial t} \mathrm{d}\vec{a} \right) = - \iint n^{\mathrm{ref}}(n_-^{\mathrm{ref}})' \frac{\partial \phi_{+-}}{\partial x_i}\Big|_{\vec{a}'}  \mathrm{d}\vec{a} \mathrm{d}\vec{a}'
\end{gathered}
\end{equation}\par

One can demonstrate that the expression for momentum in the Lagrangian approach is the same as in the Eulerian approach, as given by Eq.~\eqref{eulerian_momentum_equation}. Additionally, as in the Eulerian case, the momentum of moving particles is conserved when the neutralizing background is both uniform and infinite.\par

Now, let us explore the energy conservation law, which we will use to match functions in the variational principle to the thermodynamic quantities. The energy conservation law is based on the observation that in the proposed Lagrangian in Eq.~\eqref{lagrangian_lagrangian}, both the local part of the Lagrangian density, $\mathcal{L}_1^{\mathrm{L}}$, and the nonlocal part, $\mathcal{L}_2^{\mathrm{L}}$, do not explicitly depend on time $t$. \par

In the case where the function $F$ does not depend on $\vec{a}, \vec{a}'$, one can demonstrate that the energy law in the Lagrangian approach can be rewritten in Eulerian variables by changing coordinates from $(\vec{a}, \vec{a}')$ to $(\vec{x}, \vec{x}')$. This results in precisely the same energy law as in the Eulerian approach. \par

The energy conservation law allows us to define the energy $E^{\mathrm{L}}$ in the Lagrangian approach for the OCP and examine how it changes over time. In the following, the energy density $\mathcal{E}^{\mathrm{L}}$ is defined in Eq.~\eqref{lagrangian_energy_density}.
\begin{equation}
\label{lagrangian_energy_equation}
\begin{gathered}
\frac{\mathrm{d}E^{\mathrm{L}}}{\mathrm{d}t} = \frac{\mathrm{d}}{\mathrm{d}t} \bigg( \int \mathcal{E}^{\mathrm{L}} \mathrm{d}\vec{a} \bigg) = 0,
\end{gathered}
\end{equation}
\begin{equation}
\label{lagrangian_energy_density}
\begin{gathered}
\mathcal{E}^{\mathrm{L}} = \frac{1}{2} m n^{\mathrm{ref}} \left( \frac{\partial \vec{x}}{\partial t} \right)^2 + n^{\mathrm{ref}} f + \frac{1}{2} \int \bigg(  n^{\mathrm{ref}}(n^{\mathrm{ref}})' (F+F^T)
+ n^{\mathrm{ref}}(n_-^{\mathrm{ref}})' \phi_{+-}(|\vec{x}-\vec{a}'|) \\
+ (n^{\mathrm{ref}})' n_-^{\mathrm{ref}} \phi_{+-}(|\vec{x}'-\vec{a}|) + n_-^{\mathrm{ref}}(n_-^{\mathrm{ref}})' \phi_{--}(|\vec{a}-\vec{a}'|) \bigg) \mathrm{d}\vec{a}'.
\end{gathered}
\end{equation}\par

It can be shown that if the nonlocal function $F$ does not depend on $\vec{a}, \vec{a}'$, the expression for the conserved energy in the Lagrangian approach is identical to the one in the Eulerian approach, as given by Eq.~\eqref{eulerian_energy_equation}. \par

Now, we consider equilibrium solutions and calculate the value of the energy, $E^{\mathrm{L}}$, in a thermodynamic equilibrium. In the Lagrangian case, the only time-dependent hydrodynamic function is the displacement of the moving particles, $\vec{x}$. We choose a reference state for the moving particles where $\vec{a}$ represents the equilibrium position of a given particle, or equivalently, its initial position, as in thermal equilibrium there is no macroscopic motion. Therefore, $\vec{x} = \vec{a}$ for all particles at all times. This also implies that the number density and specific entropy of the moving particles in the reference state are their equilibrium number density and equilibrium specific entropy, which are assumed to be uniform and time-independent with values $n^{\mathrm{ref}} = n_0, s^{\mathrm{ref}} = s_0$. \par

Similar to the Eulerian case, to fully specify the value of the energy in equilibrium, we assume that the number density of the stationary neutralizing background is uniform, with a value $n_-^{\mathrm{ref}} = n_{-,0}$. We also apply the charge neutrality condition $q_+ n_0 + q_- n_{-,0} = 0$, which, as discussed before, is consistent with the experiments. In the thermodynamic limit, where the number of moving particles $N \to \infty$ and the volume of the system $V \to \infty$ while the number density is fixed, the equilibrium energy of the OCP diverges. Therefore, we consider the equilibrium energy per particle, $E_0/N$. From Eqs.~\eqref{lagrangian_energy_equation}, \eqref{lagrangian_energy_density}, we obtain the following result.
\begin{equation}
\label{lagrangian_equilibrium_energy}
\begin{gathered}
\frac{E_0}{N} = f(n_0, s_0) + \frac{N}{2V^2} \iint (F+F^T)(n_0, n_0, s_0, s_0, |\vec{a}-\vec{a}'|, \vec{a}, \vec{a}') \mathrm{d}\vec{a}\mathrm{d}\vec{a}'- \frac{n_0}{2} \int \frac{q^2_+}{4 \pi \varepsilon_0 |\vec{r}|}\mathrm{d}\vec{r}
\end{gathered}
\end{equation}\par

This expression can be compared to the expression for the thermodynamic equilibrium energy of the OCP per moving particle, $E_{\mathrm{th}}/N$, as given in Eq.~\eqref{thermodynamic_equilibrium_energy}. It is important to note that in thermodynamics, the nonlocal integral term depends solely on $\vec{r} = |\vec{a}-\vec{a}'|$. Therefore, to maintain consistency with this result, we modify our assumption from $F$ depending on $\vec{a}, \vec{a}'$ to instead depend on the combination $|\vec{a}-\vec{a}'|$. Consequently, the integral involving $F+F^T$ in Eq.~\eqref{lagrangian_equilibrium_energy} can also be simplified by changing variables to $\vec{R} = (\vec{a} + \vec{a}')/2, \vec{r} = \vec{a}-\vec{a}'$. \par

When comparing the result to Eq.~\eqref{thermodynamic_equilibrium_energy}, we derive the following constraints on the functions $f, F$. Here, we can use the fact that $F^T(n_0, n_0, s_0, s_0, |\vec{r}|, |\vec{r}|) = F(n_0, n_0, s_0, s_0, |\vec{r}|, |\vec{r}|)$, directly from the definition.
\begin{equation}
\label{lagrangian_energy_constraint_1}
\begin{gathered}
f(n_0, s_0) = \frac{3}{2} k_B T(n_0, s_0), \quad
F(n_0, n_0, s_0, s_0, |\vec{r}|, |\vec{r}|) = \frac{q^2_+}{8 \pi \varepsilon_0 |\vec{r}|}  g(n_0, s_0, |\vec{r}|).
\end{gathered}
\end{equation}\par

In the Lagrangian approach, we encounter the same issue as we did in the Eulerian approach, namely that the constraint in Eq.~\eqref{lagrangian_energy_constraint_1} uniquely determines the function $f$ but not $F$. This occurs because $F$ is specified by this constraint only for values when $n$ is equal to $n'$ and when $s$ is equal to $s'$. As discussed previously, we address this by modifying our choice of $F$ to a more restrictive one that depends solely on $n, s, |\vec{x}-\vec{x}'|, |\vec{a}-\vec{a}'|$. \par

However, in the Lagrangian approach, this still does not uniquely determine $F$ due to its dependence on $\vec{a}, \vec{a}'$ because the constraint is specified only for the case when $|\vec{x}-\vec{x}'|$ is equal to $|\vec{a}-\vec{a}'|$. Now, let us consider four simple possible options for $F$ that are consistent with the equilibrium constraint, but they differ in their dependence on $|\vec{x}-\vec{x}'|$ and $|\vec{a}-\vec{a}'|$.
\begin{equation}
\label{lagrangian_F_option_1}
\begin{gathered}
F(n, s, |\vec{x}-\vec{x}'|, |\vec{a}-\vec{a}'|) = \frac{q^2_+}{8 \pi \varepsilon_0 |\vec{x}-\vec{x}'|}  g(n, s, |\vec{x}-\vec{x}'|),
\end{gathered}
\end{equation}
\begin{equation}
\label{lagrangian_F_option_2}
\begin{gathered}
F(n, s, |\vec{x}-\vec{x}'|, |\vec{a}-\vec{a}'|) = \frac{q^2_+}{8 \pi \varepsilon_0 |\vec{x}-\vec{x}'|}  g(n, s, |\vec{a}-\vec{a}'|),
\end{gathered}
\end{equation}
\begin{equation}
\label{lagrangian_F_option_3}
\begin{gathered}
F(n, s, |\vec{x}-\vec{x}'|, |\vec{a}-\vec{a}'|) = \frac{q^2_+}{8 \pi \varepsilon_0 |\vec{a}-\vec{a}'|}  g(n, s, |\vec{x}-\vec{x}'|),
\end{gathered}
\end{equation}
\begin{equation}
\label{lagrangian_F_option_4}
\begin{gathered}
F(n, s, |\vec{x}-\vec{x}'|, |\vec{a}-\vec{a}'|) = \frac{q^2_+}{8 \pi \varepsilon_0 |\vec{a}-\vec{a}'|}  g(n, s, |\vec{a}-\vec{a}'|).
\end{gathered}
\end{equation}\par

The first option, as provided in Eq.~\eqref{lagrangian_F_option_1}, exactly corresponds to the function $F$ that we considered in the Eulerian approach and will result in the same equations of motion. In contrast, the third and fourth options, given in Eqs.~\eqref{lagrangian_F_option_3}, \eqref{lagrangian_F_option_4}, under the assumption of weak coupling, where $g(n, s, |\vec{r}|) = 1$, will yield $\partial (F+F^T)/\partial x_i = 0$. Consequently, these options will not generate the nonlocal electrostatic force between the moving particles, as described in Eq.~\eqref{lagrangian_final_EOM}. Therefore, we now turn our attention to the second option provided in Eq.~\eqref{lagrangian_F_option_2}. In this case, we can rewrite the equations of motion, as given in Eq.~\eqref{lagrangian_final_EOM}, by using the fact that $\phi_{ij}(r) = q_i q_j/4 \pi \varepsilon_0 r$, where $i, j$ represent the corresponding species of the particles, and $r$ represents the distance between the particles.
\begin{equation}
\label{lagrangian_final_EOM_with_thermo}
\begin{gathered}
m n^{\mathrm{ref}} \frac{\partial^2 x_i}{\partial t^2} = - \sum \limits_{j=1}^3 \frac{\partial (\mathrm{det} \left( \partial \vec{x} / \partial \vec{a} \right))}{\partial (\partial_j x_i)} \frac{\partial}{\partial a_j} \Bigg( \frac{3}{2} k_B n^2 \frac{\partial T}{\partial n}\Big|_s 
+ \frac{q^2_+}{8 \pi \varepsilon_0} \int \frac{n^2 (n^{\mathrm{ref}})'}{|\vec{x}-\vec{x}'|} \frac{\partial g}{\partial n}\Big|_{s, |\vec{a}-\vec{a}'|} \mathrm{d}\vec{a}' \Bigg)\\
- \frac{q_+ n^{\mathrm{ref}}}{4 \pi \varepsilon_0} \int \Bigg( q_+ (n^{\mathrm{ref}})' \frac{(g+g^T)}{2} \frac{\partial}{\partial x_i} \left( \frac{1}{|\vec{x}-\vec{x}'|}\right)\Big|_{\vec{x}'} 
+ q_- (n_-^{\mathrm{ref}})' \frac{\partial}{\partial x_i} \left( \frac{1}{|\vec{x}-\vec{a}'|}\right)\Big|_{\vec{a}'} \Bigg) \mathrm{d}\vec{a}'
\end{gathered}
\end{equation} \par

As one can observe from this equation, the equations of motion in the Lagrangian approach differ from those in the Eulerian approach, as given by Eq.~\eqref{eulerian_final_with_thermo}, due to the nonlocal force term. In this term, now $(g+g^T)/2$ is no longer inside the derivative with respect to $\vec{x}$. Therefore, it is possible to generate the vorticity required for transverse waves.

\subsection{Dispersion laws}
\label{section:dispersion_laws_lagrangian}

While discussing the energy conservation law, we explored equilibrium solutions, where the reference state for the moving particles is chosen such that $\vec{a}$ represents the position of a particle in thermal equilibrium, and as a result, $\vec{x} = \vec{a}$ at all times. Additionally, the number density and specific entropy of the moving particles in this reference state correspond to thermal equilibrium, where they are assumed to be uniform and time-independent, with values $n^{\mathrm{ref}} = n_0, s^{\mathrm{ref}} = s_0$. The stationary neutralizing background also has a uniform number density in the reference state corresponding to thermal equilibrium, with a value $n_-^{\mathrm{ref}} = n_{-,0}$. This value is related to $n_0$ through the charge neutrality assumption $q_+ n_0 + q_- n_{-,0} = 0$. \par

Now, we would like to examine the linearized equations of motion that correspond to the OCP being close to thermodynamic equilibrium. In that scenario, we assume the following form for the displacement field $\vec{x}$, where $\vec{\xi}$ denotes the first-order correction.
\begin{equation}
\begin{gathered}
\vec{x} = \vec{a} + \vec{\xi} (\vec{a}, t)
\end{gathered}
\end{equation}\par

With this assumption about $\vec{x}$, we expand the equations of motion given in Eq.~\eqref{lagrangian_final_EOM_with_thermo} up to the first order. We first checked that our proposed equilibrium solutions indeed satisfy these equations of motion. To analyze the first-order equations, we employ the Fourier transform with respect to the spatial coordinates, $\vec{a}$. The intermediate steps of the derivation are found in Appendix~\ref{appendix_lagrangian_dispersion}. We find that, in contrast to the Eulerian approach, there are now transverse modes, as given by the following transverse dispersion law. It is worth noting that this expression exactly matches the QLCA result \cite{PhysRevA.41.5516, doi:10.1063/1.873814, doi:10.1063/1.4942169}.
\begin{equation}
\label{transverse_dispersion_Lagrangian}
\begin{gathered} 
\omega^2_T(|\vec{k}|) = \frac{q^2_+ n_0}{m \varepsilon_0} \int_0^{\infty} \frac{(g - 1)_0}{|\vec{k}| r^2} \Bigg( \sin(|\vec{k}|r)  
+ 3 \frac{ \cos(|\vec{k}|r)}{|\vec{k}|r} - 3 \frac{\sin(|\vec{k}|r)}{|\vec{k}|^2r^2} \Bigg) \mathrm{d}r
\end{gathered}
\end{equation}\par

For the longitudinal direction, we find the following dispersion law. We reintroduce the notation where the subscript "$0$" signifies that a function is evaluated at equilibrium values. 
\begin{equation}
\label{longitudinal_dispersion_Lagrangian}
\begin{gathered} 
\omega^2_L(|\vec{k}|) = 
\frac{q_+^2 n_0}{m \varepsilon_0} + |\vec{k}|^2 \Bigg( \Bigg[ \frac{\partial}{\partial n} \Big( \frac{3 k_B}{2 m}  n^2 \frac{\partial T}{\partial n}\Big|_s \Big)\Big|_s \Bigg]_0 
+ \frac{q^2_+ n_0}{2 m \varepsilon_0} \int_0^{\infty} r \Bigg[ \frac{\partial}{\partial n} \Big( n^2 \frac{\partial (g-1)}{\partial n}\Big|_{s, r} \Big)\Big|_{s, r} \Bigg]_0 \mathrm{d}r \Bigg) \\
+ \frac{q^2_+ n_0^2}{m \varepsilon_0} \int_0^{\infty} \frac{1}{r}\Bigg[ \frac{\partial (g - 1)}{\partial n}\Big|_{s, r}\Bigg]_0 \left( \frac{\sin(|\vec{k}|r)}{|\vec{k}|r} - \cos(|\vec{k}|r) \right) \mathrm{d}r \\
+ \frac{2 q^2_+ n_0}{m \varepsilon_0}\int_0^{\infty} \frac{(g - 1)_0}{|\vec{k}| r^2} \Bigg( 3 \frac{\sin(|\vec{k}|r)}{|\vec{k}|^2r^2} 
-  3 \frac{ \cos(|\vec{k}|r)}{|\vec{k}|r} - \sin(|\vec{k}|r) \Bigg) \mathrm{d}r
\end{gathered}
\end{equation}\par 

When comparing this to the longitudinal dispersion law obtained in the Eulerian approach, as given in Eq.~\eqref{longitudinal_dispersion_Eulerian}, we observe that the terms independent of $|\vec{k}|$ and the terms purely proportional to $|\vec{k}|^2$ are the same. Furthermore, the terms that do not contain derivatives with respect to the number density $n$ are identical to those given by the QLCA \cite{PhysRevA.41.5516, doi:10.1063/1.873814, doi:10.1063/1.4942169}. \par

From the obtained dispersion laws given in Eqs.~\eqref{transverse_dispersion_Lagrangian}, \eqref{longitudinal_dispersion_Lagrangian}, we see that the transverse dispersion relation depends solely on the equilibrium pair distribution function, whereas the longitudinal dispersion relation also depends on its adiabatic derivatives, similar to the Eulerian approach. This is expected, as the nonlocal contribution to the energy of the OCP depends on the pair distribution function, as shown in Eq.~\eqref{thermodynamic_equilibrium_energy}. Therefore, the adiabatic derivatives of the pair distribution function in the dispersion relation are related to the adiabatic derivatives of the nonlocal energy. This parallels the usual Euler equations \cite{Landau1987Fluid} and shows that the dispersion relation accounts for both local and nonlocal contributions to the total energy of the OCP, consistent with results in thermodynamics. \par

As discussed in the Eulerian approach, it is convenient to express the obtained dispersion laws in the normalized variables. We can again use the results from thermodynamics, which state that $g(n, s, |\vec{r}|) = g(\Gamma, |\vec{r}|/a)$ and that the expressions for equilibrium energy $E_0$ and equilibrium pressure $p_0$ are given by Eq.~\eqref{equilibrium_energy_and_pressure}, which also define the excess internal energy $u$. As before, the variables $\Gamma$ and $a$ will oscillate around their respective equilibrium values $\Gamma_0, a_0$ due to their dependence on $n, s$.\par

In the normalized variables, one gets the following transverse dispersion relation that is parametrized only by the equilibrium value $\Gamma_0$. Again, we emphasize that this result exactly corresponds to the QLCA result \cite{PhysRevA.41.5516, doi:10.1063/1.873814, doi:10.1063/1.4942169}.
\begin{equation}
\label{transverse_dispersion_Lagrangian_normalized}
\begin{gathered} 
\left( \frac{\omega_T}{\omega_p} \right)^2 (\Gamma_0, |\vec{q}_0|) = \int_0^{\infty} \frac{(g - 1)(\Gamma_0, x)}{|\vec{q}_0| x^2} \Bigg( \sin(|\vec{q}_0|x)  
+ 3 \frac{ \cos(|\vec{q}_0|x)}{|\vec{q}_0|x} - 3 \frac{\sin(|\vec{q}_0|x)}{|\vec{q}_0|^2x^2} \Bigg) \mathrm{d}x
\end{gathered}
\end{equation}\par

The result for the longitudinal dispersion relation in the normalized variables that is parametrized only by the equilibrium value $\Gamma_0$ is given next in Eq.~\eqref{longitudinal_dispersion_Lagrangian_normalized}. Here, the terms appearing in the dispersion relation are given by Eqs.~\eqref{eulerian_f1}, \eqref{eulerian_f2}, \eqref{eulerian_js}, \eqref{lagrangian_bs}, \eqref{lagrangian_ells}, and the derivative at constant $s$ is given by Eq.~\eqref{constant_s_derivative}.
\begin{equation}
\label{longitudinal_dispersion_Lagrangian_normalized}
\begin{gathered} 
\left( \frac{\omega_L}{\omega_p} \right)^2 (\Gamma_0, |\vec{q}_0|) = 1 + \frac{|\vec{q}_0|^2}{\Gamma_0}f_2(\Gamma_0) + \left( n \frac{\partial j}{\partial n}\Big|_s \right) (\Gamma_0, |\vec{q}_0|) \\
+ \left( n \frac{\partial }{\partial n} \left( n \frac{\partial j}{\partial n}\Big|_s \right)\Big|_s \right) (\Gamma_0, |\vec{q}_0|) + b(\Gamma_0, |\vec{q}_0|)
+ \left( n \frac{\partial \ell}{\partial n}\Big|_s \right) (\Gamma_0, |\vec{q}_0|),
\end{gathered}
\end{equation}
\begin{equation}
\label{lagrangian_bs}
\begin{gathered}
b(\Gamma, |\vec{q}|) = 2 \int_0^{\infty} \frac{\big( g - 1 \big)(\Gamma, x)}{|\vec{q}|x^2}\Bigg( 3 \frac{\sin(|\vec{q}|x)}{|\vec{q}|^2x^2} 
-  3 \frac{ \cos(|\vec{q}|x)}{|\vec{q}|x} - \sin(|\vec{q}|x) \Bigg) \mathrm{d}x,
\end{gathered}
\end{equation}
\begin{equation}
\label{lagrangian_ells}
\begin{gathered}
\ell(\Gamma, |\vec{q}|) = \int_0^{\infty} \frac{\big( g - 1 \big)(\Gamma, x)}{x} \left( \frac{\sin(|\vec{q}|x)}{|\vec{q}|x} - \cos(|\vec{q}|x) \right) \mathrm{d}x.
\end{gathered}
\end{equation}\par

As in the Eulerian approach, in principle, the obtained dispersion laws in the normalized variables can now be computed for an arbitrary pair distribution function, $g(\Gamma, |\vec{r}|/a)$. However, as before, the longitudinal dispersion relation depends on derivatives with respect to $\Gamma$ of integrals that include the pair distribution function. This means that for a precise evaluation of the longitudinal dispersion law, a pair distribution function should be known precisely as a function of $\Gamma$. However, as discussed in detail while analyzing the Eulerian approach, such precise fits are not known in the literature. Only the precise dependence of the excess internal energy, $u(\Gamma)$, is known as a function of $\Gamma$. \par

To resolve this issue, we will use the same approximation as discussed in the Eulerian approach, where the pair distribution function, $g(\Gamma, |\vec{r}|/a)$, takes the form of a step function with a value of $0$ when $|\vec{r}|/a < (R/a)(\Gamma)$, and $1$ when $|\vec{r}|/a > (R/a)(\Gamma)$. The dependence of $R/a$ on $\Gamma$ is obtained by computing the equilibrium energy given in Eq.~\eqref{thermodynamic_equilibrium_energy} for the assumed approximation. We use Eq.~\eqref{equilibrium_energy_and_pressure} to relate it to $u(\Gamma)$, the result of which is given in Eq.~\eqref{R_over_a}. This approximation simplifies the integrals appearing in the dispersion law, leading to simpler and more practical expressions and allowing us to avoid additional numerical errors due to the computation of the integral. The transverse dispersion relation is now given in Eq.~\eqref{transverse_dispersion_Lagrangian_normalized_simplified}. In this equation, $R/a$ is taken to be the correct value for a given $\Gamma_0$ and agrees with the QLCA result for the same pair distribution function approximation \cite{doi:10.1063/1.4942169}. 
\begin{equation}
\label{transverse_dispersion_Lagrangian_normalized_simplified}
\begin{gathered} 
\left( \frac{\omega_T}{\omega_p} \right)^2 (\Gamma_0, |\vec{q}_0|) = \frac{1}{3} + \frac{\cos \left( |\vec{q}_0| R/a \right)}{(|\vec{q}_0| R/a)^2} - \frac{\sin \left( |\vec{q}_0| R/a \right)}{(|\vec{q}_0| R/a)^3}
\end{gathered}
\end{equation}\par

The integrals in the longitudinal dispersion relation, as given by Eqs.~\eqref{eulerian_js}, \eqref{lagrangian_bs}, \eqref{lagrangian_ells}, can be rewritten as shown in Eqs.~\eqref{simplified_j_h_Eulerian}, \eqref{simplified_b_ell_Lagrangian}, with $R/a$ taken to be the correct value for a given $\Gamma$.
\begin{equation}
\label{simplified_b_ell_Lagrangian}
\begin{gathered}
b(\Gamma, |\vec{q}|) = -\frac{2}{3} - 2 \frac{\cos \left( |\vec{q}| R/a \right)}{(|\vec{q}| R/a)^2} + 2 \frac{\sin \left( |\vec{q}| R/a \right)}{(|\vec{q}| R/a)^3}, \quad
\ell(\Gamma, |\vec{q}|) = \frac{\sin \left( |\vec{q}| R/a \right)}{(|\vec{q}| R/a)} - 1.
\end{gathered}
\end{equation}\par

\begin{figure*}
\includegraphics[width=5.4in, height=1.8in]{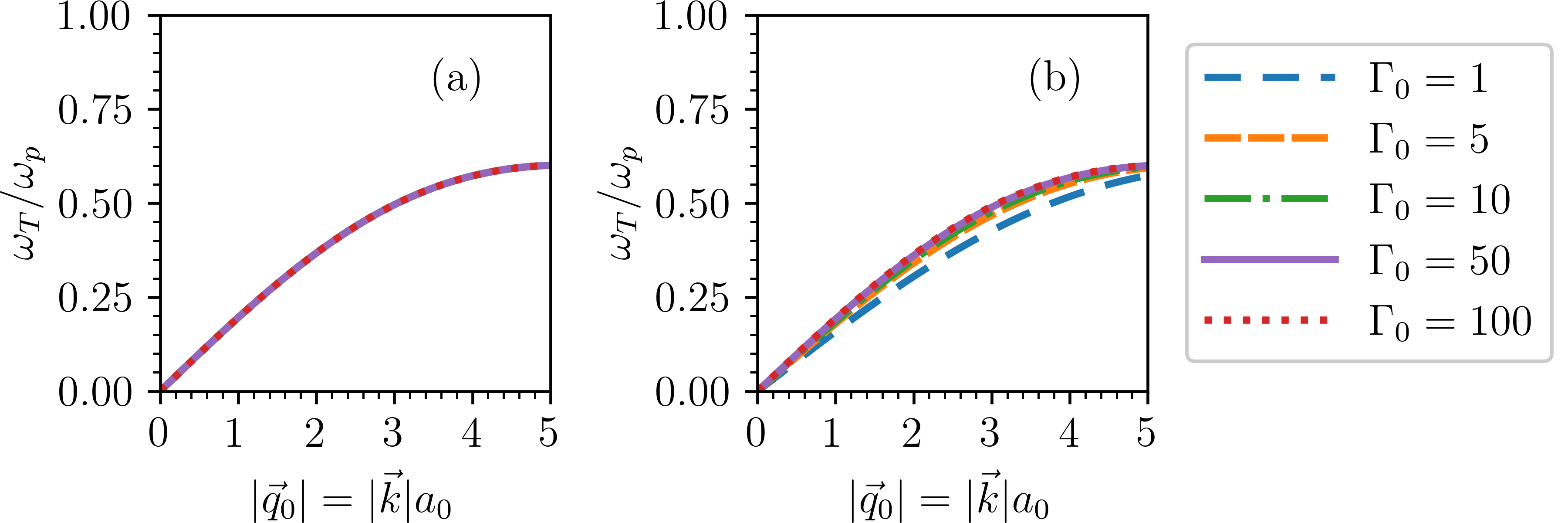} 
\caption{The transverse dispersion law in the Lagrangian approach, expressed in normalized variables, is shown for different values of $\Gamma_0$ as given by Eq.~\eqref{transverse_dispersion_Lagrangian_normalized_simplified}. We assume the step function approximation for the pair distribution function for a given $u(\Gamma)$ defined in Eq.~\eqref{equilibrium_energy_and_pressure}. In (a), $u(\Gamma) = -0.9\Gamma$. In (b), $u(\Gamma) = -0.9\Gamma + 0.5944\Gamma^{1/3} - 0.2786$.}
\label{fig:ocp_paper_lagrangian_transverse}
\end{figure*}

Now, let us consider different fits for the excess internal energy $u(\Gamma)$ as defined in Eq.~\eqref{equilibrium_energy_and_pressure}. First, let us consider a simple fit $u(\Gamma) = -0.9 \Gamma$ that is accurate for a very strong coupling where $\Gamma \gg 1$ \cite{doi:10.1063/1.4897386}. For such a fit, $R/a = \sqrt{6/5}$, independent of the value of $\Gamma$. From Eq.~\eqref{transverse_dispersion_Lagrangian_normalized_simplified} it can be observed that in this approximation the transverse dispersion relation is independent of $\Gamma$. The results for various values of $\Gamma_0$ are shown in Fig.~\ref{fig:ocp_paper_lagrangian_transverse}(a). \par

The longitudinal dispersion law given in Eq.~\eqref{longitudinal_dispersion_Lagrangian_normalized}, supplemented with Eq.~\eqref{eulerian_f2} and the simplified expressions in Eqs.~\eqref{simplified_j_h_Eulerian}, \eqref{simplified_b_ell_Lagrangian}, becomes as follows.
\begin{equation}
\label{simplified_Lagrangian_dispersion_simple_fit}
\begin{gathered}
\left( \frac{\omega_L}{\omega_p} \right)^2 (\Gamma_0, |\vec{q}_0|) = \frac{1}{3} + \left( \frac{5}{9\Gamma_0} + \frac{1}{15} \right) |\vec{q}_0|^2 \\
+ \left( \frac{1}{3} + \frac{2}{\left( \sqrt{\frac{6}{5}}|\vec{q}_0| \right)^2} \right) \left(\frac{\sin \left( \sqrt{\frac{6}{5}} |\vec{q}_0| \right)}{\sqrt{\frac{6}{5}}|\vec{q}_0|} -  \cos \left( \sqrt{\frac{6}{5}} |\vec{q}_0| \right) \right)
\end{gathered}
\end{equation}\par

\begin{figure*}
\includegraphics[width=5.4in, height=1.8in]{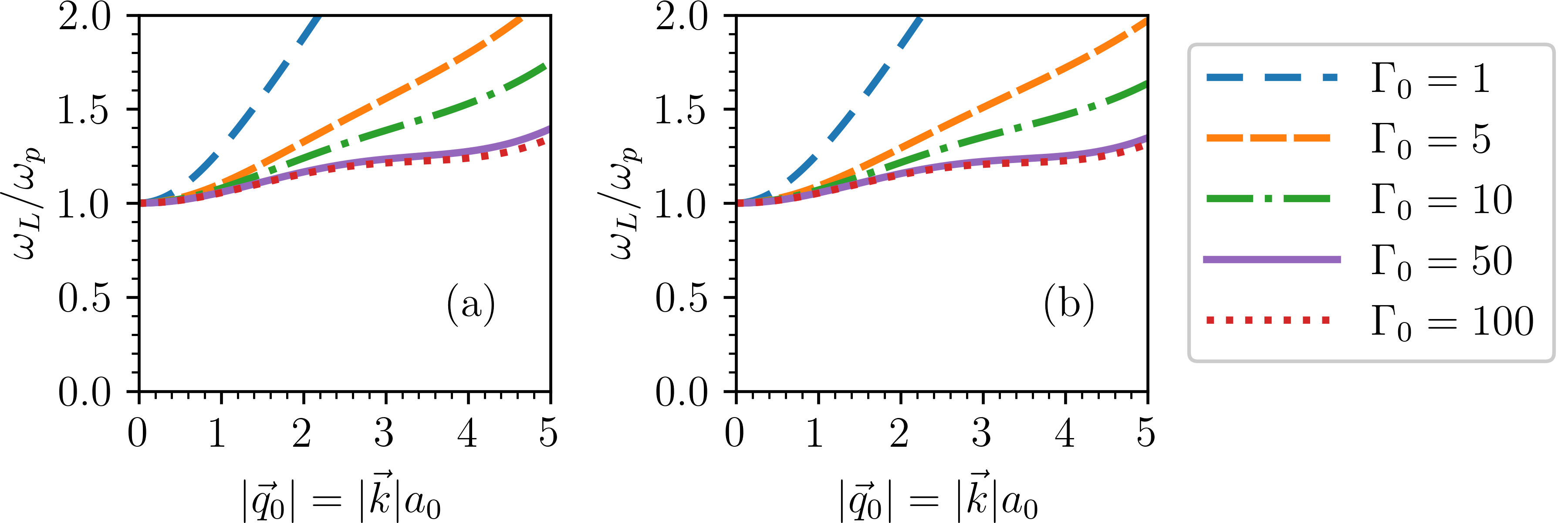} 
\caption{The longitudinal dispersion law in the Lagrangian approach, expressed in normalized variables, is shown for different values of $\Gamma_0$ as given by Eq.~\eqref{longitudinal_dispersion_Lagrangian_normalized}. We assume the step function approximation for the pair distribution function, and therefore, we use Eqs.~\eqref{eulerian_f1}, \eqref{eulerian_f2}, \eqref{R_over_a}, \eqref{simplified_j_h_Eulerian}, \eqref{simplified_b_ell_Lagrangian} for a given $u(\Gamma)$ defined in Eq.~\eqref{equilibrium_energy_and_pressure}. In (a), $u(\Gamma) = -0.9\Gamma$, and the result is provided by Eq.~\eqref{simplified_Lagrangian_dispersion_simple_fit}. In (b), $u(\Gamma) = -0.9\Gamma + 0.5944\Gamma^{1/3} - 0.2786$.}
\label{fig:lagrangian_longitudinal}
\end{figure*}

The results for different values of $\Gamma_0$ are presented in Fig.~\ref{fig:lagrangian_longitudinal}(a). Several key properties can be observed from the figure. Similar to the Eulerian approach, for weak coupling where $\Gamma_0 = 1$, the dispersion relation resembles the behavior of an ideal gas. However, as $\Gamma_0$ is increased, the behavior changes dramatically. Nonetheless, at no value of $\Gamma_0$ do we observe the onset of negative dispersion because for small values of $|\vec{q}_0|$, $\omega_L \geq \omega_p$. Furthermore, this remains true for all values of $|\vec{q}_0|$. 

 \par

The simpler fit for $u(\Gamma)$ does not account for thermal effects in the nonlocal interaction. Therefore, we also consider a more precise fit, $u(\Gamma) = -0.9\Gamma + 0.5944\Gamma^{1/3} - 0.2786$, which is accurate for $\Gamma \geq 1$ and includes thermal effects \cite{doi:10.1063/1.4897386}. The results for different values of $\Gamma_0$ for the transverse dispersion law are shown in Fig.~\ref{fig:ocp_paper_lagrangian_transverse}(b), and for the longitudinal dispersion law, they are presented in Fig.~\ref{fig:lagrangian_longitudinal}(b). Similar to the Eulerian approach, it is evident that the more complex fit for $u(\Gamma)$ yields results that are qualitatively the same as in the case of the simpler fit, with minor quantitative differences. \par

The results for the more precise fit are also compared to the results obtained by QLCA \cite{doi:10.1063/1.4942169} using the same approximation of the pair distribution function. The results are also compared to the current fluctuation spectra obtained from molecular dynamics simulations provided by the authors of Ref.~\onlinecite{https://doi.org/10.1002/ctpp.201400098}, where dispersion laws are identified by the peaks of the spectra. \par

Results for QLCA and molecular dynamics for the longitudinal dispersion law can be seen in Fig.~\ref{fig:comparison_of_longitudinal}. It can be observed that as $\Gamma_0$ increases, the predicted dispersion law starts to differ from the results of the molecular dynamics simulations. In particular, the latter shows the onset of negative dispersion in the range of values from $\Gamma_0 = 9.5$ to $\Gamma_0 = 10.0$ \cite{https://doi.org/10.1002/ctpp.201400098, doi:10.1063/1.3679586}, contrary to the predicted dispersion law that has no such onset. Moreover, the predicted longitudinal dispersion law, in the limit as $\Gamma_0 \to \infty$, does not tend to the QLCA result, which is considered accurate in such a limit \cite{doi:10.1063/1.4942169}. \par

Results for QLCA and molecular dynamics for the transverse dispersion law can be seen in Fig.~\ref{fig:comparison_of_transverse}. The predicted transverse dispersion law, at all values of $\Gamma_0$, agrees with the QLCA result for the same pair distribution function approximation. It also agrees with the results of molecular dynamics simulations at large values of wavevector and correctly predicts that the transverse dispersion law tends to the value $\omega_p/\sqrt{3}$ in the limit as $|\vec{q}_0| = |\vec{k}|a_0 \to \infty$. However, it fails to predict the numerically observed disappearance of the transverse waves at small values of wavevector \cite{10.1063/1.5088141}. Nonetheless, as in the case of QLCA, this issue can be remedied by considering relaxation in the context of generalized hydrodynamics \cite{10.1063/1.5088141}. \par

Compared to the Eulerian approach, the discussed Lagrangian approach now accurately predicts the transverse dispersion law for large values of wavevector. However, there are still issues with the longitudinal dispersion law. In the Lagrangian approach, there is no longer an onset of negative dispersion, and the results still do not agree with the results from QLCA and molecular dynamics simulations at high values of $\Gamma_0$. This motivates us to consider the final alternative variational approach, discussed in the next Section~\ref{section:dispersion_laws_modified_lagrangian}.

\subsection{Modified Lagrangian approach}
\label{section:dispersion_laws_modified_lagrangian}

The difference in the previously described Eulerian and Lagrangian variational principles lies in how the pair distribution function is assumed to behave out of equilibrium. Specifically, in the Eulerian approach, the pair distribution function is a function of number density $n$, specific entropy $s$, and the difference in positions in the laboratory frame $|\vec{x}-\vec{x}'|$. In contrast, in the Lagrangian approach, it is assumed to depend on $n, s$, and the difference in positions in the reference state $|\vec{a}-\vec{a}'|$. Such a difference results in distinct dispersion laws. For example, it determines whether there are transverse dispersion modes and whether there are regions where longitudinal dispersion modes become unstable with $\omega_L^2 < 0$. \par

Notice that there are other possible changes to how the pair distribution function behaves out of equilibrium. For example, instead of depending on $n$, it might depend on the number density $n^{\mathrm{ref}}$ in a reference state. Similarly, instead of depending on $s$, it might depend on the specific entropy $s^{\mathrm{ref}}$ in a reference state. Due to the continuity equations given by Eq.~\eqref{lagrangian_continuity_equations}, changing the dependence from $s$ to $s^{\mathrm{ref}}$ would not change the results, but changing it from $n$ to $n^{\mathrm{ref}}$ might lead to significant differences. This possibility may be of particular interest, given that the transverse dispersion law in the Lagrangian approach reasonably agrees with the molecular dynamics simulations, unlike the longitudinal dispersion law. Notice that, as shown by Eqs.~\eqref{transverse_dispersion_Lagrangian}, \eqref{longitudinal_dispersion_Lagrangian}, only the longitudinal dispersion law depends on derivatives of the pair distribution function with respect to number density. Therefore, only the longitudinal dispersion law would be sensitive to the discussed change. This might allow us to improve the longitudinal dispersion law without affecting the transverse modes. \par

In the case of the local energy term $f$, both of the previously considered variational approaches had it such that, out of all thermodynamic functions, it depends only on temperature $T$. To be consistent with the Euler equations \cite{herivel_1955, doi:10.1098/rspa.1968.0103, lin_1963}, it is necessary to have it depend on $n,s$ instead of $n^{\mathrm{ref}}, s^{\mathrm{ref}}$. This motivates the consideration of the pair distribution function out of equilibrium as a function of temperature $T$ rather than specific entropy $s$, and to assume that $T$ depends on $n,s$, and not on $n^{\mathrm{ref}}, s^{\mathrm{ref}}$. \par

However, there is a dependence on number density in the pair distribution function that does not come from the temperature, and there is no reason to believe it must depend on $n$ rather than $n^{\mathrm{ref}}$. In particular, both of our previously considered approaches assumed dependence on $n$, so we would like to explore what happens if one assumes dependence on $n^{\mathrm{ref}}$ instead. In other words, we assume that the pair distribution function out of equilibrium is a function of the form $g(n, s, |\vec{a}-\vec{a}'|) = g(n^{\mathrm{ref}}, T(n, s), |\vec{a}-\vec{a}'|)$. \par

With such an assumption regarding the pair distribution function, the equations of motion remain as given by Eq.~\eqref{lagrangian_final_EOM_with_thermo}. The momentum and energy laws also stay the same, with the same identification of functions $f, F$ as provided in Eqs.~\eqref{lagrangian_energy_constraint_1}, \eqref{lagrangian_F_option_2}. The transverse and longitudinal dispersion laws are also retained and described by Eqs.~\eqref{transverse_dispersion_Lagrangian}, \eqref{longitudinal_dispersion_Lagrangian}. However, one must be careful with computing derivatives at constant $s$. \par

As previously mentioned, it is convenient to rewrite the obtained dispersion laws in normalized variables. We again make use of results from thermodynamics, stating that in equilibrium, $g(n_0, s_0, |\vec{r}|) = g(\Gamma_0, |\vec{r}|/a_0)$, and the expressions for equilibrium energy $E_0$ and equilibrium pressure $p_0$ are provided in Eq.~\eqref{equilibrium_energy_and_pressure}, which also define the excess internal energy $u$. With our new assumption regarding how $g$ behaves out of equilibrium, we have $g(n, s, |\vec{r}|) = g(\Gamma(n_0, T(n,s)), |\vec{r}|/a_0)$. Consequently, $\Gamma$ will oscillate around its respective equilibrium value $\Gamma_0$ due to its dependence on $n,s$. However, the value of $a$ remains fixed during the motion of the system at its equilibrium value $a_0$. \par

The transverse dispersion relation in the normalized variables is the same as in the Lagrangian approach, as given by Eq.~\eqref{transverse_dispersion_Lagrangian_normalized}, as it does not depend on number density derivatives at constant entropy. However, the longitudinal dispersion relation in the normalized variables is now different. For each equilibrium value $\Gamma_0$, it is provided in Eq.~\eqref{longitudinal_dispersion_modified_Lagrangian_normalized}. The terms appearing in the dispersion relation are given by Eqs.~\eqref{eulerian_f1}, \eqref{eulerian_f2}, \eqref{eulerian_js}, \eqref{lagrangian_bs}, \eqref{lagrangian_ells}, \eqref{modified_lagrangian_j1}, \eqref{modified_lagrangian_j2}, \eqref{modified_lagrangian_l1}.
\begin{equation}
\label{longitudinal_dispersion_modified_Lagrangian_normalized}
\begin{gathered} 
\left( \frac{\omega_L}{\omega_p} \right)^2 (\Gamma_0, |\vec{q}_0|) = 1 + \frac{|\vec{q}_0|^2}{\Gamma_0}f_2(\Gamma_0) + j_1 (\Gamma_0, |\vec{q}_0|)
+ j_2(\Gamma_0, |\vec{q}_0|) + b(\Gamma_0, |\vec{q}_0|) + \ell_1 (\Gamma_0, |\vec{q}_0|),
\end{gathered}
\end{equation}
\begin{equation}
\label{modified_lagrangian_j1}
\begin{gathered} 
j_1(\Gamma, |\vec{q}|) = -\frac{2\Gamma }{3}\frac{\partial j}{\partial \Gamma}\Big|_{|\vec{q}|} (\Gamma, |\vec{q}|) \bigg( f_1(\Gamma) - \frac{2}{3}f_1^2(\Gamma)
+ \frac{\Gamma}{3} \frac{\mathrm{d}f_1}{\mathrm{d}\Gamma}(\Gamma) \Big(1 - 2f_1(\Gamma) \Big) \bigg),
\end{gathered}
\end{equation}
\begin{equation}
\label{modified_lagrangian_j2}
\begin{gathered} 
j_2(\Gamma, |\vec{q}|) = \frac{4\Gamma^2}{9} f_1^2(\Gamma)  \frac{\partial^2 j}{\partial \Gamma^2}\Big|_{|\vec{q}|}  (\Gamma, |\vec{q}|),
\end{gathered}
\end{equation}
\begin{equation}
\label{modified_lagrangian_l1}
\begin{gathered} 
\ell_1(\Gamma, |\vec{q}|) = -\frac{2\Gamma }{3} f_1(\Gamma) \frac{\partial \ell}{\partial \Gamma}\Big|_{|\vec{q}|} (\Gamma, |\vec{q}|).
\end{gathered}
\end{equation} \par

As discussed in both previous variational approaches in detail, due to insufficient numerical data regarding how the pair distribution function $g(\Gamma, |\vec{r}|/a)$ depends on $\Gamma$, we use a step function approximation to simplify integrals and make calculations more practical. The step function approximation for the pair distribution function is such that it has a value of $0$ when $|\vec{r}|/a < (R/a)(\Gamma)$, and $1$ when $|\vec{r}|/a > (R/a)(\Gamma)$. The dependence of $R/a$ on $\Gamma$ is determined by computing the equilibrium energy, as given in Eq.~\eqref{thermodynamic_equilibrium_energy}, for the given approximation. We then use Eq.~\eqref{equilibrium_energy_and_pressure} to relate it to $u(\Gamma)$. The result of this computation is provided in Eq.~\eqref{R_over_a}. \par

In this case, the transverse dispersion relation is provided in Eq.~\eqref{transverse_dispersion_Lagrangian_normalized_simplified}, where $R/a$ is the correct value for a given $\Gamma_0$. This result agrees with the QLCA result for the same pair distribution function approximation \cite{doi:10.1063/1.4942169}. The integrals found in the longitudinal dispersion relation, as given by Eqs.~\eqref{eulerian_js}, \eqref{lagrangian_bs}, \eqref{lagrangian_ells}, can be simplified as before, as described by Eqs.~\eqref{simplified_j_h_Eulerian}, \eqref{simplified_b_ell_Lagrangian}, where $R/a$ is the correct value for a given $\Gamma$. \par

Now, let us consider different fits for the excess internal energy, $u(\Gamma)$, as defined in Eq.~\eqref{equilibrium_energy_and_pressure}. First, let us examine a simple fit $u(\Gamma) = -0.9 \Gamma$, which is accurate for very strong coupling where $\Gamma \gg 1$ \cite{doi:10.1063/1.4897386}. As discussed previously, for such a fit, $R/a = \sqrt{6/5}$, independent of the value of $\Gamma$. The results for the transverse dispersion relation for different values of $\Gamma_0$ are presented as before in Fig.~\ref{fig:ocp_paper_lagrangian_transverse}(a). \par 

The longitudinal dispersion law, as given in Eq.~\eqref{longitudinal_dispersion_modified_Lagrangian_normalized} and supplemented with Eq.~\eqref{eulerian_f2} and the simplified expressions in Eqs.~\eqref{simplified_j_h_Eulerian}, \eqref{simplified_b_ell_Lagrangian}, becomes as follows. It is worth noting that this result is exactly the ideal gas speed of sound term combined with the QLCA expression for the same pair distribution function approximation \cite{doi:10.1063/1.4942169}. The idea of phenomenologically combining the ideal gas speed of sound term with the QLCA expression has appeared in the literature \cite{10.1063/1.4965903, PhysRevE.102.033207}. However, here we have provided a more rigorous justification for such a result, as it was derived from a general theoretical framework. Additionally, notice that this result only holds when $u(\Gamma) = -0.9\Gamma$, and our theory allows for a consistent generalization to the case when the functional form of $u(\Gamma)$ is more complicated.
\begin{equation}
\label{simplified_modified_Lagrangian_dispersion_simple_fit}
\begin{gathered}
\left( \frac{\omega_L}{\omega_p} \right)^2 (\Gamma_0, |\vec{q}_0|) = \frac{1}{3} + \frac{5}{9} \frac{|\vec{q}_0|^2}{\Gamma_0} - 2 \frac{\cos \left( \sqrt{\frac{6}{5}} |\vec{q}_0| \right)}{\left( \sqrt{\frac{6}{5}}|\vec{q}_0| \right)^2}
+ 2 \frac{\sin \left( \sqrt{\frac{6}{5}} |\vec{q}_0| \right)}{\left( \sqrt{\frac{6}{5}}|\vec{q}_0| \right)^3}
\end{gathered}
\end{equation}\par

\begin{figure*}
\includegraphics[width=5.4in, height=1.8in]{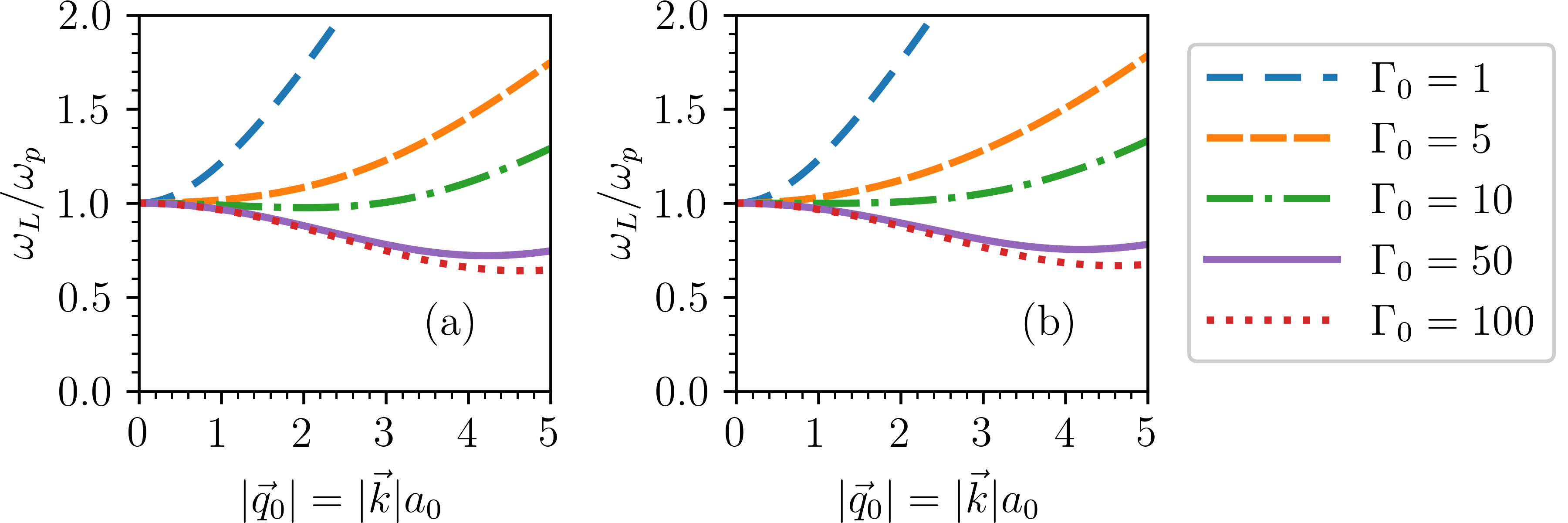} 
\caption{The longitudinal dispersion law in the modified Lagrangian approach, expressed in normalized variables, is shown for different values of $\Gamma_0$ as given by Eq.~\eqref{longitudinal_dispersion_modified_Lagrangian_normalized}. We assume the step function approximation for the pair distribution function, and therefore, we use Eqs.~\eqref{eulerian_f1}, \eqref{eulerian_f2}, \eqref{R_over_a}, \eqref{simplified_j_h_Eulerian}, \eqref{simplified_b_ell_Lagrangian} for a given $u(\Gamma)$ defined in Eq.~\eqref{equilibrium_energy_and_pressure}. In (a), $u(\Gamma) = -0.9\Gamma$, and the result is provided by Eq.~\eqref{simplified_modified_Lagrangian_dispersion_simple_fit}. In (b), $u(\Gamma) = -0.9\Gamma + 0.5944\Gamma^{1/3} - 0.2786$.}
\label{fig:modified_lagrangian_longitudinal}
\end{figure*}

The results for different values of $\Gamma_0$ are presented in Fig.~\ref{fig:modified_lagrangian_longitudinal}(a), and the following properties can be observed from the figure. As in both previous approaches, for weak coupling where $\Gamma_0 = 1$, the dispersion relation resembles the behavior of an ideal gas. However, as $\Gamma_0$ increases, the behavior undergoes a dramatic change. Similar to the Eulerian approach, but unlike the original Lagrangian approach, there is a critical value of $\Gamma_0 = 7.0$ at which we observe the onset of negative dispersion. In other words, for small values of $|\vec{q}_0|$, $\omega_L \leq \omega_p$ instead of $\omega_L \geq \omega_p$. However, unlike the Eulerian approach but similar to the original Lagrangian approach, there are no unstable regions. This means that for all $\Gamma_0$, we find that $\omega_L^2 > 0$ for all $|\vec{q}_0|$. 

 \par

The simpler fit for $u(\Gamma)$ does not account for thermal effects in the nonlocal interaction. Therefore, we also consider a more precise fit, $u(\Gamma) = -0.9\Gamma + 0.5944\Gamma^{1/3} - 0.2786$, which is accurate for $\Gamma \geq 1$ and includes thermal effects \cite{doi:10.1063/1.4897386}. \par

The results for different values of $\Gamma_0$ for the transverse dispersion law are presented in Fig.~\ref{fig:ocp_paper_lagrangian_transverse}(b), and for the longitudinal dispersion law, they are shown in Fig.~\ref{fig:modified_lagrangian_longitudinal}(b). As in both previous approaches, one can observe that the more complicated fit for $u(\Gamma)$ yields results that are qualitatively the same as in the case of the simpler fit, with small quantitative differences. For instance, the onset of negative dispersion is now observed at $\Gamma_0 = 9.5$. \par

The results for the more precise fit are also compared to the results obtained by QLCA \cite{doi:10.1063/1.4942169} for the same approximation of the pair distribution function, as well as to current fluctuation spectra obtained from molecular dynamics simulations provided by the authors of Ref.~\onlinecite{https://doi.org/10.1002/ctpp.201400098}, where dispersion laws are identified by the peaks in the spectra. \par

The comparison for the longitudinal dispersion law is presented in Fig.~\ref{fig:comparison_of_longitudinal}. The predicted dispersion law agrees with the molecular dynamics simulations. For $\Gamma_0 = 1, 5, 10$, the agreement extends through the range of $0 < |\vec{q}_0| < 5$. For the large value $\Gamma_0 = 80$, the agreement is observed for the range of $0 < |\vec{q}_0| < 2$. Additionally, the predicted onset of negative dispersion at $\Gamma_0 = 9.5$ aligns excellently with the range of values from $\Gamma_0 = 9.5$ to $\Gamma_0 = 10.0$, estimated by the molecular dynamics simulations \cite{https://doi.org/10.1002/ctpp.201400098, doi:10.1063/1.3679586}. Moreover, the predicted longitudinal dispersion law in the limit as $\Gamma_0 \to \infty$ tends to the QLCA result, which is considered accurate in such a limit \cite{doi:10.1063/1.4942169}. This can be observed for a simpler fit for $u(\Gamma)$ by taking the limit $\Gamma_0 \to \infty$ directly in Eq.~\eqref{simplified_modified_Lagrangian_dispersion_simple_fit}, while for the more complicated fit for $u(\Gamma)$, it was confirmed numerically. Consequently, in such a strong coupling limit, the predicted longitudinal dispersion law, following the QLCA result, tends towards the value $\omega_p/\sqrt{3}$ in the limit as $|\vec{q}_0| = |\vec{k}|a_0 \to \infty$. \par

The comparison for the transverse dispersion law is presented in Fig.~\ref{fig:comparison_of_transverse}. Similar to the previous case of the Lagrangian approach, the predicted dispersion law at all $\Gamma_0$ exactly agrees with the QLCA result and aligns with the results of molecular dynamics simulations at large values of wavevector. It correctly predicts that the transverse dispersion law tends to the value $\omega_p/\sqrt{3}$ in the limit as $|\vec{q}_0| = |\vec{k}|a_0 \to \infty$. However, it fails to predict the numerically observed disappearance of transverse waves at small values of the wavevector \cite{10.1063/1.5088141}. Nonetheless, similar to the case of QLCA, this can be remedied by considering relaxation in the context of generalized hydrodynamics \cite{10.1063/1.5088141}. \par

\section{Discussion and Conclusion}

We have explored variational principles for the hydrodynamics of the classical OCP in both the Eulerian approach to hydrodynamics, where the hydrodynamic functions depend on the position in the laboratory frame, and the Lagrangian approach, where the hydrodynamic functions depend on some reference position, such as the initial or equilibrium position of a fluid particle. We motivated Lagrangian densities that are used in the variational principles and showed how to obtain equations of motion and conservation laws for momentum and energy when the variational principle is nonlocal. In the Eulerian approach, one has to introduce constraint fields in the variational principle to obtain continuity equations for number density and specific entropy. Because of that, one has to be careful when eliminating the introduced constraint fields from the resulting equations, as we have shown. In the Lagrangian approach, such issues with constraint fields do not occur. \par

Consistency with equilibrium results from thermodynamics was ensured by employing the energy conservation law and calculating the energy expression from a given variational principle. This expression was evaluated in equilibrium and matched with the known thermodynamic expression. Consequently, all the considered variational principles yield the same result for equilibrium energy, even though their out-of-equilibrium behavior differs due to various assumptions regarding the pair distribution function. To fully specify each of the considered variational approaches, we also assumed that the out-of-equilibrium pair distribution function depends solely on number densities and specific entropies in terms of $n, s$, rather than using more complex models such as $(n+n')/2, (s+s')/2$ or $\sqrt{nn'}, \sqrt{ss'}$. \par

To analyze different proposed variational principles, we obtained linearized equations of motion and calculated longitudinal and transverse dispersion laws. In all of these approaches, the dispersion laws depend not only on the pair distribution function but also on its adiabatic derivatives. This is similar to the case of Euler equations, where dispersion laws depend on the adiabatic derivatives of energy. However, in our case, we have included a nonlocal contribution to the energy of the OCP that depends on the pair distribution function. Nevertheless, we are not aware of precise numerical data regarding the adiabatic derivatives of the pair distribution function. \par

Due to this lack of data, we used a simple step function approximation to the pair distribution function. We chose the radius for the transition of the step function to be consistent with the equilibrium energy expression. This approach allows us to significantly simplify integrals that appear in the dispersion laws. We hope that our accurate results will serve as motivation to perform a more careful numerical analysis of the pair distribution function in equilibrium in the future. With precise knowledge of its adiabatic derivatives, we can avoid the simple step function approximation we used and obtain even more accurate results. \par

In each of the considered variational approaches, dispersion laws were computed using two fits for the nonlocal contribution to the energy of the OCP. First, a simpler fit was considered, which is valid for large values of the plasma parameter, $\Gamma \gg 1$, and for which simple analytical expressions for the dispersion laws can be obtained, see Eqs.~\eqref{simplified_Eulerian_dispersion_simple_fit}, \eqref{transverse_dispersion_Lagrangian_normalized_simplified}, \eqref{simplified_Lagrangian_dispersion_simple_fit}, \eqref{simplified_modified_Lagrangian_dispersion_simple_fit}. Then, a more complicated fit that is valid for $\Gamma > 1$ was considered, for which dispersion laws were computed numerically. We found that compared to the simpler fit with its simple analytical expressions, the results are qualitatively the same, with some small quantitative differences. This can be seen in Figs.~\ref{fig:eulerian_longitudinal}, \ref{fig:ocp_paper_lagrangian_transverse}, \ref{fig:lagrangian_longitudinal}, \ref{fig:modified_lagrangian_longitudinal}. \par

Finally, we compared the obtained dispersion laws for different variational approaches across a wide range of equilibrium values of the plasma parameter, $\Gamma_0$, to the results of theoretical calculations using QLCA and the numerical results of molecular dynamics simulations. The results can be observed in Figs.~\ref{fig:comparison_of_longitudinal}, \ref{fig:comparison_of_transverse}. Indeed, we can see that different variational approaches yield significantly different results for the dispersion laws, emphasizing the importance of understanding how the pair distribution function behaves out of equilibrium within the variational principles. \par

From the comparison of the transverse dispersion laws in Fig.~\ref{fig:comparison_of_transverse}, we can observe that for small values of the wavevector, $|\vec{q}_0|$, the Eulerian variational approach correctly predicts the absence of transverse waves in the system. Meanwhile, for large values of the wavevector, the modified Lagrangian approach agrees with the QLCA result and correctly predicts the presence of transverse waves. As with QLCA, one can improve our models by introducing relaxation, as done in generalized hydrodynamics, which allows for the accurate prediction of the disappearance of transverse waves for small wavevector values. The transverse dispersion law does not differentiate all of the variational principles from the QLCA result. To understand the differences between theoretical models, it is, therefore, better to examine the longitudinal dispersion law, as shown in Fig.~\ref{fig:comparison_of_longitudinal}. \par

From the comparison of the longitudinal dispersion laws, we conclude that the modified Lagrangian variational principle provides very accurate results for all values of the wavevector, $|\vec{q}_0|$, and for all considered values of $\Gamma_0$ when compared to the results of molecular dynamics calculations. It is also more accurate than the other variational approaches considered and QLCA. In the regime of large wavelengths and small wavevectors, the modified Lagrangian variational approach correctly predicts the onset of negative dispersion to occur at $\Gamma_0 = 9.5$, in agreement with the range of values $9.5 \leq \Gamma_0 \leq 10.0$ predicted by the molecular dynamics simulations. In the limit of strong coupling, where $\Gamma_0 \to \infty$, the modified Lagrangian variational approach tends towards the QLCA result. Thus, in the regime of small wavelengths and large wavevectors, it predicts that both longitudinal and transverse dispersion laws tend towards the finite value $\omega_p/\sqrt{3}$ as $|\vec{q}_0| = |\vec{k}|a_0 \to \infty$. \par

The excellent agreement of the dispersion laws predicted by the modified Lagrangian variational principle, as compared to the molecular dynamics simulations, shows that our variational hydrodynamic approach can be used to describe motion on small length scales where the wavelength is comparable to the average distance between the particles. This suggests that one can apply this general variational approach with reasonable confidence in the future to nonlinear problems of the OCP, such as the motion of vortices and adiabatic expansion, where molecular dynamics data might not be available. It also suggests that one can apply it to other long-range systems, for example, Yukawa fluids, two-component plasma, two-dimensional OCP, and fluids consisting of electric or magnetic dipoles. \par

Determining the correct action indeed requires experimental input, and we have shown how the procedure of developing theory using the variational principle works. But once the correct input has been achieved, and the complete set of variables has been assumed, one gets the payoff that a self-consistent theory that can be applied to nonlinear processes has been obtained. \par

In our analysis, we have ignored the effects of viscosity, relaxation, and heat transfer. In the future, one can consider adding them by including additional terms in the equations of motion produced by the variational principle. Viscosity can be included by adding a term related to the velocity gradient, as seen in the case of the Navier-Stokes equations. Relaxation can be included by adding a linear in velocity friction force between moving particles and the stationary neutralizing background, as in the two-fluid plasma equations. Alternatively, relaxation can be introduced for the hydrodynamic variables for their approach to local thermodynamic equilibrium values, as in generalized hydrodynamics. Heat transfer can be accounted for by adding a term related to the temperature gradient to the energy conservation law obtained from the variational principle, similar to the case of the Navier-Stokes equations. 

\begin{acknowledgments}
This research has been funded by the AFOSR (FA9550-21-1-0295). The views, opinions, and/or findings expressed are those of the authors and should not be interpreted as representing the official
views or policies of the Department of Defense or the U.S. Government. We thank T. Chuna, Z. Johnson, M. S. Murillo, L. G. Silvestri for many valuable discussions and Z. Donk\'{o}, I. Korolov for the provided numerical data from the molecular dynamics simulations.
\end{acknowledgments}


\appendix
\section{Eulerian approach}
\label{appendix_Eulerian}

\subsection{Equations of motion}

To perform the variations of the nonlocal terms in the Lagrangian given by Eq.~\eqref{eulerian_lagrangian}, change variables under the integral from $(\vec{x}, \vec{x}')$ to $(\vec{x}', \vec{x})$, which has a unit Jacobian. Also, for any field that is being varied, denoted as $\Psi$, and any functions $f,g$, we have that $(fg)^T = f^T g^T$, and $(\partial_{\Psi'} f)^T = \partial_{\Psi} (f^T)$. \par

Using these results, we obtain the following equations of motion from the variational principle.
\begin{equation}
\label{eulerian_variation_n}
\begin{gathered}
\delta n: \quad \frac{1}{2} m\vec{v}^2 - f(n,s) - n\frac{\partial f}{\partial n}\Big|_s - \frac{\partial \alpha}{\partial t} - \vec{v} \cdot \vec{\nabla} \alpha \\
= \int \bigg( n'(F+F^T) + nn' \frac{\partial (F+F^T)}{\partial n}\Big|_{n', s, s', |\vec{x}-\vec{x}'|} 
+ n'_- \phi_{+-}(|\vec{x} - \vec{x}'|) \bigg) \mathrm{d}\vec{x}',
\end{gathered}
\end{equation}
\begin{equation}
\label{eulerian_variation_v}
\begin{gathered}
\delta \vec{v}: \quad mn\vec{v} + \beta \vec{\nabla}s - n\vec{\nabla}\alpha = \vec{0},
\end{gathered}
\end{equation}
\begin{equation}
\label{eulerian_variation_s}
\begin{gathered}
\delta s: \quad n \frac{\partial f}{\partial s}\Big|_n + \frac{\partial \beta}{\partial t} + \vec{\nabla} \cdot (\beta \vec{v}) 
= - \int nn' \frac{\partial (F+F^T)}{\partial s}\Big|_{n,n',s',|\vec{x}-\vec{x}'|} \mathrm{d}\vec{x}'.
\end{gathered}
\end{equation} \par

Our goal with the variational principle is to understand the equation of motion for $\partial \vec{v} / \partial t$. However, when examining the obtained Eqs.~\eqref{eulerian_variation_n}, \eqref{eulerian_variation_v}, \eqref{eulerian_variation_s}, one can see one of the disadvantages of the Eulerian formulation of variational principles. In this formulation, the resulting equations depend on the constraint fields $\alpha, \beta$, which then must be solved by using clever manipulations in terms of $n,s, \vec{v}$. To achieve this, we repeat computations similar to those for Euler equations \cite{herivel_1955}. \par

From Eq.~\eqref{eulerian_variation_v}, one can solve for $\vec{\nabla} \alpha$ and not for $\alpha$ itself. Therefore, one takes the gradient of Eq.~\eqref{eulerian_variation_n}, expands the gradients of $f, (F+F^T)$ using the chain rule, and substitutes the expression for $\vec{\nabla} \alpha$. Then, the resulting equation does not depend on $\alpha$ anymore, but depends on derivatives of $(\beta \vec{\nabla} s)/ n$. The time derivative of $(\beta \vec{\nabla} s)/ n$ can be simplified by using the product rule and then using the time derivatives of $n, s, \beta$ as given in Eqs.~\eqref{eulerian_continuity_equations}, \eqref{eulerian_variation_s}. The remaining terms with $\beta$ can be simplified by using the product rule and the identity $\vec{a} \times (\vec{b} \times \vec{c}) = (\vec{a} \cdot \vec{c})\vec{b} - (\vec{a}\cdot \vec{b})\vec{c}$ with $\vec{a} = n\vec{v}, \vec{b} = \vec{\nabla}(\beta / n), \vec{c} = \vec{\nabla}s$. Finally, to simplify the terms with cross products and curls, use the mathematical identity $\vec{\nabla}(\beta / n) \times \vec{\nabla}s = \vec{\nabla} \times (\beta \vec{\nabla s}/n)$, and use Eq.~\eqref{eulerian_variation_v} to obtain $ \vec{\nabla} \times (\beta \vec{\nabla s}/n) = -m\vec{\nabla} \times \vec{v}$. Finally, use the following identity for derivatives of the velocity field.
\begin{equation}
\label{eulerian_time_derivative}
\begin{gathered}
\frac{\partial \vec{v}}{\partial t} + (\vec{v} \cdot \vec{\nabla})\vec{v} = \frac{\partial \vec{v}}{\partial t} + \vec{\nabla} \left( \frac{\vec{v}^2}{2} \right) - \vec{v} \times (\vec{\nabla} \times \vec{v} ) 
\end{gathered}
\end{equation}

\subsection{Conservation laws}

To compute the momentum conservation law for our nonlocal variational principle, we use an approach analogous to that used in the case of local variational principles \cite{Landau1980Classical, goldstein:mechanics}. We apply the chain rule to expand the full derivatives of $\mathcal{L}_1$ and, motivated by the equation of motion in Eq.~\eqref{eulerian_final_EOM}, $(1/2)\int (\mathcal{L}_2 + \mathcal{L}_2^T) \mathrm{d}\vec{x}'$ with respect to $x_i$. Due to nonlocality, we also use the chain rule to expand the full derivative of $(\mathcal{L}_2 + \mathcal{L}_2^T)/2$ with respect to $x'_i$. It is important to note that the integral over $\vec{x}'$ of the latter is zero, as can be shown by integration by parts. We add these calculations and simplify terms using all of the equations of motion obtained from the variational principle, and then apply the product rule. Finally, to simplify the obtained equation in our particular case, we use the chain rule, integration by parts, and our assumptions regarding $\mathcal{L}_1, F, \phi_{+-}, \phi_{--}$ to obtain the following equation that nearly has the form of a conservation law. 
\begin{equation}
\label{eulerian_momentum_law}
\begin{gathered}
\frac{\partial}{\partial t}\left( \alpha \frac{\partial n}{\partial x_i} + \beta \frac{\partial s}{\partial x_i} \right) - \frac{\partial}{\partial x_i} \left( \frac{1}{2} mn\vec{v}^2 - nf \right)
+ \sum \limits_{j = 1}^3 \frac{\partial}{\partial x_j} \left( \alpha \frac{\partial (n v_j)}{\partial x_i} + \beta v_j \frac{\partial s}{\partial x_i} \right)\\
 = - \int n'_- \frac{\partial n}{\partial x_i} \phi_{+-}\mathrm{d}\vec{x}' - \int n' \Bigg( \frac{\partial n}{\partial x_i} (F+F^T) 
+ n \frac{\partial n}{\partial x_i} \frac{\partial (F+F^T)}{\partial n}\Big|_{n',s,s',|\vec{x}-\vec{x}'|}\\
+ n\frac{\partial s}{\partial x_i} \frac{\partial (F+F^T)}{\partial s}\Big|_{n,n',s',|\vec{x}-\vec{x}'|} \Bigg) \mathrm{d}\vec{x}'
\end{gathered}
\end{equation}

Similar to the equations of motion obtained from the variational principle, the resulting momentum conservation law depends on the constraint fields $\alpha, \beta$, which highlights a disadvantage of Eulerian formulation. We will now rewrite it so that it is expressed solely in terms of $n, s, \vec{v}$. \par

As $\alpha$ is only given in the equations of motion in Eqs.~\eqref{eulerian_variation_n}, \eqref{eulerian_variation_v} through its derivatives, we use the product rule to introduce an additional spatial derivative to $\alpha$ in all of the terms containing it. This enables us to employ Eqs.~\eqref{eulerian_variation_n}, \eqref{eulerian_variation_v}, along with the continuity equation for number density given in Eq.~\eqref{eulerian_continuity_equations}, to eliminate all of the terms containing $\alpha$. After this simplification, all of the terms with $\beta$ also disappear. Finally, to simplify the nonlocal terms that contain $F, \phi_{+-}$, we expand the derivatives using the chain rule, and then use assumptions about the explicit dependence of $\vec{x}, \vec{x}'$ in $F, \phi_{+-}$, and apply integration by parts. The final result is given next. 
\begin{equation}
\label{eulerian_momentum_law_final}
\begin{gathered}
\frac{\partial (mnv_i)}{\partial t} + \vec{\nabla} \cdot (m n v_i \vec{v}) + \frac{\partial}{\partial x_i} \bigg( n^2 \frac{\partial f}{\partial n}\Big|_s
+ \int n^2 n' \frac{\partial (F+F^T)}{\partial n}\Big|_{n', s, s', |\vec{x}-\vec{x}'|} \mathrm{d}\vec{x}'  \bigg) \\
= - \int \Bigg( nn'\frac{\partial (F+F^T)}{\partial x_i}\Big|_{n,n',s,s',\vec{x}'} + nn'_- \frac{\partial \phi_{+-}}{\partial x_i}\Big|_{\vec{x}'} \Bigg) \mathrm{d}\vec{x}'
\end{gathered}
\end{equation} \par

Notice that this equation could also be obtained directly by adding the equation of motion for $\vec{v}$ given by Eq.~\eqref{eulerian_final_EOM} to the continuity equation for number density in Eq.~\eqref{eulerian_continuity_equations} and using the product rule. However, this method requires the knowledge that the correct expression for the momentum density in our theory is indeed $mn \vec{v}$. \par

The momentum conservation law given by Eq.~\eqref{eulerian_momentum_law_final} allows us to define the full momentum $\vec{P}$ of the particles that move in the OCP and consider how it changes over time. In particular, we integrate Eq.~\eqref{eulerian_momentum_law_final} with respect to $\vec{x}$ and use integration by parts. To further simplify nonlocal terms, we use the properties that for all functions $f,g$, we have $(f^T)^T = f, (f + g)^T = f^T + g^T, (fg)^T = f^T g^T$. Additionally, we note that the integral of $f$ over $\vec{x}, \vec{x}'$ is the same as the integral of $f^T$. We also use the following identity. 
\begin{equation}
\label{identity:F_FT}
\left( \frac{\partial (F+F^T)}{\partial x_i}\Big|_{n,n',s,s',\vec{x}'} \right)^T = \frac{\partial ((F+F^T)^T)}{\partial x'_i}\Big|_{n,n',s,s',\vec{x}} 
\end{equation} \par

These identities together with the assumption that $F$ depends explicitly on $\vec{x}, \vec{x}'$ only through the combination $\vec{x}-\vec{x}'$ are used to show the following integral is zero.
\begin{equation}
\begin{gathered}
\iint nn'\frac{\partial (F+F^T)}{\partial x_i}\Big|_{n,n',s,s',\vec{x}'} \mathrm{d}\vec{x} \mathrm{d}\vec{x}' = 0
\end{gathered}
\end{equation} \par

To compute the energy conservation law for our nonlocal variational principle, we employ an analogous approach as in the case of local variational principles \cite{Landau1980Classical, goldstein:mechanics}. We use the chain rule to expand the full derivatives of $\mathcal{L}_1$ and, motivated by the equation of motion in Eq.~\eqref{eulerian_final_EOM}, $(1/2)\int (\mathcal{L}_2 + \mathcal{L}_2^T) \mathrm{d}\vec{x}'$ with respect to $t$. We add these calculations and simplify terms using all of the equations of motion obtained from the variational principle and the product rule. To further simplify the obtained equation in our particular case, we rewrite time derivatives of $n, s$ and $n',s'$ in the nonlocal terms using the continuity equations given in Eq.~\eqref{eulerian_continuity_equations}, and then apply integration by parts to the nonlocal terms with derivatives with respect to $\vec{x}'$.
\begin{equation}
\label{eulerian_energy_law}
\begin{gathered}
\frac{\partial}{\partial t}\Bigg( -\frac{1}{2} nm\vec{v}^2 + nf - \alpha \vec{\nabla} \cdot (n\vec{v}) - \beta \vec{v} \cdot \vec{\nabla}s 
+ \int \Bigg( \frac{n n'}{2}(F+F^T) + \frac{nn'_-}{2} \phi_{+-} + \frac{n'n_-}{2} \phi_{+-} \\
+ \frac{n_- n'_-}{2} \phi_{--} \Bigg) \mathrm{d}\vec{x}'\Bigg) + \sum \limits_{j = 1}^3 \frac{\partial}{\partial x_j} \left( \alpha \frac{\partial (n v_j)}{\partial t} + \beta v_j \frac{\partial s}{\partial t} \right)
= \int \Bigg( \frac{n'}{2}(F+F^T) \vec{\nabla} \cdot (n\vec{v}) \\
+ \frac{nn'}{2} \frac{\partial (F+F^T)}{\partial n}\Big|_{n',s,s',|\vec{x}-\vec{x}'|} \vec{\nabla} \cdot (n\vec{v}) 
+ \frac{nn'}{2} \frac{\partial (F+F^T)}{\partial s}\Big|_{n,n',s',|\vec{x}-\vec{x}'|} \vec{v} \cdot \vec{\nabla}s \\
+ \frac{nn'}{2}\vec{v}' \cdot \frac{\partial (F+F^T)}{\partial \vec{x}'}\Big|_{n,n',s,s',\vec{x}}
- \frac{n (n')^2}{2} \frac{\partial (F+F^T)}{\partial n'}\Big|_{n,s,s',|\vec{x}-\vec{x}'|} \vec{\nabla}' \cdot \vec{v}' \Bigg) \mathrm{d}\vec{x}' \\
+ \int \Bigg( \frac{n'_-}{2} \phi_{+-} \vec{\nabla} \cdot (n\vec{v}) + \frac{n' n_-}{2} \vec{v}' \cdot \frac{\partial \phi_{+-}}{\partial \vec{x}'}\Big|_{\vec{x}}  \Bigg) \mathrm{d}\vec{x}'
\end{gathered}
\end{equation} \par

As before, we observe a disadvantage of the Eulerian formulation of the variational principles, as the obtained energy conservation law depends on the constraint fields $\alpha, \beta$. We now rewrite it so that it is given only in terms of $n, s, \vec{v}$, using a similar strategy as when analyzing the momentum conservation law. \par

As $\alpha$ appears in the equations of motion in Eqs.~\eqref{eulerian_variation_n}, \eqref{eulerian_variation_v} only through its derivatives, we use the product rule to introduce an additional spatial derivative to $\alpha$ in all terms containing it. This allows to utilize Eqs.~\eqref{eulerian_variation_n}, \eqref{eulerian_variation_v}, together with the continuity equation for specific entropy in Eq.~\eqref{eulerian_continuity_equations}, to eliminate all terms that include $\alpha$. Following this simplification, all terms with $\beta$ also vanish. Finally, for simplifying nonlocal terms, we employ the chain rule and integration by parts. The final result is provided next.
\begin{equation}
\label{eulerian_energy_law_final}
\begin{gathered}
\frac{\partial \mathcal{E}}{\partial t} + \sum \limits_{j=1}^3 \frac{\partial J_j}{\partial x_j} = \sigma,
\end{gathered}
\end{equation}
\begin{equation}
\label{eulerian_energy_density}
\begin{gathered}
\mathcal{E} = \frac{1}{2} m n \vec{v}^2 + nf+ \int \bigg( \frac{n n'}{2}(F+F^T) + \frac{nn'_-}{2} \phi_{+-}
+ \frac{n'n_-}{2} \phi_{+-} + \frac{n_- n'_-}{2} \phi_{--} \bigg) \mathrm{d}\vec{x}',
\end{gathered}
\end{equation}
\begin{equation}
\label{eulerian_energy_flux}
\begin{gathered}
J_j =  v_j \Bigg( \frac{1}{2} m n \vec{v}^2 + nf + n^2 \frac{\partial f}{\partial n}\Big|_s
+ \int \Bigg( \frac{n n'}{2}(F+F^T)\\
 + n^2 n' \frac{\partial (F+F^T)}{\partial n}\Big|_{n', s, s', |\vec{x}-\vec{x}'|}
+ \frac{nn'_-}{2} \phi_{+-}  \Bigg) \mathrm{d}\vec{x}' \Bigg),
\end{gathered}
\end{equation}
\begin{equation}
\label{eulerian_energy_nonlocal_term}
\begin{gathered}
\sigma = \int \frac{n n'}{2} \bigg( n(\vec{\nabla} \cdot \vec{v}) \frac{\partial (F+F^T)}{\partial n}\Big|_{n', s, s', |\vec{x}-\vec{x}'|}
- n'(\vec{\nabla}' \cdot \vec{v}') \frac{\partial (F+F^T)}{\partial n'}\Big|_{n, s, s', |\vec{x}-\vec{x}'|} \bigg) \mathrm{d}\vec{x}' \\
+\int \frac{n n'}{2} \bigg( \vec{v}' \cdot \frac{\partial (F+F^T)}{\partial \vec{x}'}\Big|_{n, n', s, s', \vec{x}}
- \vec{v} \cdot \frac{\partial (F+F^T)}{\partial \vec{x}}\Big|_{n, n', s, s', \vec{x}'}  \bigg) \mathrm{d}\vec{x}' \\
+ \int \bigg( \frac{n' n_-}{2} \vec{v}' \cdot \frac{\partial \phi_{+-}}{\partial \vec{x}'}\Big|_{\vec{x}} - \frac{n n'_-}{2} \vec{v} \cdot \frac{\partial \phi_{+-}}{\partial \vec{x}}\Big|_{\vec{x}'} \bigg) \mathrm{d}\vec{x}'.
\end{gathered}
\end{equation}\par

Notice that this equation could also be obtained by directly computing the time derivative of the correct energy density as given in the energy conservation law. Then, one can use the product rule and time derivatives found in the equations of motion given by Eqs.~\eqref{eulerian_continuity_equations}, \eqref{eulerian_final_EOM}. To simplify, one employs the product rule, the chain rule, integration by parts, and the identity $\vec{v} \cdot (\vec{v} \cdot \vec{\nabla})\vec{v} = \vec{v} \cdot \vec{\nabla}(\vec{v}^2/2)$. This approach avoids issues with constraint fields $\alpha, \beta$ but requires knowledge of the correct expression for the energy density in our theory. \par

The energy conservation law, given by Eq.~\eqref{eulerian_energy_law_final}, allows us to define the energy $E$ of the OCP and consider how it changes over time. To achieve this, we integrate Eq.~\eqref{eulerian_energy_law_final} with respect to $\vec{x}$ and use integration by parts. To further simplify nonlocal terms, we employ the following properties: for all functions $f,g$, we have $(f^T)^T = f, (f + g)^T = f^T + g^T, (fg)^T = f^T g^T$, and the integral of $f$ over $\vec{x}, \vec{x}'$ is the same as the integral of $f^T$. To simplify nonlocal terms involving $\phi_{+-}$, we use the fact that $(\partial \phi_{+-}/\partial \vec{x})^T = \partial (\phi^T_{+-})/\partial \vec{x}'$, and that $\phi^T_{+-} = \phi_{+-}$ due to the assumption that it depends on $\vec{x}, \vec{x}'$ through $|\vec{x} - \vec{x}'|$. To simplify nonlocal terms with $F$, we use the identity in Eq.~\eqref{identity:F_FT}, and analogous identities, where in Eq.~\eqref{identity:F_FT}, $x_i, x_i'$ are replaced by $n, n'$ and $s, s'$, respectively.\par

To obtain from Eq.~\eqref{eulerian_energy_equation} the equilibrium energy per particle, $E_0/N$, we change variables inside the integrals to $\vec{R} = (\vec{x} + \vec{x}')/2, \vec{r} = \vec{x}-\vec{x}'$, which has a unit Jacobian. This change of variables allows us to simplify the integrals, as we assume that $F, \phi_{+-}, \phi_{--}$ depend explicitly on $\vec{x}, \vec{x}'$ only through the combination $\vec{x}-\vec{x}'$. Finally, we can further simplify the expression using the assumed form of the potential $\phi_{ij}(r) = q_i q_j/4 \pi \varepsilon_0 r$, where $i, j$ are the corresponding species of the particles, and $r$ is the distance between the particles.

\vspace{\baselineskip}
\vspace{\baselineskip}
\subsection{Dispersion laws}
\label{appendix_eulerian_dispersion}
We first consider Eqs.~\eqref{eulerian_continuity_equations}, \eqref{eulerian_final_with_thermo}, and determine whether our proposed equilibrium solutions satisfy them. \par

The continuity equations in Eq.~\eqref{eulerian_continuity_equations} are satisfied since all hydrodynamic functions do not depend on either spatial coordinates or time, resulting in their derivatives being zero.\par

In Eq.~\eqref{eulerian_final_with_thermo}, the terms on the left-hand side are zero because $\vec{v} = \vec{0}$ in equilibrium. The first term under the gradient on the right-hand side is also zero since it does not depend on spatial coordinates due to its dependence on $n_0, s_0$. As for the second term under the gradient, we can use the fact that $n = n_0, n' = n_0$, so the integral is only over a function that depends on $\vec{x}, \vec{x}'$ through $\vec{x} - \vec{x}'$. This allows us to change variables under the integral to $\vec{r} = \vec{x}-\vec{x}'$, making it independent of $\vec{x}$. Finally, let us consider the remaining nonlocal terms on the right-hand side. We can use the fact that $n, s, n_-$ are all uniform in space, so the integrals become over a full derivative with respect to $\vec{x}$, which can be integrated by parts to become zero. \par

We now expand equations of motion given by Eqs.~\eqref{eulerian_continuity_equations}, \eqref{eulerian_final_with_thermo} up to the first order. The expanded continuity equations up to the first order are as follows. 
\begin{equation}
\label{linearized_continuity_Eulerian}
\begin{gathered}
\frac{\partial n_1}{\partial t} + n_0 (\vec{\nabla} \cdot \vec{v}_1) = 0, \quad \frac{\partial s_1}{\partial t}= 0.
\end{gathered}
\end{equation}\par

To expand the equation of motion for $\vec{v}$, given by Eq.~\eqref{eulerian_final_with_thermo}, up to the first order, we use the fact that the functions inside the integrals often depend on $\vec{x}, \vec{x}'$ solely through $\vec{x} - \vec{x}'$. This allows us to switch from derivatives with respect to $\vec{x}$ to derivatives with respect to $\vec{x}'$, enabling us to apply integration by parts. The resulting linearized equation for $\vec{v}_1$ shows that $n_1, s_1$ only appear in it through gradients.\par

To combine the linearized equations of motion, we take the gradient of Eq.~\eqref{linearized_continuity_Eulerian} and an additional time derivative of the linearized equation of motion for $\vec{v}_1$. This yields the following equation, which depends solely on $\vec{v}_1$, with the subscript "$0$" indicating that a function is evaluated at equilibrium values. We also simplify using $g_0 = (g^T)_0$ and $(\partial g/\partial n)_0 = (\partial g^T / \partial n')_0$. 
\begin{equation}
\label{linearized_velocity_final_Eulerian}
\begin{gathered}
m \frac{\partial^2 \vec{v}_1}{\partial t^2} = \vec{\nabla}(\vec{\nabla} \cdot \vec{v}_1) \Bigg( \Bigg[ \frac{\partial}{\partial n} \Big( \frac{3}{2} k_B n^2 \frac{\partial T}{\partial n}\Big|_s \Big)\Big|_s \Bigg]_0 
+ \frac{q^2_+ n_0}{8 \pi \varepsilon_0} \int \frac{1}{|\vec{x}-\vec{x}'|} \Bigg[ \frac{\partial}{\partial n} \Big( n^2 \frac{\partial g}{\partial n}\Big|_{s, |\vec{x}-\vec{x}'|} \Big)\Big|_{s, |\vec{x}-\vec{x}'|} \Bigg]_0 \mathrm{d}\vec{x}' \Bigg)\\
+ \frac{q^2_+ n_0}{4 \pi \varepsilon_0} \int \frac{\vec{\nabla}'(\vec{\nabla}' \cdot \vec{v}'_1)}{|\vec{x}-\vec{x}'|} \Bigg( g_0 + n_0 \Bigg[ \frac{\partial g}{\partial n}\Big|_{s, |\vec{x}-\vec{x}'|}\Bigg]_0  \Bigg) \mathrm{d}\vec{x}'
\end{gathered}
\end{equation}\par

To solve this equation, we use the Fourier transform with respect to the spatial coordinates, $\vec{x}$, and use the following convention.
\begin{equation}
\label{eulerian_fourier_transform_convention}
\vec{V}(\vec{k}, t) = \widehat{\vec{v}_1}(\vec{k}, t) = \int \vec{v}_1 (\vec{x}, t) e^{-i\vec{k} \cdot \vec{x}} \mathrm{d}\vec{x}
\end{equation}\par

With this convention, we find that for any functions $f, g$ that depend on $\vec{x}$, $\widehat{\partial f/ \partial x_j} = i k_j \widehat{f}$ and $\widehat{ f * g} = \widehat{f} \widehat{g}$, where $*$ denotes convolution. Using these results and changing variables under the first integral from $\vec{x}'$ to $\vec{r} = \vec{x} - \vec{x}'$, which has a unit Jacobian, one can take the Fourier transform of Eq.~\eqref{linearized_velocity_final_Eulerian} to obtain a differential equation for each value of $\vec{k}$. \par

One must be careful with the convergence of integrals. From thermodynamics \cite{doi:10.1063/1.1727895}, it is known that for the equilibrium pair distribution function, $g(n_0, s_0, |\vec{r}|) \to 1$ as $|\vec{r}| \to \infty$. Therefore, it is convenient to rewrite everything in terms of $g - 1$ instead of $g$. Additionally, we use that the Fourier transform of $1/|\vec{r}|$ is \cite{berestetskii1982quantum} $4\pi/|\vec{k}|^2$. 
\begin{equation}
\label{FT_velocity_Eulerian}
\begin{gathered}
m \frac{\partial^2 \vec{V}}{\partial t^2} = -\vec{k}(\vec{k} \cdot \vec{V}) \Bigg( \Bigg[ \frac{\partial}{\partial n} \Big( \frac{3}{2} k_B n^2 \frac{\partial T}{\partial n}\Big|_s \Big)\Big|_s \Bigg]_0
+  \frac{q^2_+ n_0}{8 \pi \varepsilon_0} \int \frac{1}{|\vec{r}|} \Bigg[ \frac{\partial}{\partial n} \Big( n^2 \frac{\partial (g-1)}{\partial n}\Big|_{s, |\vec{r}|} \Big)\Big|_{s, |\vec{r}|} \Bigg]_0 \mathrm{d}\vec{r} \\
+ \frac{q_+^2 n_0}{\varepsilon_0 |\vec{k}|^2} + \frac{q^2_+ n_0}{4 \pi \varepsilon_0} \int \frac{e^{-i\vec{k} \cdot \vec{r}}}{|\vec{r}|} \bigg( (g - 1)_0 
+ n_0 \Bigg[ \frac{\partial (g - 1)}{\partial n}\Big|_{s, |\vec{r}|}\Bigg]_0  \bigg) \mathrm{d}\vec{r} \Bigg)
\end{gathered}
\end{equation}\par

To simplify the analysis of the longitudinal and transverse dispersion modes, it is convenient to rotate the coordinate system so that, in the rotated coordinate system, $\vec{k} = (0, 0, |\vec{k}|)$. In this case, the transverse modes correspond to the $x,y$ directions, while the longitudinal mode corresponds to the $z$ direction. To do this, we use the fact that the differential equation is linear, consider the properties of the dot product under a rotation, and also rotate the integration variable. Moreover, we can simplify further by switching to spherical coordinates and performing the angular integrals. \par

After rotating the coordinate system, we find from Eq.~\eqref{FT_velocity_Eulerian} that for the transverse directions, $\partial^2 V_x / \partial t^2 = \partial^2 V_y / \partial t^2 = 0$, indicating the absence of transverse modes. For the longitudinal direction, we have the dispersion law given by Eq.~\eqref{longitudinal_dispersion_Eulerian}.\par

The result given by Eq.~\eqref{eulerian_f1} allows us to take derivatives at constant $s$ of functions of $\Gamma$. In particular, we have the following, derived from the definition of $\Gamma$.
\begin{equation}
\label{adiabatic_gamma_derivative}
\begin{gathered}
n \frac{\partial \Gamma}{\partial n}\Big|_s = n \left( \frac{\partial \Gamma}{\partial n}\Big|_T + \frac{\partial \Gamma}{\partial T}\Big|_n \frac{\partial T}{\partial n}\Big|_s \right) = \frac{\Gamma}{3} - \frac{2 \Gamma}{3} f_1(\Gamma)
\end{gathered}
\end{equation}\par

To analyze the terms in Eq.~\eqref{longitudinal_dispersion_Eulerian} with integrals of $g-1$, in addition to the previous identities, we use the fact that $g = g(\Gamma, |\vec{r}|/a)$. To do this, we move the number density derivatives outside of the integrals, change variables inside the integrals to $x = r/a$, and introduce the normalized wavevector $\vec{q} = \vec{k}a$. In this case, the integrals are rewritten in terms of functions of $\Gamma$ and $|\vec{q}|$, allowing us to use the identity given by Eq.~\eqref{constant_s_derivative}, where we used Eq.~\eqref{adiabatic_gamma_derivative} and $\partial |\vec{q}|/\partial n|_s = -|\vec{q}|/3n$.

To relate $(R/a)(\Gamma)$ and $u(\Gamma)$ for the assumed step function approximation of the pair distribution function, in Eq.~\eqref{thermodynamic_equilibrium_energy}, we go to spherical coordinates, change variables to $x = |\vec{r}|/a$, and then apply the assumed form for $g(\Gamma, |\vec{r}|/a)$, along with the definitions of $a$ and $\Gamma$. This yields the following result.
\begin{equation}
\begin{gathered}
\frac{E_0}{N} = \frac{3}{2} k_B T - \frac{3}{4} k_B T \Gamma \left( \frac{R}{a} \right)^2(\Gamma)
\end{gathered}
\end{equation}\par

When comparing this expression to Eq.~\eqref{equilibrium_energy_and_pressure}, one can derive the Eq.~\eqref{R_over_a}. \par

For the simple fit $u(\Gamma) = -0.9 \Gamma$, where $u(\Gamma)$ is linear in $\Gamma$, we can observe from Eq.~\eqref{R_over_a} that $R/a$ is independent of $\Gamma$. In this case, $R/a = \sqrt{6/5}$. This means that, according to Eq.~\eqref{simplified_j_h_Eulerian}, the functions $j, h$ are both independent of $\Gamma$. This greatly simplifies taking number density derivatives at constant $s$ as described in Eq.~\eqref{constant_s_derivative}. Additionally, it is worth noting that for such a fit, $f_1(\Gamma) = 1$ based on Eq.~\eqref{eulerian_f1}, as $u(\Gamma)/\Gamma$ is independent of $\Gamma$. \par

\section{Lagrangian approach}
\label{appendix_Lagrangian}
\subsection{Equations of motion}

As in the case of Eulerian coordinates, one must be cautious when performing variations of the nonlocal terms in the Lagrangian given by Eq.~\eqref{lagrangian_lagrangian}, but the strategy is analogous. To carry out the variations, change variables under the integral from $(\vec{a}, \vec{a}')$ to $(\vec{a}', \vec{a})$, which has a unit Jacobian. Also, for any argument $\Psi$ of a nonlocal function and any functions $f,g$, we have that $(fg)^T = f^T g^T$, and $(\partial_{\Psi'} f)^T = \partial_{\Psi} (f^T)$. Additionally, we can use the following property of determinants \cite{herivel_1955} that holds for all directions $i, j$. 
\begin{equation}
\label{lagrangian_determinant_property_1}
\sum \limits_{j=1}^3 \frac{\partial}{\partial a_j} \left( \frac{\partial (\mathrm{det} \left( \partial \vec{x} / \partial \vec{a} \right))}{\partial (\partial_j x_i)} \right) = 0
\end{equation}\par

Combining these results, we obtain the equation of motion for $\vec{x}$ given by Eq.~\eqref{lagrangian_final_EOM}. Again, notice that, unlike in the Eulerian case, there are no constraint functions in the Lagrangian given in Eq.~\eqref{lagrangian_lagrangian}. Therefore, the resulting equations of motion are already in their final form, and no additional manipulations are needed. \par

To check the consistency with the Eulerian approach when $F$ does not depend on  $\vec{a}, \vec{a}'$, Eq.~\eqref{lagrangian_continuity_equations} is used, along with the following identity for an arbitrary function $f$, which is obtained from the chain rule \cite{herivel_1955} and holds for all directions $i, j$.
\begin{equation}
\label{property_determinants_2}
\sum \limits_{j=1}^3 \frac{\partial (\mathrm{det} \left( \partial \vec{x} / \partial \vec{a} \right))}{\partial (\partial_j x_i)} \frac{\partial f}{\partial a_j} = \mathrm{det} \left( \frac{\partial \vec{x}}{\partial \vec{a}} \right) \frac{\partial f}{\partial x_i}
\end{equation}

\subsection{Conservation laws}

Consider the momentum conservation law in each of the directions, $i = 1,2,3$. In the Lagrangian variables, the variational principle is applied to the displacement field $\vec{x}$ to obtain the equations of motion. Therefore, the momentum conservation law is determined by the equations of motion obtained in Eq.~\eqref{lagrangian_final_EOM}. To show that it nearly takes the form of a conservation law, we rewrite it by considering that $n^{\mathrm{ref}}$ does not depend on time and applying the product rule along with Eq.~\eqref{lagrangian_determinant_property_1}.
\begin{equation}
\label{lagrangian_momentum_conservation_law}
\begin{gathered}
\frac{\partial}{\partial t}\left( m n^{\mathrm{ref}} \frac{\partial x_i}{\partial t} \right) +\sum \limits_{j=1}^3 \frac{\partial}{\partial a_j} \Bigg( \frac{\partial (\mathrm{det} \left( \partial \vec{x} / \partial \vec{a} \right))}{\partial (\partial_j x_i)} n^2 \frac{\partial f}{\partial n}\Big|_s \\
+ \frac{\partial (\mathrm{det} \left( \partial \vec{x} / \partial \vec{a} \right))}{\partial (\partial_j x_i)}  \int n^2 (n^{\mathrm{ref}})' \frac{\partial (F+F^T)}{\partial n}\Big|_{n', s, s', |\vec{x}-\vec{x}'|, \vec{a}, \vec{a}'} \mathrm{d}\vec{a}' \Bigg)\\
= - \int \bigg( n^{\mathrm{ref}}(n^{\mathrm{ref}})'\frac{\partial (F+F^T)}{\partial x_i}\Big|_{n,n',s,s',\vec{x}', \vec{a}, \vec{a}'}
+ n^{\mathrm{ref}}(n_-^{\mathrm{ref}})' \frac{\partial \phi_{+-}}{\partial x_i}\Big|_{\vec{a}'} \bigg) \mathrm{d}\vec{a}'
\end{gathered}
\end{equation} \par

The momentum conservation law, as given in Eq.~\eqref{lagrangian_momentum_conservation_law}, allows us to define the total momentum $\vec{P}$ of the particles moving in the OCP and examine how it changes over time. To accomplish this, integrate Eq.~\eqref{lagrangian_momentum_conservation_law} with respect to $\vec{a}$ and use integration by parts. For further simplification of nonlocal terms, employ the properties that hold for all functions $f,g$, $(f^T)^T = f, (f + g)^T = f^T + g^T, (fg)^T = f^T g^T$, and note that the integral of $f$ over $\vec{a}, \vec{a}'$ is the same as the integral of $f^T$. Additionally, use an analogous identity as in Eq.~\eqref{identity:F_FT}, where $F$ now depends also on $\vec{a}, \vec{a}'$ as well. By combining these results with the assumption that $F$ depends explicitly on $\vec{x}, \vec{x}'$ only through the combination $\vec{x} - \vec{x}'$, we get that the following integral is zero.
\begin{equation}
\begin{gathered}
\iint n^{\mathrm{ref}}(n^{\mathrm{ref}})'\frac{\partial (F+F^T)}{\partial x_i}\Big|_{n,n',s,s',\vec{x}', \vec{a}, \vec{a}'} \mathrm{d}\vec{a} \mathrm{d}\vec{a}' = 0
\end{gathered}
\end{equation}\par

One can demonstrate that the expression for momentum in the Lagrangian approach is the same as in the Eulerian approach, as given by Eq.~\eqref{eulerian_momentum_equation}, by changing the coordinates under the integral from $\vec{a}$ to $\vec{x}$ and using the definition of $n$ provided by Eq.~\eqref{lagrangian_continuity_equations}. To show that the momentum of moving particles is conserved when the neutralizing background is both uniform and infinite, use Eq.~\eqref{lagrangian_momentum_equation} by noting that $\phi_{+-}$ depends on $\vec{x}, \vec{a}'$ only through the combination $\vec{x}-\vec{a}'$ and apply integration by parts.\par

To derive the energy conservation law for our nonlocal variational principle, we follow a similar approach to that used in the case of local variational principles \cite{Landau1980Classical, goldstein:mechanics} and in the Eulerian approach. We employ the chain rule to expand the full derivatives of $\mathcal{L}_1^{\mathrm{L}}$ and, guided by the equation of motion provided in Eq.~\eqref{lagrangian_final_EOM}, $(1/2)\int (\mathcal{L}_2^{\mathrm{L}} + (\mathcal{L}_2^{\mathrm{L}})^T) \mathrm{d}\vec{a}'$ with respect to $t$. These calculations are then combined and simplified using all the equations of motion obtained from the variational principle, along with the product rule. Similar to the case of equations of motion, computations in the Lagrangian coordinates offer the advantage over Eulerian coordinates as they require no additional manipulations for the resulting energy law, as there are no constraint fields. Here, the energy density is defined in Eq.~\eqref{lagrangian_energy_density}.
\begin{equation}
\label{lagrangian_energy_law}
\begin{gathered}
\frac{\partial \mathcal{E}^{\mathrm{L}}}{\partial t} + \sum \limits_{j=1}^3 \frac{\partial J^{\mathrm{L}}_j}{\partial a_j} = \sigma^{\mathrm{L}},
\end{gathered}
\end{equation}
\begin{equation}
\label{lagrangian_energy_flux}
\begin{gathered}
J^{\mathrm{L}}_j =  \sum \limits_{i=1}^3 \frac{\partial x_i}{\partial t} \frac{\partial (\mathrm{det} \left( \partial \vec{x} / \partial \vec{a} \right))}{\partial (\partial_j x_i)} \bigg( n^2 \frac{\partial f}{\partial n}\Big|_s
+ \int n^2 (n^{\mathrm{ref}})' \frac{\partial (F+F^T)}{\partial n}\Big|_{n', s, s', |\vec{x}-\vec{x}'|, \vec{a}, \vec{a}'} \mathrm{d}\vec{a}'  \bigg),
\end{gathered}
\end{equation}
\begin{equation}
\label{lagrangian_energy_nonlocal_term}
\begin{gathered}
\sigma^{\mathrm{L}} = \frac{1}{2} \sum \limits_{i=1}^3 \sum \limits_{j=1}^3 \int \bigg( n^2 (n^{\mathrm{ref}})' \frac{\partial (F+F^T)}{\partial n}\Big|_{n', s, s', |\vec{x}-\vec{x}'|, \vec{a}, \vec{a}'} 
\frac{\partial (\mathrm{det} \left( \partial \vec{x} / \partial \vec{a} \right))}{\partial (\partial_j x_i)} \frac{\partial^2 x_i}{\partial a_j \partial t} \\
- n^{\mathrm{ref}}(n')^2 \frac{\partial (F+F^T)}{\partial n'}\Big|_{n, s, s', |\vec{x}-\vec{x}'|, \vec{a}, \vec{a}'}
\frac{\partial (\mathrm{det} \left( \partial \vec{x}' / \partial \vec{a}' \right))}{\partial (\partial'_j x'_i)} \frac{\partial^2 x'_i}{\partial a'_j \partial t} \bigg) \mathrm{d}\vec{a}' \\
+ \frac{1}{2} \sum \limits_{i=1}^3 \int n^{\mathrm{ref}} (n^{\mathrm{ref}})' \bigg( \frac{\partial (F+F^T)}{\partial x'_i}\Big|_{n,n',s,s',\vec{x}, \vec{a}, \vec{a}'} \frac{\partial x'_i}{\partial t}
- \frac{\partial (F+F^T)}{\partial x_i}\Big|_{n,n',s,s',\vec{x}', \vec{a}, \vec{a}'} \frac{\partial x_i}{\partial t} \bigg) \mathrm{d}\vec{a}' \\
+ \frac{1}{2} \sum \limits_{i=1}^3 \int \bigg( (n^{\mathrm{ref}})' n_-^{\mathrm{ref}}\frac{\partial \phi_{+-}}{\partial x'_i}\Big|_{\vec{a}}\frac{\partial x'_i}{\partial t} 
- n^{\mathrm{ref}}(n_-^{\mathrm{ref}})' \frac{\partial \phi_{+-}}{\partial x_i}\Big|_{\vec{a}'}\frac{\partial x_i}{\partial t} \bigg) \mathrm{d}\vec{a}'.
\end{gathered}
\end{equation}\par

Note that the energy conservation law in Eq.~\eqref{lagrangian_energy_law} can also be derived by directly computing the time derivative of the correct energy density given in Eq.~\eqref{lagrangian_energy_density} using the equations of motion, as shown in Eq.~\eqref{lagrangian_final_EOM}. To simplify, we apply the chain rule, use the definition of $n$ as given in Eq.~\eqref{lagrangian_continuity_equations}, and consider that in the reference state $n^{\mathrm{ref}}, s^{\mathrm{ref}}, n_-^{\mathrm{ref}}$ are all independent of time. Afterward, we can combine terms using the product rule and apply the identity regarding determinants, as shown in Eq.~\eqref{lagrangian_determinant_property_1}. Similar to the Eulerian case, the drawback of this approach is the requirement to know the correct expression for the energy density. \par

We now demonstrate that in the case where the function $F$ does not depend on $\vec{a}, \vec{a}'$, the energy law in the Lagrangian approach can be rewritten in Eulerian variables by changing coordinates from $(\vec{a}, \vec{a}')$ to $(\vec{x}, \vec{x}')$. This results in precisely the same energy law as in the Eulerian approach, as given by Eqs.~\eqref{eulerian_energy_law_final}, \eqref{eulerian_energy_density}, \eqref{eulerian_energy_flux}, and \eqref{eulerian_energy_nonlocal_term}. One has to be careful as $\vec{a}$ might represent the reference state position of either moving particles or the neutralizing stationary background.\par

For the term in Eq.~\eqref{lagrangian_energy_density} with $\phi_{--}$, it is worth noting that due to $n_-^{\mathrm{ref}}$ being independent of time, its time derivative is zero, just like for the analogous term in the Eulerian energy density in Eq.~\eqref{eulerian_energy_density}. That means that terms with $\phi_{--}$ can be added or removed as needed. For the term in Eq.~\eqref{lagrangian_energy_density} with $(n^{\mathrm{ref}})' n_-^{\mathrm{ref}}$, one can compute the time derivative using the chain rule. This computation results in the following, which corresponds to a term in Eq.~\eqref{lagrangian_energy_nonlocal_term}.
\begin{equation}
\begin{gathered}
\frac{\partial}{\partial t} \left(\frac{1}{2} \int \bigg( (n^{\mathrm{ref}})' n_-^{\mathrm{ref}} \phi_{+-}(|\vec{x}'-\vec{a}|) \bigg) \mathrm{d}\vec{a}'\right) 
= \frac{1}{2} \sum \limits_{i=1}^3 \int \bigg( (n^{\mathrm{ref}})' n_-^{\mathrm{ref}}\frac{\partial \phi_{+-}}{\partial x'_i}\Big|_{\vec{a}}\frac{\partial x'_i}{\partial t}\bigg) \mathrm{d}\vec{a}'
\end{gathered}
\end{equation}\par

Due to this identity, change of variables can be carried out independently for these two terms, separate from the others. To achieve this, switch from $\vec{a}$ to $\vec{x}$ by using the assumed displacement field for the neutralizing stationary background, where $\vec{x} = \vec{a}$. Consequently, we also have $n_- = n_-^{\mathrm{ref}}$. As for the integrals, perform variable transformations from $\vec{a}'$ to $\vec{x}'$ using the displacement field of the moving particles and make use of Eq.~\eqref{lagrangian_continuity_equations}. \par

For all the other terms, replace $n^{\mathrm{ref}}$ in terms of $n$ using Eq.~\eqref{lagrangian_continuity_equations}. Change from $\vec{a}$ to $\vec{x}$ by using the displacement field of the moving particles, and switch from $\vec{a}'$ to $\vec{x}'$ using either the displacement field of the moving particles or that of the neutralizing stationary background, where $\vec{x}'=\vec{a}'$ and $n_-^{\mathrm{ref}} = n_-$. To simplify these terms, apply the chain and product rules, along with the properties of determinants given in Eqs.~\eqref{lagrangian_determinant_property_1}, \eqref{property_determinants_2}, and the following property \cite{chorin2013mathematical}, where $\vec{v}$ represents Eulerian velocity, and $\vec{\nabla}$ denotes the gradient with respect to $\vec{x}$. 
\begin{equation}
\label{property_determinants_3}
\frac{\partial}{\partial t} \left( \mathrm{det} \left( \frac{\partial \vec{x}}{\partial \vec{a}} \right) \right) =  \mathrm{det} \left( \frac{\partial \vec{x}}{\partial \vec{a}} \right) (\vec{\nabla} \cdot \vec{v}) 
\end{equation}\par

The energy conservation law, as given in Eq.~\eqref{lagrangian_energy_law}, allows us to define the energy $E^{\mathrm{L}}$ in the Lagrangian approach for the OCP and examine how it changes over time. To achieve this, we integrate this equation with respect to $\vec{a}$ and employ integration by parts. To further simplify nonlocal terms, we use the following properties: for all functions $f,g$, we have $(f^T)^T = f, (f + g)^T = f^T + g^T, (fg)^T = f^T g^T$. Additionally, the integral of $f$ over $\vec{a}, \vec{a}'$ is the same as the integral of $f^T$. Moreover, for any argument $\Psi$ of a nonlocal function, $(\partial_{\Psi'} f)^T = \partial_{\Psi} (f^T)$. To simplify nonlocal terms with $\phi_{+-}$, we use the definition of transpose. \par

It can be shown that if the nonlocal function $F$ does not depend on $\vec{a}, \vec{a}'$, the expression for the conserved energy in the Lagrangian approach is identical to the one in the Eulerian approach, as given by Eq.~\eqref{eulerian_energy_equation}. To demonstrate this, change the variables inside the integrals from $(\vec{a}, \vec{a}')$ to $(\vec{x}, \vec{x}')$, apply the continuity equation given in Eq.~\eqref{lagrangian_continuity_equations}, and note that for the neutralizing stationary background, the reference state was selected so that $\vec{x} = \vec{a}$ and, consequently, $n_- = n_-^{\mathrm{ref}}$. \par

To obtain from Eqs.~\eqref{lagrangian_energy_equation}, \eqref{lagrangian_energy_density}, the equilibrium energy per particle, $E_0/N$, notice that simplification occurs when one uses the assumed form of the potential $\phi_{ij}(r) = q_i q_j/4 \pi \varepsilon_0 r$, where $i, j$ are the corresponding species of the particles, and $r$ is the distance between the particles. For integrals involving $\phi_{+-}, \phi_{--}$, one can change variables to $\vec{R} = (\vec{a} + \vec{a}')/2, \vec{r} = \vec{a}-\vec{a}'$, which has a unit Jacobian. \par

\vspace{\baselineskip}
\subsection{Dispersion laws}
\label{appendix_lagrangian_dispersion}

We first consider Eq.~\eqref{lagrangian_final_EOM_with_thermo} and determine whether it is satisfied for the proposed equilibrium solution. \par

The left-hand side of Eq.~\eqref{lagrangian_final_EOM_with_thermo} is zero because $\vec{x} = \vec{a}$ does not depend on time. On the right-hand side, the sum over derivatives with respect to $a_j$ is zero because the function for which this derivative is taken does not depend on $\vec{a}$. To see this, we can use the continuity equations provided in Eq.~\eqref{lagrangian_continuity_equations}, along with the result that for $\vec{x} = \vec{a}$, we have $\mathrm{det} \left( \partial \vec{x} / \partial \vec{a} \right) = 1$. This leads to the conclusion that $n = n_0, s = s_0$. \par

For the integral, we notice that the functions inside it only depend on $\vec{r} = \vec{a} - \vec{a}'$, so we can eliminate the dependence on $\vec{a}$ by changing the integration variable to $\vec{r}$. To demonstrate that the nonlocal force is zero, we directly perform derivatives with respect to $x_i$ and use the equilibrium results along with $n = n_0, s = s_0$ to show that the functions inside the integral depend on $\vec{a}, \vec{a}'$ only through the combination $\vec{r} = \vec{a} - \vec{a}'$. After changing the integration variable to $\vec{r}$, we can use the fact that the function inside the integral is odd with respect to $r_i$. \par

Let us now consider the linearized equations. Since the equations of motion given in Eq.~\eqref{lagrangian_final_EOM_with_thermo} depend on $n,s$, it is useful to expand the continuity equations, as given in  Eq.~\eqref{lagrangian_continuity_equations}, up to the first order.
\begin{equation}
\label{linearized_continuity_Lagrangian}
\begin{gathered}
n(\vec{x}, t) = n_0(1 - \vec{\nabla} \cdot \vec{\xi}) + ... , \quad s(\vec{x}, t) = s_0.
\end{gathered}
\end{equation}\par

Now, let us expand the equations of motion provided in Eq.~\eqref{lagrangian_final_EOM_with_thermo} up to the first order. The derivatives of $\mathrm{det} \left( \partial \vec{x} / \partial \vec{a} \right)$ can be expanded through direct computation. Other terms can be simplified using Eq.~\eqref{linearized_continuity_Lagrangian}, $q_+ n_0 + q_- n_{-,0} = 0$, as well as the fact that for any function $f$ depending solely on $\vec{x}, \vec{x}'$ through the combination $\vec{x}-\vec{x}'$, we have $\partial f/ \partial x_i = -\partial f/\partial x'_i$. For some integrals, it is useful to change the integration variable to $\vec{r} = \vec{a}-\vec{a}'$. Additionally, we reintroduce the notation where the subscript "$0$" signifies that a function is evaluated at equilibrium values. Using this notation, we can further simplify the result by using $g_0 = (g^T)_0$ and $(\partial g/\partial n)_0 = (\partial g^T / \partial n')_0$. 
\begin{equation}
\label{linearized_velocity_final_Lagrangian}
\begin{gathered}
m \frac{\partial^2 \xi_i}{\partial t^2} = \frac{\partial (\vec{\nabla} \cdot \vec{\xi})}{\partial a_i} \Bigg(\Bigg[ \frac{\partial}{\partial n} \Big( \frac{3}{2} k_B n^2 \frac{\partial T}{\partial n}\Big|_s \Big)\Big|_s \Bigg]_0 
+ \frac{q^2_+ n_0}{8 \pi \varepsilon_0} \int \frac{1}{|\vec{r}|} \Bigg[ \frac{\partial}{\partial n} \Big( n^2 \frac{\partial g}{\partial n}\Big|_{s, |\vec{r}|} \Big)\Big|_{s, |\vec{r}|} \Bigg]_0 \mathrm{d}\vec{r} \Bigg) \\
- \frac{q^2_+ n_0^2}{8 \pi \varepsilon_0}\sum \limits_{j=1}^3 \frac{\partial}{\partial a_i} \bigg( \int \bigg[ \frac{\partial g}{\partial n}\Big|_{s, |\vec{a}-\vec{a}'|} \bigg]_0 \frac{\partial}{\partial a_j} \left( \frac{1}{|\vec{a}-\vec{a}'|}\right)\Big|_{\vec{a}'} 
\cdot (\xi_j - \xi'_j ) \mathrm{d}\vec{a}' \bigg) \\
+ \frac{q^2_+ n_0^2}{8 \pi \varepsilon_0} \int \bigg[ \frac{\partial g}{\partial n}\Big|_{s, |\vec{a}-\vec{a}'|} \bigg]_0 \frac{\partial}{\partial a_i} \left( \frac{1}{|\vec{a}-\vec{a}'|}\right)\Big|_{\vec{a}'} \cdot \big( (\vec{\nabla} \cdot \vec{\xi}) + (\vec{\nabla}' \cdot \vec{\xi}') \big) \mathrm{d}\vec{a}' \\
- \frac{q^2_+ n_0}{4 \pi \varepsilon_0} \sum \limits_{j=1}^3 \int \frac{\partial^2}{\partial a_i \partial a_j} \left( \frac{1}{|\vec{a}-\vec{a}'|}\right)\Big|_{\vec{a}'} \Big( (g-1)_0 \xi_j - g_0 \xi'_j  \Big) \mathrm{d}\vec{a}'
\end{gathered}
\end{equation}
\par

To solve this equation, we employ the Fourier transform with respect to the spatial coordinates $\vec{a}$ and use the same convention as we did in the Eulerian approach, as given in Eq.~\eqref{eulerian_fourier_transform_convention}. 
\begin{equation}
\label{lagrangian_fourier_transform_convention}
\vec{\Psi}(\vec{k}, t) = \widehat{\vec{\xi}}(\vec{k}, t) = \int \vec{\xi} (\vec{a}, t) e^{-i\vec{k} \cdot \vec{a}} \mathrm{d}\vec{a}
\end{equation}\par

With this convention, we once again obtain the results that, for any functions $f,g$ that depend on $\vec{a}$, $\widehat{\partial f/ \partial a_j} = i k_j \widehat{f}$ and $\widehat{ f * g} = \widehat{f} \widehat{g}$, where $*$ denotes convolution. Using these results, we take the Fourier transform of Eq.~\eqref{linearized_velocity_final_Lagrangian} to obtain a differential equation for each value of $\vec{k}$. As in the Eulerian approach, we simplify by changing the integration variable to $\vec{r} = \vec{a}-\vec{a}'$ when it is convenient. \par

Once again, it is important to be careful regarding the convergence of integrals. As before, we rely on the result from thermodynamics \cite{doi:10.1063/1.1727895} that $g(n_0, s_0, |\vec{r}|) \to 1$ as $|\vec{r}| \to \infty$. Therefore, it is convenient to express everything in terms of $g - 1$ instead of $g$. In addition to using the Fourier transform of $1/|\vec{r}|$ being\cite{berestetskii1982quantum} $4\pi/|\vec{k}|^2$, we use the following results \cite{10.1119/1.13127} for the derivatives of $1/|\vec{r}|$, where $\delta(\vec{r})$ represents the Dirac delta distribution centered at the origin. 
\begin{equation}
\begin{gathered}
\frac{\partial}{\partial r_j} \left( \frac{1}{|\vec{r}|} \right) = -\frac{r_j}{|\vec{r}|^3}, \quad
\frac{\partial^2}{\partial r_i \partial r_j} \left( \frac{1}{|\vec{r}|} \right) = \frac{3 r_i r_j - \delta_{ij} |\vec{r}|^2}{|\vec{r}|^5} - \frac{4\pi}{3} \delta_{ij} \delta(\vec{r}).
\end{gathered}
\end{equation}\par

Combining these results, we can express the Fourier transform of Eq.~\eqref{linearized_velocity_final_Lagrangian} as follows.
\begin{equation}
\label{FT_linearized_equations_Lagrangian}
\begin{gathered}
m \frac{\partial^2 \vec{\Psi}}{\partial t^2} = -\vec{k}(\vec{k}\cdot \vec{\Psi}) \Bigg( \frac{q_+^2 n_0}{\varepsilon_0 |\vec{k}|^2} + \Bigg[ \frac{\partial}{\partial n} \Big( \frac{3}{2} k_B n^2 \frac{\partial T}{\partial n}\Big|_s \Big)\Big|_s \Bigg]_0 \\
 + \frac{q^2_+ n_0}{8 \pi \varepsilon_0} \int \frac{1}{|\vec{r}|} \Bigg[ \frac{\partial}{\partial n} \Big( n^2 \frac{\partial (g-1)}{\partial n}\Big|_{s, |\vec{r}|} \Big)\Big|_{s, |\vec{r}|} \Bigg]_0 \mathrm{d}\vec{r} \Bigg) \\
+  \frac{i q^2_+ n_0^2 \vec{k}}{8 \pi \varepsilon_0} \int \bigg[ \frac{\partial (g-1)}{\partial n}\Big|_{s, |\vec{r}|} \bigg]_0 \frac{(\vec{r} \cdot \vec{\Psi})}{|\vec{r}|^3} \left(1 - e^{-i\vec{k} \cdot \vec{r}} \right) \mathrm{d}\vec{r} \\
- i(\vec{k} \cdot \vec{\Psi})\frac{q^2_+ n_0^2}{8 \pi \varepsilon_0} \int \bigg[ \frac{\partial (g-1)}{\partial n}\Big|_{s, |\vec{r}|} \bigg]_0 \frac{\vec{r}}{|\vec{r}|^3}\left(1 + e^{-i\vec{k} \cdot \vec{r}} \right) \mathrm{d}\vec{r} \\
- \frac{q^2_+ n_0}{4 \pi \varepsilon_0} \int (g-1)_0 \left( \frac{3\vec{r}(\vec{r} \cdot \vec{\Psi})}{|\vec{r}|^5} - \frac{\vec{\Psi}}{|\vec{r}|^3} \right)\left(1 - e^{-i\vec{k} \cdot \vec{r}} \right) \mathrm{d}\vec{r}
\end{gathered}
\end{equation}\par

To simplify the analysis of the longitudinal and transverse dispersion modes, we employ the same strategy used in the Eulerian approach, where the coordinate system is rotated in such a way that, in the rotated coordinate system, $\vec{k} = (0, 0, |\vec{k}|)$. In this case, the transverse modes correspond to the $x,y$ directions, while the longitudinal mode corresponds to the $z$ direction. To achieve this, we use the fact that the differential equation is linear, properties of the dot product under a rotation, and rotate the integration variable. It is worth noting that due to some of the functions under the integrals being odd with respect to the integration variables, certain integrals become zero. Further simplification for the remaining integrals can be achieved by going to spherical coordinates and performing the angular integrals. \par

After rotating the coordinate system, we find from Eq.~\eqref{FT_linearized_equations_Lagrangian} that for the transverse directions, $\partial^2 \Psi_x / \partial t^2 = -\omega_T^2(|\vec{k}|) \Psi_x$ and $\partial^2 \Psi_y / \partial t^2 = -\omega_T^2(|\vec{k}|) \Psi_y$, with transverse dispersion law given by Eq.~\eqref{transverse_dispersion_Lagrangian}. For the longitudinal direction, we find $\partial^2 \Psi_z / \partial t^2 = -\omega_L^2(|\vec{k}|) \Psi_z$, along with the longitudinal dispersion law given by Eq.~\eqref{longitudinal_dispersion_Lagrangian}.

To compute the term with the temperature derivatives at constant entropy in the longitudinal dispersion law given in Eq.~\eqref{longitudinal_dispersion_Lagrangian}, one uses previously discussed results on the temperature derivative as in Eq.~\eqref{eulerian_f1} and on the derivative at constant entropy of a function of $\Gamma$ as in Eq.~\eqref{constant_s_derivative}.\par

To analyze terms with integrals of $g-1$ in both of the dispersion laws, in addition to the previous identities, use the strategy already discussed in the Eulerian approach. Using $g = g(\Gamma, |\vec{r}|/a)$, take number density derivatives outside of the integrals, change variables inside the integrals to $x = r/a$, and introduce normalized wavevector $\vec{q} = \vec{k}a$. In that case, integrals are rewritten in terms of functions of $\Gamma$ and $|\vec{q}|$, and we can use the identity given in Eq.~\eqref{constant_s_derivative}.\par

For the simple fit $u(\Gamma) = -0.9 \Gamma$, where $u(\Gamma)$ is linear in $\Gamma$, we can observe from Eq.~\eqref{R_over_a} that $R/a$ is independent of $\Gamma$. In this case, $R/a = \sqrt{6/5}$. Moreover, from Eqs.~\eqref{simplified_j_h_Eulerian}, \eqref{simplified_b_ell_Lagrangian}, this implies that the functions $j, b, \ell$ appearing in the longitudinal dispersion relation given in Eq.~\eqref{longitudinal_dispersion_Lagrangian_normalized} are also independent of $\Gamma$. These results greatly simplify the calculations of taking number density derivatives at constant $s$, as shown in Eq.~\eqref{constant_s_derivative}. Furthermore, as before, for such a fit, we have $f_1(\Gamma) = 1$ by Eq.~\eqref{eulerian_f1}. \par

\subsection{Modified Lagrangian approach}

The computation of the term in the longitudinal dispersion law given in Eq.~\eqref{longitudinal_dispersion_Lagrangian} involving the temperature derivatives at constant entropy is the same as in both previous approaches. One uses results on the temperature derivative, as shown in Eq.~\eqref{eulerian_f1}, and on the derivative at constant entropy of a function of $\Gamma$, as demonstrated in Eq.~\eqref{constant_s_derivative}. \par

To analyze terms with integrals of $g-1$ in both of the dispersion laws, in addition to the previous identities, use the following strategy, which is slightly different from the strategy used in the two previous approaches. First, take number density derivatives outside of the integrals. Then, change variables inside the integrals to $x = r/a_0$ and introduce the normalized wavevector $\vec{q}_0 = \vec{k}a_0$. In this case, integrals are rewritten in terms of functions of $\Gamma(n_0, T(n,s))$ and $|\vec{q}_0|$. Next, one can use the following identity for an arbitrary function $f(\Gamma(n_0, T(n,s)), |\vec{q}_0|)$ on the derivative of number density at constant entropy, where we use the chain rule, Eq.~\eqref{eulerian_f1}, and the definition of $\Gamma$. 
\begin{equation}
\label{modified_constant_s_derivative}
\begin{gathered}
n \frac{\partial f}{\partial n}\Big|_s \big(\Gamma(n_0, T(n,s)), |\vec{q}_0|\big) = -\frac{2}{3}  \Gamma(n_0, T(n,s))  \frac{\partial f}{\partial \Gamma} \Big|_{|\vec{q}_0|}f_1\big(\Gamma(n, T) \big) 
\end{gathered}
\end{equation}\par

For the simple fit $u(\Gamma) = -0.9 \Gamma$, where $u(\Gamma)$ is linear in $\Gamma$, we can observe from Eq.~\eqref{R_over_a} that $R/a$ is independent of $\Gamma$. In this case, $R/a = \sqrt{6/5}$. As before, from Eqs.~\eqref{simplified_j_h_Eulerian}, \eqref{simplified_b_ell_Lagrangian}, it follows that the functions $j, b, \ell$ appearing in the longitudinal dispersion relation, as given in Eq.~\eqref{longitudinal_dispersion_modified_Lagrangian_normalized}, are also independent of $\Gamma$. This significantly simplifies the dispresion relation, and from Eqs.~\eqref{modified_lagrangian_j1}, \eqref{modified_lagrangian_j2}, \eqref{modified_lagrangian_l1}, we deduce that $j_1, j_2, \ell_1$ are all zero. Additionally, as previously discussed, for such a fit, we have $f_1(\Gamma) = 1$ according to Eq.~\eqref{eulerian_f1}. \par

\bibliography{article_source_file}

\begin{thebibliography}{46}%
\makeatletter
\providecommand \@ifxundefined [1]{%
 \@ifx{#1\undefined}
}%
\providecommand \@ifnum [1]{%
 \ifnum #1\expandafter \@firstoftwo
 \else \expandafter \@secondoftwo
 \fi
}%
\providecommand \@ifx [1]{%
 \ifx #1\expandafter \@firstoftwo
 \else \expandafter \@secondoftwo
 \fi
}%
\providecommand \natexlab [1]{#1}%
\providecommand \enquote  [1]{``#1''}%
\providecommand \bibnamefont  [1]{#1}%
\providecommand \bibfnamefont [1]{#1}%
\providecommand \citenamefont [1]{#1}%
\providecommand \href@noop [0]{\@secondoftwo}%
\providecommand \href [0]{\begingroup \@sanitize@url \@href}%
\providecommand \@href[1]{\@@startlink{#1}\@@href}%
\providecommand \@@href[1]{\endgroup#1\@@endlink}%
\providecommand \@sanitize@url [0]{\catcode `\\12\catcode `\$12\catcode
  `\&12\catcode `\#12\catcode `\^12\catcode `\_12\catcode `\%12\relax}%
\providecommand \@@startlink[1]{}%
\providecommand \@@endlink[0]{}%
\providecommand \url  [0]{\begingroup\@sanitize@url \@url }%
\providecommand \@url [1]{\endgroup\@href {#1}{\urlprefix }}%
\providecommand \urlprefix  [0]{URL }%
\providecommand \Eprint [0]{\href }%
\providecommand \doibase [0]{https://doi.org/}%
\providecommand \selectlanguage [0]{\@gobble}%
\providecommand \bibinfo  [0]{\@secondoftwo}%
\providecommand \bibfield  [0]{\@secondoftwo}%
\providecommand \translation [1]{[#1]}%
\providecommand \BibitemOpen [0]{}%
\providecommand \bibitemStop [0]{}%
\providecommand \bibitemNoStop [0]{.\EOS\space}%
\providecommand \EOS [0]{\spacefactor3000\relax}%
\providecommand \BibitemShut  [1]{\csname bibitem#1\endcsname}%
\let\auto@bib@innerbib\@empty
\bibitem [{\citenamefont {Brush}, \citenamefont {Sahlin},\ and\ \citenamefont
  {Teller}(1966)}]{doi:10.1063/1.1727895}%
  \BibitemOpen
  \bibfield  {author} {\bibinfo {author} {\bibfnamefont {S.~G.}\ \bibnamefont
  {Brush}}, \bibinfo {author} {\bibfnamefont {H.~L.}\ \bibnamefont {Sahlin}},\
  and\ \bibinfo {author} {\bibfnamefont {E.}~\bibnamefont {Teller}},\
  }\bibfield  {title} {\enquote {\bibinfo {title} {Monte {C}arlo {S}tudy of a
  {O}ne‐{C}omponent plasma. {I}},}\ }\href
  {https://doi.org/10.1063/1.1727895} {\bibfield  {journal} {\bibinfo
  {journal} {J. Chem. Phys.}\ }\textbf {\bibinfo {volume} {45}},\ \bibinfo
  {pages} {2102--2118} (\bibinfo {year} {1966})}\BibitemShut {NoStop}%
\bibitem [{\citenamefont {Hansen}(1973)}]{PhysRevA.8.3096}%
  \BibitemOpen
  \bibfield  {author} {\bibinfo {author} {\bibfnamefont {J.~P.}\ \bibnamefont
  {Hansen}},\ }\bibfield  {title} {\enquote {\bibinfo {title} {Statistical
  {M}echanics of {D}ense {I}onized {M}atter. {I}. {E}quilibrium {P}roperties of
  the {C}lassical {O}ne-{C}omponent {P}lasma},}\ }\href
  {https://doi.org/10.1103/PhysRevA.8.3096} {\bibfield  {journal} {\bibinfo
  {journal} {Phys. Rev. A}\ }\textbf {\bibinfo {volume} {8}},\ \bibinfo {pages}
  {3096--3109} (\bibinfo {year} {1973})}\BibitemShut {NoStop}%
\bibitem [{\citenamefont {van Horn}(1968)}]{1968ApJ151227V}%
  \BibitemOpen
  \bibfield  {author} {\bibinfo {author} {\bibfnamefont {H.~M.}\ \bibnamefont
  {van Horn}},\ }\bibfield  {title} {\enquote {\bibinfo {title}
  {Crystallization of {W}hite {D}warfs},}\ }\href
  {https://doi.org/10.1086/149432} {\bibfield  {journal} {\bibinfo  {journal}
  {Astrophys. J.}\ }\textbf {\bibinfo {volume} {151}},\ \bibinfo {pages} {227}
  (\bibinfo {year} {1968})}\BibitemShut {NoStop}%
\bibitem [{\citenamefont {van Horn}(2019)}]{VanHorn_2019}%
  \BibitemOpen
  \bibfield  {author} {\bibinfo {author} {\bibfnamefont {H.~M.}\ \bibnamefont
  {van Horn}},\ }\bibfield  {title} {\enquote {\bibinfo {title} {The
  crystallization of white dwarf stars},}\ }\href
  {https://doi.org/10.1038/s41550-019-0695-1} {\bibfield  {journal} {\bibinfo
  {journal} {Nat. Astron.}\ }\textbf {\bibinfo {volume} {3}},\ \bibinfo {pages}
  {129--130} (\bibinfo {year} {2019})}\BibitemShut {NoStop}%
\bibitem [{\citenamefont {Demmel}\ \emph {et~al.}(2005)\citenamefont {Demmel},
  \citenamefont {Hosokawa}, \citenamefont {Pilgrim},\ and\ \citenamefont
  {Tsutsui}}]{DEMMEL200598}%
  \BibitemOpen
  \bibfield  {author} {\bibinfo {author} {\bibfnamefont {F.}~\bibnamefont
  {Demmel}}, \bibinfo {author} {\bibfnamefont {S.}~\bibnamefont {Hosokawa}},
  \bibinfo {author} {\bibfnamefont {W.-C.}\ \bibnamefont {Pilgrim}},\ and\
  \bibinfo {author} {\bibfnamefont {S.}~\bibnamefont {Tsutsui}},\ }\bibfield
  {title} {\enquote {\bibinfo {title} {Evidences for optic modes in molten
  {NaI}},}\ }\href {https://doi.org/https://doi.org/10.1016/j.nimb.2005.06.025}
  {\bibfield  {journal} {\bibinfo  {journal} {Nucl. Instrum. Methods Phys. Res.
  B}\ }\textbf {\bibinfo {volume} {238}},\ \bibinfo {pages} {98--101} (\bibinfo
  {year} {2005})}\BibitemShut {NoStop}%
\bibitem [{\citenamefont {Demmel}, \citenamefont {Hosokawa},\ and\
  \citenamefont {Pilgrim}(2021)}]{Demmel_2021}%
  \BibitemOpen
  \bibfield  {author} {\bibinfo {author} {\bibfnamefont {F.}~\bibnamefont
  {Demmel}}, \bibinfo {author} {\bibfnamefont {S.}~\bibnamefont {Hosokawa}},\
  and\ \bibinfo {author} {\bibfnamefont {W.-C.}\ \bibnamefont {Pilgrim}},\
  }\bibfield  {title} {\enquote {\bibinfo {title} {Collective particle dynamics
  of molten {NaCl} by inelastic x-ray scattering},}\ }\href
  {https://doi.org/10.1088/1361-648X/ac101c} {\bibfield  {journal} {\bibinfo
  {journal} {J. Phys. Condens. Matter}\ }\textbf {\bibinfo {volume} {33}},\
  \bibinfo {pages} {375103} (\bibinfo {year} {2021})}\BibitemShut {NoStop}%
\bibitem [{\citenamefont {Hansen}\ and\ \citenamefont
  {McDonald}(1975)}]{PhysRevA.11.2111}%
  \BibitemOpen
  \bibfield  {author} {\bibinfo {author} {\bibfnamefont {J.~P.}\ \bibnamefont
  {Hansen}}\ and\ \bibinfo {author} {\bibfnamefont {I.~R.}\ \bibnamefont
  {McDonald}},\ }\bibfield  {title} {\enquote {\bibinfo {title} {Statistical
  mechanics of dense ionized matter. {IV}. {D}ensity and charge fluctuations in
  a simple molten salt},}\ }\href {https://doi.org/10.1103/PhysRevA.11.2111}
  {\bibfield  {journal} {\bibinfo  {journal} {Phys. Rev. A}\ }\textbf {\bibinfo
  {volume} {11}},\ \bibinfo {pages} {2111--2123} (\bibinfo {year}
  {1975})}\BibitemShut {NoStop}%
\bibitem [{\citenamefont {Bataller}\ \emph {et~al.}(2019)\citenamefont
  {Bataller}, \citenamefont {Latshaw}, \citenamefont {Koulakis},\ and\
  \citenamefont {Putterman}}]{Bataller2019DynamicsOS}%
  \BibitemOpen
  \bibfield  {author} {\bibinfo {author} {\bibfnamefont {A.}~\bibnamefont
  {Bataller}}, \bibinfo {author} {\bibfnamefont {A.}~\bibnamefont {Latshaw}},
  \bibinfo {author} {\bibfnamefont {J.~P.}\ \bibnamefont {Koulakis}},\ and\
  \bibinfo {author} {\bibfnamefont {S.}~\bibnamefont {Putterman}},\ }\bibfield
  {title} {\enquote {\bibinfo {title} {Dynamics of strongly coupled
  two-component plasma via ultrafast spectroscopy},}\ }\href
  {https://doi.org/10.1364/OL.44.005832} {\bibfield  {journal} {\bibinfo
  {journal} {Opt. Lett.}\ }\textbf {\bibinfo {volume} {44}},\ \bibinfo {pages}
  {5832--5835} (\bibinfo {year} {2019})}\BibitemShut {NoStop}%
\bibitem [{\citenamefont {Killian}\ \emph {et~al.}(2007)\citenamefont
  {Killian}, \citenamefont {Pattard}, \citenamefont {Pohl},\ and\ \citenamefont
  {Rost}}]{KILLIAN200777}%
  \BibitemOpen
  \bibfield  {author} {\bibinfo {author} {\bibfnamefont {T.}~\bibnamefont
  {Killian}}, \bibinfo {author} {\bibfnamefont {T.}~\bibnamefont {Pattard}},
  \bibinfo {author} {\bibfnamefont {T.}~\bibnamefont {Pohl}},\ and\ \bibinfo
  {author} {\bibfnamefont {J.}~\bibnamefont {Rost}},\ }\bibfield  {title}
  {\enquote {\bibinfo {title} {Ultracold neutral plasmas},}\ }\href
  {https://doi.org/https://doi.org/10.1016/j.physrep.2007.04.007} {\bibfield
  {journal} {\bibinfo  {journal} {Phys. Rep.}\ }\textbf {\bibinfo {volume}
  {449}},\ \bibinfo {pages} {77--130} (\bibinfo {year} {2007})}\BibitemShut
  {NoStop}%
\bibitem [{\citenamefont {Maxson}\ \emph {et~al.}(2013)\citenamefont {Maxson},
  \citenamefont {Bazarov}, \citenamefont {Wan}, \citenamefont {Padmore},\ and\
  \citenamefont {Coleman-Smith}}]{Maxson_2013}%
  \BibitemOpen
  \bibfield  {author} {\bibinfo {author} {\bibfnamefont {J.~M.}\ \bibnamefont
  {Maxson}}, \bibinfo {author} {\bibfnamefont {I.~V.}\ \bibnamefont {Bazarov}},
  \bibinfo {author} {\bibfnamefont {W.}~\bibnamefont {Wan}}, \bibinfo {author}
  {\bibfnamefont {H.~A.}\ \bibnamefont {Padmore}},\ and\ \bibinfo {author}
  {\bibfnamefont {C.~E.}\ \bibnamefont {Coleman-Smith}},\ }\bibfield  {title}
  {\enquote {\bibinfo {title} {Fundamental photoemission brightness limit from
  disorder induced heating},}\ }\href
  {https://doi.org/10.1088/1367-2630/15/10/103024} {\bibfield  {journal}
  {\bibinfo  {journal} {New J. Phys.}\ }\textbf {\bibinfo {volume} {15}},\
  \bibinfo {pages} {103024} (\bibinfo {year} {2013})}\BibitemShut {NoStop}%
\bibitem [{\citenamefont {Morfill}\ and\ \citenamefont
  {Ivlev}(2009)}]{RevModPhys.81.1353}%
  \BibitemOpen
  \bibfield  {author} {\bibinfo {author} {\bibfnamefont {G.~E.}\ \bibnamefont
  {Morfill}}\ and\ \bibinfo {author} {\bibfnamefont {A.~V.}\ \bibnamefont
  {Ivlev}},\ }\bibfield  {title} {\enquote {\bibinfo {title} {Complex plasmas:
  An interdisciplinary research field},}\ }\href
  {https://doi.org/10.1103/RevModPhys.81.1353} {\bibfield  {journal} {\bibinfo
  {journal} {Rev. Mod. Phys.}\ }\textbf {\bibinfo {volume} {81}},\ \bibinfo
  {pages} {1353--1404} (\bibinfo {year} {2009})}\BibitemShut {NoStop}%
\bibitem [{\citenamefont {Landau}\ and\ \citenamefont
  {Lifshitz}(1987)}]{Landau1987Fluid}%
  \BibitemOpen
  \bibfield  {author} {\bibinfo {author} {\bibfnamefont {L.~D.}\ \bibnamefont
  {Landau}}\ and\ \bibinfo {author} {\bibfnamefont {E.~M.}\ \bibnamefont
  {Lifshitz}},\ }\href@noop {} {\emph {\bibinfo {title} {Fluid Mechanics:
  Volume 6 (Course of Theoretical Physics)}}},\ \bibinfo {edition} {2nd}\ ed.\
  (\bibinfo  {publisher} {Butterworth-Heinemann},\ \bibinfo {year}
  {1987})\BibitemShut {NoStop}%
\bibitem [{\citenamefont {Desbiens}, \citenamefont {Arnault},\ and\
  \citenamefont {Clérouin}(2016)}]{doi:10.1063/1.4963388}%
  \BibitemOpen
  \bibfield  {author} {\bibinfo {author} {\bibfnamefont {N.}~\bibnamefont
  {Desbiens}}, \bibinfo {author} {\bibfnamefont {P.}~\bibnamefont {Arnault}},\
  and\ \bibinfo {author} {\bibfnamefont {J.}~\bibnamefont {Clérouin}},\
  }\bibfield  {title} {\enquote {\bibinfo {title} {Parametrization of pair
  correlation function and static structure factor of the one component plasma
  across coupling regimes},}\ }\href {https://doi.org/10.1063/1.4963388}
  {\bibfield  {journal} {\bibinfo  {journal} {Phys. Plasmas}\ }\textbf
  {\bibinfo {volume} {23}},\ \bibinfo {pages} {092120} (\bibinfo {year}
  {2016})}\BibitemShut {NoStop}%
\bibitem [{\citenamefont {DeWitt}(1978)}]{refId0}%
  \BibitemOpen
  \bibfield  {author} {\bibinfo {author} {\bibfnamefont {H.~E.}\ \bibnamefont
  {DeWitt}},\ }\bibfield  {title} {\enquote {\bibinfo {title} {Statistical
  mechanics of dense plasmas: Numerical simulation and theory},}\ }\href
  {https://doi.org/10.1051/jphyscol:1978132} {\bibfield  {journal} {\bibinfo
  {journal} {J. Phys. Colloq.}\ }\textbf {\bibinfo {volume} {39}},\ \bibinfo
  {pages} {C1--173--C1--180} (\bibinfo {year} {1978})}\BibitemShut {NoStop}%
\bibitem [{\citenamefont {Khrapak}\ and\ \citenamefont
  {Khrapak}(2014)}]{doi:10.1063/1.4897386}%
  \BibitemOpen
  \bibfield  {author} {\bibinfo {author} {\bibfnamefont {S.~A.}\ \bibnamefont
  {Khrapak}}\ and\ \bibinfo {author} {\bibfnamefont {A.~G.}\ \bibnamefont
  {Khrapak}},\ }\bibfield  {title} {\enquote {\bibinfo {title} {Simple
  thermodynamics of strongly coupled one-component-plasma in two and three
  dimensions},}\ }\href {https://doi.org/10.1063/1.4897386} {\bibfield
  {journal} {\bibinfo  {journal} {Phys. Plasmas}\ }\textbf {\bibinfo {volume}
  {21}},\ \bibinfo {pages} {104505} (\bibinfo {year} {2014})}\BibitemShut
  {NoStop}%
\bibitem [{\citenamefont {Hamaguchi}, \citenamefont {Farouki},\ and\
  \citenamefont {Dubin}(1996)}]{10.1063/1.472802}%
  \BibitemOpen
  \bibfield  {author} {\bibinfo {author} {\bibfnamefont {S.}~\bibnamefont
  {Hamaguchi}}, \bibinfo {author} {\bibfnamefont {R.~T.}\ \bibnamefont
  {Farouki}},\ and\ \bibinfo {author} {\bibfnamefont {D.~H.~E.}\ \bibnamefont
  {Dubin}},\ }\bibfield  {title} {\enquote {\bibinfo {title} {{Phase diagram of
  Yukawa systems near the one‐component‐plasma limit revisited}},}\ }\href
  {https://doi.org/10.1063/1.472802} {\bibfield  {journal} {\bibinfo  {journal}
  {J. Chem. Phys.}\ }\textbf {\bibinfo {volume} {105}},\ \bibinfo {pages}
  {7641--7647} (\bibinfo {year} {1996})}\BibitemShut {NoStop}%
\bibitem [{\citenamefont {Murillo}(2006)}]{PhysRevLett.96.165001}%
  \BibitemOpen
  \bibfield  {author} {\bibinfo {author} {\bibfnamefont {M.~S.}\ \bibnamefont
  {Murillo}},\ }\bibfield  {title} {\enquote {\bibinfo {title} {Ultrafast
  {D}ynamics of {S}trongly {C}oupled {P}lasmas},}\ }\href
  {https://doi.org/10.1103/PhysRevLett.96.165001} {\bibfield  {journal}
  {\bibinfo  {journal} {Phys. Rev. Lett.}\ }\textbf {\bibinfo {volume} {96}},\
  \bibinfo {pages} {165001} (\bibinfo {year} {2006})}\BibitemShut {NoStop}%
\bibitem [{\citenamefont {Acciarri}, \citenamefont {Moore},\ and\ \citenamefont
  {Baalrud}(2022)}]{Acciarri_2022}%
  \BibitemOpen
  \bibfield  {author} {\bibinfo {author} {\bibfnamefont {M.~D.}\ \bibnamefont
  {Acciarri}}, \bibinfo {author} {\bibfnamefont {C.}~\bibnamefont {Moore}},\
  and\ \bibinfo {author} {\bibfnamefont {S.~D.}\ \bibnamefont {Baalrud}},\
  }\bibfield  {title} {\enquote {\bibinfo {title} {Strong {C}oulomb coupling
  influences ion and neutral temperatures in atmospheric pressure plasmas},}\
  }\href {https://doi.org/10.1088/1361-6595/aca69c} {\bibfield  {journal}
  {\bibinfo  {journal} {Plasma Sources Sci. Technol.}\ }\textbf {\bibinfo
  {volume} {31}},\ \bibinfo {pages} {125005} (\bibinfo {year}
  {2022})}\BibitemShut {NoStop}%
\bibitem [{\citenamefont {Hansen}, \citenamefont {McDonald},\ and\
  \citenamefont {Pollock}(1975)}]{PhysRevA.11.1025}%
  \BibitemOpen
  \bibfield  {author} {\bibinfo {author} {\bibfnamefont {J.~P.}\ \bibnamefont
  {Hansen}}, \bibinfo {author} {\bibfnamefont {I.~R.}\ \bibnamefont
  {McDonald}},\ and\ \bibinfo {author} {\bibfnamefont {E.~L.}\ \bibnamefont
  {Pollock}},\ }\bibfield  {title} {\enquote {\bibinfo {title} {Statistical
  mechanics of dense ionized matter. {III}. {D}ynamical properties of the
  classical one-component plasma},}\ }\href
  {https://doi.org/10.1103/PhysRevA.11.1025} {\bibfield  {journal} {\bibinfo
  {journal} {Phys. Rev. A}\ }\textbf {\bibinfo {volume} {11}},\ \bibinfo
  {pages} {1025--1039} (\bibinfo {year} {1975})}\BibitemShut {NoStop}%
\bibitem [{\citenamefont {Korolov}\ \emph {et~al.}(2015)\citenamefont
  {Korolov}, \citenamefont {Kalman}, \citenamefont {Silvestri},\ and\
  \citenamefont {Donkó}}]{https://doi.org/10.1002/ctpp.201400098}%
  \BibitemOpen
  \bibfield  {author} {\bibinfo {author} {\bibfnamefont {I.}~\bibnamefont
  {Korolov}}, \bibinfo {author} {\bibfnamefont {G.~J.}\ \bibnamefont {Kalman}},
  \bibinfo {author} {\bibfnamefont {L.}~\bibnamefont {Silvestri}},\ and\
  \bibinfo {author} {\bibfnamefont {Z.}~\bibnamefont {Donkó}},\ }\bibfield
  {title} {\enquote {\bibinfo {title} {The {D}ynamical {S}tructure {F}unction
  of the {O}ne-{C}omponent {P}lasma {R}evisited},}\ }\href
  {https://doi.org/https://doi.org/10.1002/ctpp.201400098} {\bibfield
  {journal} {\bibinfo  {journal} {Contrib. Plasma Phys.}\ }\textbf {\bibinfo
  {volume} {55}},\ \bibinfo {pages} {421--427} (\bibinfo {year}
  {2015})}\BibitemShut {NoStop}%
\bibitem [{\citenamefont {Landau}\ and\ \citenamefont
  {Lifshitz}(1976)}]{Landau1976Mechanics}%
  \BibitemOpen
  \bibfield  {author} {\bibinfo {author} {\bibfnamefont {L.~D.}\ \bibnamefont
  {Landau}}\ and\ \bibinfo {author} {\bibfnamefont {E.~M.}\ \bibnamefont
  {Lifshitz}},\ }\href@noop {} {\emph {\bibinfo {title} {Mechanics: Volume 1
  (Course of Theoretical Physics)}}},\ \bibinfo {edition} {3rd}\ ed.\ (\bibinfo
   {publisher} {Butterworth-Heinemann},\ \bibinfo {year} {1976})\BibitemShut
  {NoStop}%
\bibitem [{\citenamefont {Landau}\ and\ \citenamefont
  {Lifshitz}(1980{\natexlab{a}})}]{Landau1980Classical}%
  \BibitemOpen
  \bibfield  {author} {\bibinfo {author} {\bibfnamefont {L.~D.}\ \bibnamefont
  {Landau}}\ and\ \bibinfo {author} {\bibfnamefont {E.~M.}\ \bibnamefont
  {Lifshitz}},\ }\href@noop {} {\emph {\bibinfo {title} {The Classical Theory
  of Fields: Volume 2 (Course of Theoretical Physics)}}},\ \bibinfo {edition}
  {4th}\ ed.\ (\bibinfo  {publisher} {Butterworth-Heinemann},\ \bibinfo {year}
  {1980})\BibitemShut {NoStop}%
\bibitem [{\citenamefont {DeWitt}\ and\ \citenamefont
  {Rosenfeld}(1979)}]{DEWITT197979}%
  \BibitemOpen
  \bibfield  {author} {\bibinfo {author} {\bibfnamefont {H.}~\bibnamefont
  {DeWitt}}\ and\ \bibinfo {author} {\bibfnamefont {Y.}~\bibnamefont
  {Rosenfeld}},\ }\bibfield  {title} {\enquote {\bibinfo {title} {Derivation of
  the one component plasma fluid equation of state in strong coupling},}\
  }\href {https://doi.org/https://doi.org/10.1016/0375-9601(79)90283-4}
  {\bibfield  {journal} {\bibinfo  {journal} {Phys. Lett. A}\ }\textbf
  {\bibinfo {volume} {75}},\ \bibinfo {pages} {79--80} (\bibinfo {year}
  {1979})}\BibitemShut {NoStop}%
\bibitem [{\citenamefont {Herivel}(1955)}]{herivel_1955}%
  \BibitemOpen
  \bibfield  {author} {\bibinfo {author} {\bibfnamefont {J.~W.}\ \bibnamefont
  {Herivel}},\ }\bibfield  {title} {\enquote {\bibinfo {title} {The derivation
  of the equations of motion of an ideal fluid by {H}amilton's principle},}\
  }\href {https://doi.org/10.1017/S0305004100030267} {\bibfield  {journal}
  {\bibinfo  {journal} {Math. Proc. Camb. Philos. Soc.}\ }\textbf {\bibinfo
  {volume} {51}},\ \bibinfo {pages} {344–349} (\bibinfo {year}
  {1955})}\BibitemShut {NoStop}%
\bibitem [{\citenamefont {Seliger}\ and\ \citenamefont
  {Whitham}(1968)}]{doi:10.1098/rspa.1968.0103}%
  \BibitemOpen
  \bibfield  {author} {\bibinfo {author} {\bibfnamefont {R.~L.}\ \bibnamefont
  {Seliger}}\ and\ \bibinfo {author} {\bibfnamefont {G.~B.}\ \bibnamefont
  {Whitham}},\ }\bibfield  {title} {\enquote {\bibinfo {title} {Variational
  principles in continuum mechanics},}\ }\href
  {https://doi.org/10.1098/rspa.1968.0103} {\bibfield  {journal} {\bibinfo
  {journal} {Proc. R. Soc. A}\ }\textbf {\bibinfo {volume} {305}},\ \bibinfo
  {pages} {1--25} (\bibinfo {year} {1968})}\BibitemShut {NoStop}%
\bibitem [{\citenamefont {Lin}(1963)}]{lin_1963}%
  \BibitemOpen
  \bibfield  {author} {\bibinfo {author} {\bibfnamefont {C.~C.}\ \bibnamefont
  {Lin}},\ }\bibfield  {title} {\enquote {\bibinfo {title} {Hydrodynamics of
  helium {II}},}\ }in\ \href@noop {} {\emph {\bibinfo {booktitle} {Liquid
  helium: Proceedings of the international school of physics “Enrico
  Fermi”}}},\ \bibinfo {series and number} {\bibinfo {number} {Course
  {XXI}}}\ (\bibinfo  {publisher} {Academic Press},\ \bibinfo {year} {1963})\
  pp.\ \bibinfo {pages} {93--146}\BibitemShut {NoStop}%
\bibitem [{\citenamefont {Putterman}(1982)}]{PUTTERMAN1982146}%
  \BibitemOpen
  \bibfield  {author} {\bibinfo {author} {\bibfnamefont {S.}~\bibnamefont
  {Putterman}},\ }\bibfield  {title} {\enquote {\bibinfo {title} {Comments on
  the variational principle and superfluid mechanics},}\ }\href
  {https://doi.org/https://doi.org/10.1016/0375-9601(82)90877-5} {\bibfield
  {journal} {\bibinfo  {journal} {Phys. Lett. A}\ }\textbf {\bibinfo {volume}
  {89}},\ \bibinfo {pages} {146--148} (\bibinfo {year} {1982})}\BibitemShut
  {NoStop}%
\bibitem [{\citenamefont {Khrapak}\ \emph {et~al.}(2019)\citenamefont
  {Khrapak}, \citenamefont {Khrapak}, \citenamefont {Kryuchkov},\ and\
  \citenamefont {Yurchenko}}]{10.1063/1.5088141}%
  \BibitemOpen
  \bibfield  {author} {\bibinfo {author} {\bibfnamefont {S.~A.}\ \bibnamefont
  {Khrapak}}, \bibinfo {author} {\bibfnamefont {A.~G.}\ \bibnamefont
  {Khrapak}}, \bibinfo {author} {\bibfnamefont {N.~P.}\ \bibnamefont
  {Kryuchkov}},\ and\ \bibinfo {author} {\bibfnamefont {S.~O.}\ \bibnamefont
  {Yurchenko}},\ }\bibfield  {title} {\enquote {\bibinfo {title} {{Onset of
  transverse (shear) waves in strongly-coupled {Y}ukawa fluids}},}\ }\href
  {https://doi.org/10.1063/1.5088141} {\bibfield  {journal} {\bibinfo
  {journal} {J. Chem. Phys.}\ }\textbf {\bibinfo {volume} {150}},\ \bibinfo
  {pages} {104503} (\bibinfo {year} {2019})}\BibitemShut {NoStop}%
\bibitem [{\citenamefont {Mithen}, \citenamefont {Daligault},\ and\
  \citenamefont {Gregori}(2012)}]{doi:10.1063/1.3679586}%
  \BibitemOpen
  \bibfield  {author} {\bibinfo {author} {\bibfnamefont {J.~P.}\ \bibnamefont
  {Mithen}}, \bibinfo {author} {\bibfnamefont {J.}~\bibnamefont {Daligault}},\
  and\ \bibinfo {author} {\bibfnamefont {G.}~\bibnamefont {Gregori}},\
  }\bibfield  {title} {\enquote {\bibinfo {title} {Onset of negative dispersion
  in the one-component plasma},}\ }\href {https://doi.org/10.1063/1.3679586}
  {\bibfield  {journal} {\bibinfo  {journal} {AIP Conf. Proc.}\ }\textbf
  {\bibinfo {volume} {1421}},\ \bibinfo {pages} {68--72} (\bibinfo {year}
  {2012})}\BibitemShut {NoStop}%
\bibitem [{\citenamefont {Khrapak}\ \emph {et~al.}(2016)\citenamefont
  {Khrapak}, \citenamefont {Klumov}, \citenamefont {Couëdel},\ and\
  \citenamefont {Thomas}}]{doi:10.1063/1.4942169}%
  \BibitemOpen
  \bibfield  {author} {\bibinfo {author} {\bibfnamefont {S.~A.}\ \bibnamefont
  {Khrapak}}, \bibinfo {author} {\bibfnamefont {B.}~\bibnamefont {Klumov}},
  \bibinfo {author} {\bibfnamefont {L.}~\bibnamefont {Couëdel}},\ and\
  \bibinfo {author} {\bibfnamefont {H.~M.}\ \bibnamefont {Thomas}},\ }\bibfield
   {title} {\enquote {\bibinfo {title} {On the long-waves dispersion in
  {Y}ukawa systems},}\ }\href {https://doi.org/10.1063/1.4942169} {\bibfield
  {journal} {\bibinfo  {journal} {Phys. Plasmas}\ }\textbf {\bibinfo {volume}
  {23}},\ \bibinfo {pages} {023702} (\bibinfo {year} {2016})}\BibitemShut
  {NoStop}%
\bibitem [{\citenamefont {Hansen}(1981)}]{Hansen1981}%
  \BibitemOpen
  \bibfield  {author} {\bibinfo {author} {\bibfnamefont {J.~P.}\ \bibnamefont
  {Hansen}},\ }\bibfield  {title} {\enquote {\bibinfo {title} {Plasmon
  dispersion of the strongly coupled one component plasma in two and three
  dimensions},}\ }\href {https://doi.org/10.1051/jphyslet:019810042017039700}
  {\bibfield  {journal} {\bibinfo  {journal} {J. Phys. Lett.}\ }\textbf
  {\bibinfo {volume} {42}},\ \bibinfo {pages} {397--400} (\bibinfo {year}
  {1981})}\BibitemShut {NoStop}%
\bibitem [{\citenamefont {Kalman}\ and\ \citenamefont
  {Golden}(1990)}]{PhysRevA.41.5516}%
  \BibitemOpen
  \bibfield  {author} {\bibinfo {author} {\bibfnamefont {G.}~\bibnamefont
  {Kalman}}\ and\ \bibinfo {author} {\bibfnamefont {K.~I.}\ \bibnamefont
  {Golden}},\ }\bibfield  {title} {\enquote {\bibinfo {title} {Response
  function and plasmon dispersion for strongly coupled {C}oulomb liquids},}\
  }\href {https://doi.org/10.1103/PhysRevA.41.5516} {\bibfield  {journal}
  {\bibinfo  {journal} {Phys. Rev. A}\ }\textbf {\bibinfo {volume} {41}},\
  \bibinfo {pages} {5516--5527} (\bibinfo {year} {1990})}\BibitemShut {NoStop}%
\bibitem [{\citenamefont {Golden}\ and\ \citenamefont
  {Kalman}(2000)}]{doi:10.1063/1.873814}%
  \BibitemOpen
  \bibfield  {author} {\bibinfo {author} {\bibfnamefont {K.~I.}\ \bibnamefont
  {Golden}}\ and\ \bibinfo {author} {\bibfnamefont {G.~J.}\ \bibnamefont
  {Kalman}},\ }\bibfield  {title} {\enquote {\bibinfo {title} {Quasilocalized
  charge approximation in strongly coupled plasma physics},}\ }\href
  {https://doi.org/10.1063/1.873814} {\bibfield  {journal} {\bibinfo  {journal}
  {Phys. Plasmas}\ }\textbf {\bibinfo {volume} {7}},\ \bibinfo {pages} {14--32}
  (\bibinfo {year} {2000})}\BibitemShut {NoStop}%
\bibitem [{\citenamefont {Khrapak}(2016)}]{10.1063/1.4965903}%
  \BibitemOpen
  \bibfield  {author} {\bibinfo {author} {\bibfnamefont {S.~A.}\ \bibnamefont
  {Khrapak}},\ }\bibfield  {title} {\enquote {\bibinfo {title} {{Onset of
  negative dispersion in one-component-plasma revisited}},}\ }\href
  {https://doi.org/10.1063/1.4965903} {\bibfield  {journal} {\bibinfo
  {journal} {Phys. Plasmas}\ }\textbf {\bibinfo {volume} {23}},\ \bibinfo
  {pages} {104506} (\bibinfo {year} {2016})}\BibitemShut {NoStop}%
\bibitem [{\citenamefont {Khrapak}\ and\ \citenamefont
  {Cou\"edel}(2020)}]{PhysRevE.102.033207}%
  \BibitemOpen
  \bibfield  {author} {\bibinfo {author} {\bibfnamefont {S.}~\bibnamefont
  {Khrapak}}\ and\ \bibinfo {author} {\bibfnamefont {L.}~\bibnamefont
  {Cou\"edel}},\ }\bibfield  {title} {\enquote {\bibinfo {title} {Dispersion
  relations of {Y}ukawa fluids at weak and moderate coupling},}\ }\href
  {https://doi.org/10.1103/PhysRevE.102.033207} {\bibfield  {journal} {\bibinfo
   {journal} {Phys. Rev. E}\ }\textbf {\bibinfo {volume} {102}},\ \bibinfo
  {pages} {033207} (\bibinfo {year} {2020})}\BibitemShut {NoStop}%
\bibitem [{\citenamefont {Murillo}(1998)}]{doi:10.1063/1.873037}%
  \BibitemOpen
  \bibfield  {author} {\bibinfo {author} {\bibfnamefont {M.~S.}\ \bibnamefont
  {Murillo}},\ }\bibfield  {title} {\enquote {\bibinfo {title} {Static local
  field correction description of acoustic waves in strongly coupling dusty
  plasmas},}\ }\href {https://doi.org/10.1063/1.873037} {\bibfield  {journal}
  {\bibinfo  {journal} {Phys. Plasmas}\ }\textbf {\bibinfo {volume} {5}},\
  \bibinfo {pages} {3116--3121} (\bibinfo {year} {1998})}\BibitemShut {NoStop}%
\bibitem [{\citenamefont {Kaw}\ and\ \citenamefont
  {Sen}(1998)}]{doi:10.1063/1.873073}%
  \BibitemOpen
  \bibfield  {author} {\bibinfo {author} {\bibfnamefont {P.~K.}\ \bibnamefont
  {Kaw}}\ and\ \bibinfo {author} {\bibfnamefont {A.}~\bibnamefont {Sen}},\
  }\bibfield  {title} {\enquote {\bibinfo {title} {Low frequency modes in
  strongly coupled dusty plasmas},}\ }\href {https://doi.org/10.1063/1.873073}
  {\bibfield  {journal} {\bibinfo  {journal} {Phys. Plasmas}\ }\textbf
  {\bibinfo {volume} {5}},\ \bibinfo {pages} {3552--3559} (\bibinfo {year}
  {1998})}\BibitemShut {NoStop}%
\bibitem [{\citenamefont {Diaw}\ and\ \citenamefont
  {Murillo}(2015)}]{PhysRevE.92.013107}%
  \BibitemOpen
  \bibfield  {author} {\bibinfo {author} {\bibfnamefont {A.}~\bibnamefont
  {Diaw}}\ and\ \bibinfo {author} {\bibfnamefont {M.~S.}\ \bibnamefont
  {Murillo}},\ }\bibfield  {title} {\enquote {\bibinfo {title} {Generalized
  hydrodynamics model for strongly coupled plasmas},}\ }\href
  {https://doi.org/10.1103/PhysRevE.92.013107} {\bibfield  {journal} {\bibinfo
  {journal} {Phys. Rev. E}\ }\textbf {\bibinfo {volume} {92}},\ \bibinfo
  {pages} {013107} (\bibinfo {year} {2015})}\BibitemShut {NoStop}%
\bibitem [{\citenamefont {Jou}\ \emph {et~al.}(1985)\citenamefont {Jou},
  \citenamefont {P\'erez-Garc\'{\i}a}, \citenamefont {Garc\'{\i}a-Col\'{\i}n},
  \citenamefont {L\'opez~de Haro},\ and\ \citenamefont
  {Rodr\'{\i}guez}}]{PhysRevA.31.2502}%
  \BibitemOpen
  \bibfield  {author} {\bibinfo {author} {\bibfnamefont {D.}~\bibnamefont
  {Jou}}, \bibinfo {author} {\bibfnamefont {C.}~\bibnamefont
  {P\'erez-Garc\'{\i}a}}, \bibinfo {author} {\bibfnamefont {L.~S.}\
  \bibnamefont {Garc\'{\i}a-Col\'{\i}n}}, \bibinfo {author} {\bibfnamefont
  {M.}~\bibnamefont {L\'opez~de Haro}},\ and\ \bibinfo {author} {\bibfnamefont
  {R.~F.}\ \bibnamefont {Rodr\'{\i}guez}},\ }\bibfield  {title} {\enquote
  {\bibinfo {title} {Generalized hydrodynamics and extended irreversible
  thermodynamics},}\ }\href {https://doi.org/10.1103/PhysRevA.31.2502}
  {\bibfield  {journal} {\bibinfo  {journal} {Phys. Rev. A}\ }\textbf {\bibinfo
  {volume} {31}},\ \bibinfo {pages} {2502--2508} (\bibinfo {year}
  {1985})}\BibitemShut {NoStop}%
\bibitem [{\citenamefont {Chen}(2019)}]{chen2015introduction}%
  \BibitemOpen
  \bibfield  {author} {\bibinfo {author} {\bibfnamefont {F.~F.}\ \bibnamefont
  {Chen}},\ }\href@noop {} {\emph {\bibinfo {title} {Introduction to Plasma
  Physics and Controlled Fusion}}},\ \bibinfo {edition} {3rd}\ ed.\ (\bibinfo
  {publisher} {Springer},\ \bibinfo {year} {2019})\BibitemShut {NoStop}%
\bibitem [{\citenamefont {Landau}\ and\ \citenamefont
  {Lifshitz}(1980{\natexlab{b}})}]{landau2013statistical}%
  \BibitemOpen
  \bibfield  {author} {\bibinfo {author} {\bibfnamefont {L.~D.}\ \bibnamefont
  {Landau}}\ and\ \bibinfo {author} {\bibfnamefont {E.~M.}\ \bibnamefont
  {Lifshitz}},\ }\href@noop {} {\emph {\bibinfo {title} {Statistical Physics:
  Volume 5 (Course of Theoretical Physics)}}},\ \bibinfo {edition} {3rd}\ ed.\
  (\bibinfo  {publisher} {Butterworth-Heinemann},\ \bibinfo {year}
  {1980})\BibitemShut {NoStop}%
\bibitem [{\citenamefont {Fairushin}, \citenamefont {Khrapak},\ and\
  \citenamefont {Mokshin}(2020)}]{FAIRUSHIN2020103359}%
  \BibitemOpen
  \bibfield  {author} {\bibinfo {author} {\bibfnamefont {I.}~\bibnamefont
  {Fairushin}}, \bibinfo {author} {\bibfnamefont {S.}~\bibnamefont {Khrapak}},\
  and\ \bibinfo {author} {\bibfnamefont {A.}~\bibnamefont {Mokshin}},\
  }\bibfield  {title} {\enquote {\bibinfo {title} {Direct evaluation of the
  physical characteristics of {Y}ukawa fluids based on a simple approximation
  for the radial distribution function},}\ }\href
  {https://doi.org/https://doi.org/10.1016/j.rinp.2020.103359} {\bibfield
  {journal} {\bibinfo  {journal} {Results Phys.}\ }\textbf {\bibinfo {volume}
  {19}},\ \bibinfo {pages} {103359} (\bibinfo {year} {2020})}\BibitemShut
  {NoStop}%
\bibitem [{\citenamefont {Goldstein}, \citenamefont {Poole},\ and\
  \citenamefont {Safko}(2001)}]{goldstein:mechanics}%
  \BibitemOpen
  \bibfield  {author} {\bibinfo {author} {\bibfnamefont {H.}~\bibnamefont
  {Goldstein}}, \bibinfo {author} {\bibfnamefont {C.}~\bibnamefont {Poole}},\
  and\ \bibinfo {author} {\bibfnamefont {J.}~\bibnamefont {Safko}},\
  }\href@noop {} {\emph {\bibinfo {title} {Classical Mechanics}}},\ \bibinfo
  {edition} {3rd}\ ed.\ (\bibinfo  {publisher} {Pearson},\ \bibinfo {year}
  {2001})\BibitemShut {NoStop}%
\bibitem [{\citenamefont {Berestetskii}, \citenamefont {Lifshitz},\ and\
  \citenamefont {Pitaevskii}(1982)}]{berestetskii1982quantum}%
  \BibitemOpen
  \bibfield  {author} {\bibinfo {author} {\bibfnamefont {V.~B.}\ \bibnamefont
  {Berestetskii}}, \bibinfo {author} {\bibfnamefont {E.~M.}\ \bibnamefont
  {Lifshitz}},\ and\ \bibinfo {author} {\bibfnamefont {L.~P.}\ \bibnamefont
  {Pitaevskii}},\ }\href@noop {} {\emph {\bibinfo {title} {Quantum
  Electrodynamics: Volume 4 (Course of Theoretical Physics)}}},\ \bibinfo
  {edition} {2nd}\ ed.\ (\bibinfo  {publisher} {Pergamon},\ \bibinfo {year}
  {1982})\BibitemShut {NoStop}%
\bibitem [{\citenamefont {Chorin}\ and\ \citenamefont
  {Marsden}(1993)}]{chorin2013mathematical}%
  \BibitemOpen
  \bibfield  {author} {\bibinfo {author} {\bibfnamefont {A.~J.}\ \bibnamefont
  {Chorin}}\ and\ \bibinfo {author} {\bibfnamefont {J.~E.}\ \bibnamefont
  {Marsden}},\ }\href@noop {} {\emph {\bibinfo {title} {A Mathematical
  Introduction to Fluid Mechanics}}},\ \bibinfo {edition} {3rd}\ ed.\ (\bibinfo
   {publisher} {Springer},\ \bibinfo {year} {1993})\BibitemShut {NoStop}%
\bibitem [{\citenamefont {Frahm}(1983)}]{10.1119/1.13127}%
  \BibitemOpen
  \bibfield  {author} {\bibinfo {author} {\bibfnamefont {C.~P.}\ \bibnamefont
  {Frahm}},\ }\bibfield  {title} {\enquote {\bibinfo {title} {{Some novel
  delta‐function identities}},}\ }\href {https://doi.org/10.1119/1.13127}
  {\bibfield  {journal} {\bibinfo  {journal} {Am. J. Phys.}\ }\textbf {\bibinfo
  {volume} {51}},\ \bibinfo {pages} {826--829} (\bibinfo {year}
  {1983})}\BibitemShut {NoStop}%
\end{thebibliography}%

\end{document}